\documentclass[twocolumn,astrosymb]{aastex701}
\hypersetup{linkcolor=red,citecolor=blue,filecolor=cyan,urlcolor=magenta}

\newcommand{\cMpch}{$h^{-1}$~cMpc}
\newcommand{\msol}{M_{\odot}}
\newcommand{\dFsm}{\delta_F / \sigma_{\rm map}}

\received{2024 Nov 27}
\revised{2025 May 21}
\accepted{2025 May 26}

\shorttitle{LATIS: IGM-Selected Overdensities}
\shortauthors{Newman et al.}

\begin{document}

\title{LATIS: A Sample of IGM-Selected Protoclusters and Protogroups at $z \sim 2.5$}

\correspondingauthor{Andrew B. Newman}
\email{anewman@carnegiescience.edu}
\author[0000-0001-7769-8660]{Andrew B. Newman}
\affiliation{Observatories of the Carnegie Institution for Science, 813 Santa Barbara Street, Pasadena, CA 91101, USA}
\email{anewman@carnegiescience.edu}

\author[0000-0001-7066-1240]{Mahdi Qezlou}
\affiliation{The University of Texas at Austin, 2515 Speedway Boulevard, Stop C1400, Austin, Texas 78712, USA}
\email{qezlou@austin.utexas.edu}

\author[0000-0002-8459-5413]{Gwen C. Rudie}
\affiliation{Observatories of the Carnegie Institution for Science, 813 Santa Barbara Street, Pasadena, CA 91101, USA}
\email{gwen@carnegiescience.edu}

\author[0000-0003-3691-937X]{Nima Chartab}
\affiliation{Observatories of the Carnegie Institution for Science, 813 Santa Barbara Street, Pasadena, CA 91101, USA}
\affiliation{Caltech/IPAC, 1200 E. California Boulevard, Pasadena, CA 91125, USA}
\email{nchartab@ipac.caltech.edu}

\author[0000-0003-4218-3944]{Guillermo A. Blanc}
\affiliation{Observatories of the Carnegie Institution for Science, 813 Santa Barbara Street, Pasadena, CA 91101, USA}
\affiliation{Departamento de Astronomía, Universidad de Chile, Camino del Observatorio 1515, Las Condes, Santiago, Chile}
\email{gblanc@carnegiescience.edu}

\author[0000-0003-4727-4327]{Daniel D. Kelson}
\affiliation{Observatories of the Carnegie Institution for Science, 813 Santa Barbara Street, Pasadena, CA 91101, USA}
\email{kelson@carnegiescience.edu}

\author[0000-0001-5803-5490]{Simeon Bird}
\affiliation{Department of Physics and Astronomy, University of California Riverside, 900 University Avenue, Riverside, CA 92521, USA}
\email{sbird@ucr.edu}

\author[0000-0002-0930-6466]{Caitlin Casey}
\affiliation{The University of Texas at Austin, 2515 Speedway Boulevard, Stop C1400, Austin, Texas 78712, USA}
\email{cmcasey@utexas.edu}

\author{Enrico Congiu}
\affiliation{European Southern Observatory (ESO), Alonso de Córdova 3107, Casilla 19, Santiago 19001, Chile}
\email{econgiu@eso.org}

\author[0000-0002-9336-7551]{Olga Cucciati}
\affiliation{INAF - Osservatorio di Astrofisica e Scienza dello Spazio di Bologna, via Gobetti 93/3, 40129 Bologna, Italy}
\email{olga.cucciati@inaf.it}

\author[0000-0001-7523-140X]{Denise Hung}
\affiliation{University of Hawai’i, Institute for Astronomy, 2680 Woodlawn Drive, Honolulu, HI 96822, USA}
\affiliation{Gemini Observatory, NSF NOIRLab, 670 N. A’ohoku Place, Hilo,
Hawai’i, 96720, USA}
\email{denise.hung@noirlab.edu}

\author[0000-0002-1428-7036]{Brian C. Lemaux}
\affiliation{Gemini Observatory, NSF NOIRLab, 670 N. A’ohoku Place, Hilo,
Hawai’i, 96720, USA}
\affiliation{Department of Physics and Astronomy, University of California, Davis, One Shields Avenue, Davis, CA 95616, USA}
\email{brian.lemaux@noirlab.edu}

\author{Victoria P\'{e}rez}
\affiliation{Departamento de Astronomía, Universidad de Chile, Camino del Observatorio 1515, Las Condes, Santiago, Chile}
\email{victoriapaz.perezgonzalez@gmail.com}

\author[0000-0002-7051-1100]{Jorge Zavala}
\affiliation{National Astronomical Observatory of Japan, 2-21-1 Osawa, Mitaka, Tokyo 181-8588, Japan}
\email{jorge.zavala@nao.ac.jp}

\begin{abstract}
The Ly$\alpha$ Tomography IMACS Survey (LATIS) has produced large 3D maps of the intergalactic medium (IGM), providing a new window on the cosmic web at $z\sim2.5$. A key advantage of Ly$\alpha$ tomography is that it enables the discovery of overdense regions without the need to detect their galaxy members in spectroscopic surveys, circumventing possible selection biases. We use these maps to identify 37 IGM-selected overdensities as regions of strong and spatially coherent Ly$\alpha$ absorption. Simulations indicate that 85\% of these are protoclusters, defined as the progenitors of $z=0$ halos with mass $M_{\rm desc} > 10^{14} \msol$, and that nearly all of the rest are protogroups ($10^{13.5} < M_{\rm desc} / \msol < 10^{14}$). We estimate the masses and space densities of the IGM-selected overdensities and show they are in accordance with mock surveys. We investigate the LATIS counterparts of some previously reported protoclusters, including the proto-supercluster Hyperion. We identify a new component of Hyperion beyond its previously known extent. We show that the Ly$\alpha$ transmission of the galaxy density peaks within Hyperion is consistent with a simple physical model (the fluctuating Gunn--Peterson approximation), suggesting that active galactic nucleus feedback or other processes have not affected the large-scale gas ionization within this structure as whole. The LATIS catalog represents an order-of-magnitude increase in the number of IGM-selected protogroups and protoclusters and will enable new investigations of the connections between galaxies and their large-scale environments at cosmic noon.
\end{abstract}

\section{Introduction}
\label{sec:intro}

Protoclusters, the diffuse regions at $z \gtrsim 2$ destined to collapse into galaxy clusters today, play several interesting roles in the study of early galaxy evolution. They are expected to contain the most massive halos, those where galaxy formation began the earliest and, later, where the mode of gas accretion transitioned first \citep{Overzier16}. Protoclusters contain a rapidly increasing fraction of the cosmic star-formation rate density toward higher redshifts and likely played a key role in the reionization of the intergalactic medium (IGM; \citealt{Chiang17,Hu21,Yajima22}). Toward cosmic noon, models anticipate that star formation began declining earlier and more quickly in protoclusters \citep{Muldrew18}. By $z \sim 2$, at least some systems show the hallmarks of galaxy clusters in the nearby universe: a red sequence of galaxies and a hot gaseous medium \citep{Newman14,Willis20}, whose thermodynamic state in turn is affected by the galaxies evolving within it. Observations of protoclusters can inform when, where, and why galaxies and their Mpc-scale environments began to affect one another's evolution.

In the broadest terms, such a project requires locating representative samples of protoclusters, assembling a representative inventory of their galaxy contents, measuring the physical properties of the member galaxies, and comparing these to control samples in other environments and to theoretical models. Each of these steps rapidly becomes more difficult with increasing redshift. Spectroscopic galaxy surveys can provide the most reliable determination of a galaxy's membership in a protocluster. But particularly at $z \gtrsim 1.5$, spectroscopic samples of galaxies are generally far from mass-limited, because the identification of a redshift relies on spectral features (e.g., ultraviolet continuum, nebular line emission) whose detectability depends mainly on a galaxy's star formation activity and dust attenuation.

Such limitations clearly make it difficult to perform a complete spectroscopic census of protocluster galaxies, including obscured galaxies and those with little or no star formation. The remarkable sensitivity of JWST can greatly improve this situation through targeted follow-up of select regions. However, wide-area surveys are still needed to locate sizable samples of protoclusters, and selection effects could also cause these surveys to miss or mischaracterize those with a substantial galaxy population that is not included in the spectroscopic selection. It is instructive that even within COSMOS, the most thoroughly observed extragalactic field for almost two decades, two protoclusters at $z \sim 3$ with prominent quiescent galaxy populations were only recently uncovered \citep{McConachie22,Ito23}.

These considerations motivate the need for methods to discover and characterize protoclusters that complement galaxy surveys. At $z \lesssim 2$, the detection of hot gas in an intracluster medium through its X-ray emission or the Sunyaev--Zel'dovich effect has provided such a means \citep[e.g.,][]{Brodwin11,Brodwin12,Gobat11,Andreon14,Andreon23,Mantz14,Mantz18,Mantz20,Willis20,DiMascolo23}. But beyond $z \sim 2.2$ these methods have been much less productive, turning up one controversial example at $z = 2.5$ \citep{Wang16,Champagne21}. 

Fortunately, the redshifted Ly$\alpha$ transition becomes observable from the ground at $z \gtrsim 2$. The Ly$\alpha$ forest absorption provides a tracer of diffuse neutral hydrogen in the IGM. By observing the Ly$\alpha$ transmission in a dense network of sight lines, the 3D structure of the IGM can be mapped and related to the density field, a technique known as IGM or Ly$\alpha$ forest tomography \citep{Pichon01,Caucci08}. Mapping the IGM with a resolution of several comoving megaparsecs (cMpc) in radius, roughly the typical extent of protoclusters \citep{Chiang14}, requires observing the Ly$\alpha$ forest in the spectra of faint Lyman-break galaxies (LBGs; \citealt{Lee14A}) to achieve a commensurate sight line density. Simulations show that the resulting Ly$\alpha$ tomographic maps can effectively be used to locate protoclusters and to estimate their masses and evolution \citep{Stark15,Qezlou22}. 

The COSMOS Lyman-Alpha Mapping and Tomography Observations (CLAMATO) survey was the first to implement the Ly$\alpha$ tomography technique using galaxy spectra \citep{Lee14B}. The CLAMATO IGM maps revealed filaments, nodes, sheets, and voids within the $z\sim2.3$ cosmic web over a volume of $4 \times 10^5$ $h^{-3}$ cMpc${}^3$ \citep{Lee18,Horowitz22}. Several previously known protoclusters were detected through Ly$\alpha$ absorption \citep{Lee16,Lee18}. The CLAMATO maps have also been used to study the frequency and structure of cosmic voids \citep{Krolewski18}. 

Another approach is to use quasar spectra to sparsely sample the IGM. \citet{Cai16} showed that strong Ly$\alpha$ absorption spanning 15 \cMpch~along a single line of sight is expected to be associated with large-scale matter overdensities. Contamination by high column density (HCD) absorption lines, which trace dense gas near galaxies rather than the diffuse IGM, is a major concern that can be mitigated by locating groups of several quasars that show coherent absorption. This method is expected to miss all but a tiny fraction of protoclusters \citep{Miller19}, but its advantage is that enormous volumes can be surveyed to access the rarest overdensities. Indeed, the MAMMOTH survey \citep{Cai16} based on this method has been quite successful in discovering several extremely rich protoclusters around $z \sim 2.3$ \citep{Cai17B,ArrigoniBattaia18,Zheng21,Shi21}, although follow-up observations have not always revealed a significant galaxy overdensity \citep{Liang21}. 

The Lyman-$\alpha$ Tomography IMACS Survey (LATIS; \citealt{Newman20}) builds on the techniques demonstrated by CLAMATO by extending them to a $10\times$ larger volume while maintaining a comparable density of sight lines. In this paper, we will use the complete LATIS maps to identify and characterize 37 IGM-selected overdensities at $z=2.2$-2.8, most of which are expected to be protoclusters. This catalog represents a roughly tenfold increase in the number of protoclusters identified independently of their galaxy populations.

Using a partially complete LATIS data set, \citet{Newman22} investigated the regions of strongest Ly$\alpha$ absorption and showed that they contained fewer LBGs than expected, a deficit large enough to prevent the recognition of many of these structures in previous galaxy spectroscopic surveys. The present paper will focus on the sample of IGM-selected overdensities in the full LATIS survey and their connection to known protoclusters. In a companion paper (\citealt{Newman25}, hereafter N25), we will quantify trends in the LBG content of the IGM-selected overdensities and update the \citet{Newman22} results using the full sample.

The paper is organized as follows. We first construct IGM maps along with galaxy density maps in the three LATIS survey fields (Section~\ref{sec:data}). We then locate IGM-selected overdensities within these maps and characterize their masses and future evolution using tailored methods developed using cosmological simulations \citep{Qezlou22}. We validate the resulting catalog of 37 IGM-selected overdensities and present several visualizations of these structures (Section~\ref{sec:catalog}). We then investigate previously reported protoclusters within our maps, especially the COSMOS field (Section~\ref{sec:litcompare}) and the Hyperion proto-supercluster (\citealt{Cucciati18}, and see references in Section~\ref{sec:hyperion}). Throughout we adopt the \citet{Planck15} cosmological parameters and report magnitudes in the AB system. 

\section{Data}
\label{sec:data}

\subsection{Observations}

The LATIS maps used in this paper cover the full survey area described by \citet{Newman20}, which spans 1.65 deg${}^2$ across the three fields: COSMOS and the Canada--France--Hawaii Legacy Survey (CFHTLS) D1 and D4 fields. (We also refer to COSMOS as D2 in our naming scheme since it contains the CFHTLS-D2 field.) We measure the Ly$\alpha$ forest in 3012 sight lines toward LBGs and QSOs comprising $4.7 \times 10^5$ spectral pixels. Each pixel is a measure of the Ly$\alpha$ flux contrast, or transmission fluctuation, $\delta_F = F / \langle F \rangle - 1$, where $\langle F \rangle$ represents the mean transmitted flux at each redshift. Maps are created by applying a Wiener filter to these pixels, using the {\tt dachshund} code by \citet{Stark15}. We note that the output of the Wiener filter depends on the model used to describe the signal covariance, and we use the same parameters as \citet{Newman20} and \citet{Qezlou22}. We smooth the Wiener filter output using a Gaussian kernel with $\sigma_{\rm sm} = 4$~\cMpch, as in our previous work \citep{Newman20,Newman22}. This kernel is somewhat larger than the mean sight line separation, which varies with redshift from $\langle d_{\perp} \rangle \approx 2.5$-4~\cMpch, which serves to increase the signal-to-noise ratio and to better match the spatial scale of protoclusters (see Section~\ref{sec:catalog}). We normalize the resulting 3D maps of $\delta_F$ by their standard deviation $\sigma_{\rm map}$.\footnote{$\sigma_{\rm map} = 0.050$, 0.048, and 0.046 in the D1, D2, and D4 fields, respectively. To mitigate edge effects, $\sigma_{\rm map}$ is computed excluding voxels within 4~\cMpch~of a map edge.} The LATIS maps maps span $z=2.2$-2.8 with cubical voxels of 1 \cMpch~on a side. The total volume enclosed by the survey footprint between $z=2.2$-2.8 is $3.96 \times 10^6$ $h^{-3}$ cMpc${}^3 \approx 10^7$ cMpc${}^{3}$. 

\begin{figure*}
    \centering
    \includegraphics[width=5in]{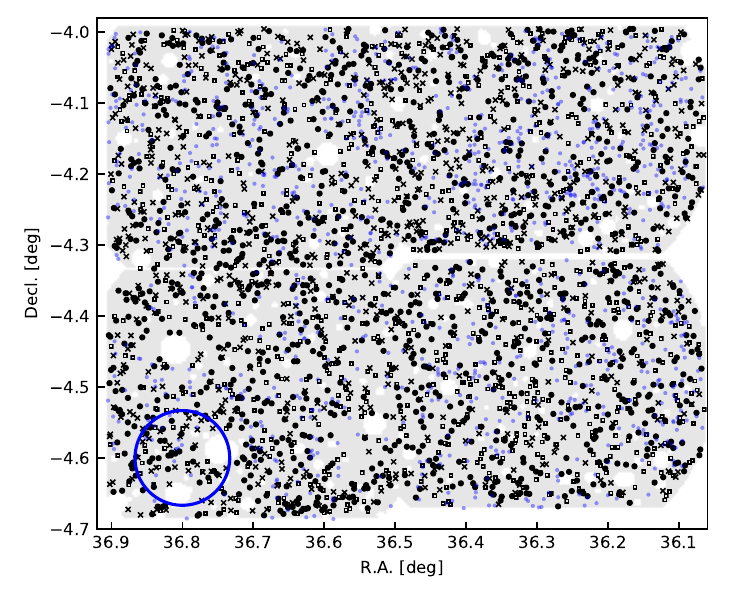}
    \caption{Map of LATIS sight lines and galaxies/QSOs in the D1 survey field. Black symbols show the 2832 targets with symbols encoding redshift range and confidence. Filled circles show Ly$\alpha$ sight lines along with galaxies contained within the $z=2.2$-2.8 range of the tomographic maps; open squares show other galaxies with high-confidence redshifts, mostly at $z < 2.2$; crosses show observed galaxies for which a high-confidence redshift was not determined; and blue circles show unobserved galaxies in the photometric parent sample with $r < 24.4$, in the high-priority magnitude range. As discussed in \citet{Newman20}, the target sampling rate for the second-tier $r = 24.4$-24.8 targets is much lower. The gray background shows the survey mask, including the boundaries of the four IMACS footprints as well as untargeted regions around bright stars. The blue circle in the lower-left corner shows the FWHM of the our 4~\cMpch~smoothing kernel at $z=2.5$.\label{fig:galdist}}
\end{figure*}

In addition to the IGM tomographic maps, we map the galaxy distribution. There are 2570 LBGs and QSOs with LATIS redshifts that lie within the IGM maps. Fig.~\ref{fig:galdist} shows the distribution of LATIS sight lines and galaxies in the D1 field in order to illustrate the level of sampling and completeness (see also \citealt{Newman20} for initial targeting statistics and Newman et al. 2025, in preparation, for full-survey statistics). To measure galaxy densities, we exclude sources without a high-confidence redshift (requiring {\tt zqual} = 3 or 4; \citealt{Newman20}), those that were observed only as part of ``bright target'' masks intended for poor observing conditions, and broad absorption line (BAL) QSOs. The LBG systemic redshifts are calibrated against transverse Ly$\alpha$ absorption \citep{Newman24}. Galaxy density maps are created by weighting each galaxy by the inverse of the effective sampling rate (ESR), which is a function of the field position \citep{Newman24}. We then convolve the resulting map by a Gaussian with $\sigma_{\rm sm} = 4$ $h^{-1}$ cMpc, matching the kernel used for the IGM maps. The galaxy space density $n$ is finally converted to an overdensity $\delta_{\rm LBG} = n / \langle n \rangle - 1$ by computing the mean density $\langle n \rangle$ within each survey field separately. This is appropriate because we found that the mean density does not vary significantly over the redshift range $z=2.2$-2.8. In prior work we referred to the galaxy overdensity as $\delta_{\rm gal}$, but here and in the companion paper we use the more specific symbol $\delta_{\rm LBG}$, because distinctions among subpopulations are of interest. (Non-BAL quasars are also included in our maps and $\delta_{\rm LBG}$ measures, but they comprise only 2\% of the sample.) Both the IGM and LBG maps are in redshift space.

For quantitative estimates of $\delta_{\rm LBG}$, we must account for the survey boundary and unobserved regions around bright stars. We do this by convolving the survey mask by the same kernel and dividing the galaxy space density $n$ map by the result. If this missing-volume correction factor exceeds 2, as is the case near the edges, we report no $\delta_{\rm LBG}$ at the position. We neglect the survey mask when creating 3D visualizations of $\delta_{\rm LBG}$ contours, since otherwise the increased noise near the survey boundary can obscure other regions.

\subsection{Mock Surveys}
\label{sec:mocks}

Mock surveys that carefully mimic LATIS within cosmological simulations are essential to properly interpret the observed maps. As described by \citet{Newman20,Newman22,Newman24}, we primarily rely on the MultiDark Planck 2 (MDPL2) 1 $h^{-3}$ Gpc${}^3$ N-body simulation \citep{Klypin16}, which offers a large volume but does not track gas. We estimate the Ly$\alpha$ transmission using the fluctuating Gunn-Peterson approximation (FGPA; \citealt{Gunn65,Croft98,Weinberg99}). Although the FGPA is a highly simplified treatment of the IGM physics, it works well at the resolution of our tomographic maps, producing a distribution of transmission fluctuations that agrees with LATIS observations \citep{Newman20,Newman22} and matching the relationship between Ly$\alpha$ transmission and matter density in magnetohydrodynamical simulations \citep{Qezlou22}.

The MDPL2 mocks are created as described by \citet{Newman22}. Briefly, we use the snapshot at $z=2.535$ near the midpoint of the LATIS maps. We create a suite of mock surveys by selecting nearly independent subvolumes that match the dimensions of the 3 LATIS survey fields. Sight lines are selected following the exact spatial distribution of the LATIS sight lines, and spectra are generated mimicking the LATIS spectral resolution, noise properties, and processing steps (see \citealt{Newman24}). IGM tomographic maps are then reconstructed using a Wiener filter following the same procedures applied to the observations. We generate 100 mocks per field (COSMOS, D1, D4) and can randomly sample combinations of these to generate a larger number of survey realizations.

In addition to the mock IGM maps, we also produce mock galaxy density fields. We select mock galaxies by randomly drawing halos among those with masses $\log M_{\rm vir} / M_{\odot} > 11.56$; this threshold matches the halo autocorrelation function to that of the observed LATIS galaxies \citep{Newman24}. The probability of a halo being selected is proportional to the ESR, evaluated at the corresponding R.A.~and decl.~of the survey field being mocked. Probabilities are scaled such that, on average, we select the same number of halos as LATIS galaxies that were observed in a given survey field. Halo overdensities $\delta_{\rm halo}$ in redshift space are then computed like the galaxy overdensities. 

One application of the mock surveys is to estimate noise in the maps, which we will use to quantify the detection significance of Ly$\alpha$ absorption peaks and the uncertainties in some of their properties. We compare the reconstructed flux fields to a noiseless one, which is computed by directly convolving the full grid of noiseless sight lines by a Gaussian kernel ($\sigma = 4$~\cMpch) with no Wiener filtering. (We consider this ``noiseless'' in the sense that no observational noise is injected and the sight lines are not subsampled.) The comparison is slightly complicated by the fact that the reconstructed and noiseless $\delta_F$ do not have an exactly one-to-one relationship, even in conditional expectation, i.e., $\langle \delta_F^{\rm noiseless} | \delta_F^{\rm rec} \rangle \neq \delta_F^{\rm rec}$ and $\langle \delta_F^{\rm rec} | \delta_F^{\rm noiseless} \rangle \neq \delta_F^{\rm noiseless}$ \citep[e.g.,][]{Lee14A, Newman20}. We find $\langle \delta_F^{\rm rec} / \sigma | \delta_F^{\rm noiseless} / \sigma \rangle \approx 0.73 \times \delta_F^{\rm noiseless} / \sigma$, where $\sigma$ refers to the standard deviation of the true or reconstructed map. Using this relation, we compute the difference between the reconstructed $\delta_F^{\rm rec}$ and its expectation value at each voxel of each mock survey. Each mock survey thereby yields one noise realization. To evaluate the noise in the map at a given voxel, we compute the standard deviation of these differences (at the same voxel position) among all mock survey realizations.

\section{A Catalog of IGM-selected Overdensities}
\label{sec:catalog}

We identify matter overdensities in the LATIS IGM maps by locating regions of strong, spatially coherent Ly$\alpha$ absorption. Perhaps the simplest method, which we use, is to smooth the Wiener-filtered maps (Section~\ref{sec:data}) and apply a threshold, selecting local minima of the transmitted Ly$\alpha$ flux where $\dFsm$ falls below the threshold. (Note that more negative values of $\delta_F$ indicate more absorption.) We refer to these local minima as Ly$\alpha$ absorption peaks or IGM-selected overdensities. More sophisticated techniques have aimed to reconstruct the density field and its evolution using fast simulations or analytic techniques, taking Ly$\alpha$ forest or galaxy survey observations as the constraints (e.g., \citealt{Horowitz19, Ata21}). Although future application of such techniques to LATIS is promising, in this initial exploration we rely on the Wiener filtering method, which has the virtues of being simple and easily reproducible. In this section we discuss the nature of Ly$\alpha$ absorption peaks in simulated surveys, select a sample from the LATIS maps, quantify and evaluate their masses and densities, visualize the overdensities within the IGM and galaxy maps, and consider the robustness of the sample.

\subsection{Ly$\alpha$ Absorption Peaks in Simulated Surveys}

\citet{Stark15} showed that smoothing and thresholding is an effective technique for identifying protoclusters in Ly$\alpha$ tomographic maps, given appropriate choices of the smoothing scale and the threshold $\delta_F$. As mentioned in Section~\ref{sec:data}, we smooth the LATIS maps with a Gaussian having $\sigma_{\rm kern} = 4$~\cMpch. This both increases the signal-to-noise ratio of the maps and approximates the average radial profile of Ly$\alpha$ absorption in simulated protoclusters derived by \citet{Stark15}, thus acting as a rough matched filter. \citet{Stark15} proposed a threshold $\delta_F / \sigma_{\rm map} < -3.5$ for identifying protoclusters. They found this to strike a good balance between purity and completeness. For a LATIS-like mean sight line density, mock surveys by \citet{Stark15} indicated that 89\% of the selected absorption peaks are protoclusters, defined as the progenitors of halos at $z=0$ with masses $> 10^{14} h^{-1}$ ${\rm M}_{\odot}$, and that most of the rest are nearly protoclusters. 

\citet{Lee16} used a looser threshold $\dFsm < -3$ to investigate structures in the CLAMATO maps. \citet{Qezlou22} explored an even lower threshold $\dFsm < -2.35$ in LATIS-like mock surveys within the IllustrisTNG300 simulation \citep{Nelson19}. These surveys mimicked the main observational characteristics of LATIS ($z = 2.5$, representative sight line density with $\langle d_{\rm perp} \rangle = 2.7$~\cMpch, representative noise in $\delta_F$) and analysis methods (including Wiener filtering and smoothing). Qezlou et al.~found that the strongest absorption peaks with $\dFsm < -3.5$ are associated with $z=0$ descendant halos that have a median virial mass $M_{\rm desc} = 10^{14.6}~\msol$. Of these, 93\% can be considered protoclusters based on a definition that $M_{\rm desc} > 10^{14}~\msol$, indicating a high level of purity. For slightly weaker absorption peaks with $-3.5 < \dFsm < -3.0$, the median is $M_{\rm desc} = 10^{14.4}~\msol$ and 81\% are protoclusters. Most of the peaks that ``fail'' to be protoclusters are nonetheless the progenitors of massive galaxy groups with $M_{\rm desc} = 10^{13.5-14.0}~{\rm M}_{\odot}$. 

We construct our catalog of IGM-selected overdensities using Ly$\alpha$ absorption peaks with $\dFsm < -3$. Based on the results described above, we expect this sample to be dominated by protoclusters (85\%) with nearly all (98\%) being protoclusters or protogroups. Rather than artificially dividing the sample into protogroup and protocluster candidates, we will present for each structure an estimate of the descendant mass at $z=0$, along with the matter overdensity and mass at the observed redshift. 

Although we expect the IGM-selected overdensities to comprise a reasonably pure sample of protoclusters, and a highly pure one if protogroups are included, the sample is far less complete. \citet{Qezlou22} also investigated completeness using LATIS-like mock surveys: for a selection threshold of $\dFsm < -2.35$, they found a  detection completeness of 10\%, 50\%, and 90\% for the progenitors of massive $z=0$ halos having $M_{\rm desc}/\msol = 10^{13.8}$, $10^{14.6}$, and $10^{15.0}$, respectively. 
For our purposes of building a robust catalog, we have adopted a stricter threshold, shifting the balance toward lower impurity (non-protocluster fraction) but higher incompleteness. By imposing our threshold $\dFsm < -3$ on the \citet{Qezlou22} analysis, we estimate a completeness of 10\%, 50\% and 90\% at $M_{\rm desc}/\msol = 10^{14.2}$, $10^{14.8}$, and $10^{15.1}$. The sources of incompleteness are discussed by \citet{Stark15} and \citet{Qezlou22}. Noiseless surveys fare only slightly better at the $\delta_F$ threshold used by Qezlou et al., so much of the incompleteness reflects the complex connection between the $z=2.5$ Ly$\alpha$ flux field, as viewed 4~\cMpch~resolution, and the assembled mass at $z=0$. Our stricter $\delta_F$ threshold cut reduces completeness further, e.g., by $2\times$ at $M_{\rm desc} \approx 10^{14.5} \msol$. We emphasize that the LATIS maps contain many more significantly detected Ly$\alpha$ absorption peaks (e.g., $-2.35 < \dFsm < -3$, corresponding to $\simeq3$-$5\sigma$ detection significance; see Section~\ref{sec:odprops}) than the strongest 37 peaks cataloged in Table~\ref{tab:catalog}, and many of these (54\%) are also expected to be protoclusters.

\begin{figure*}
    \centering
    \includegraphics[width=3.25in]{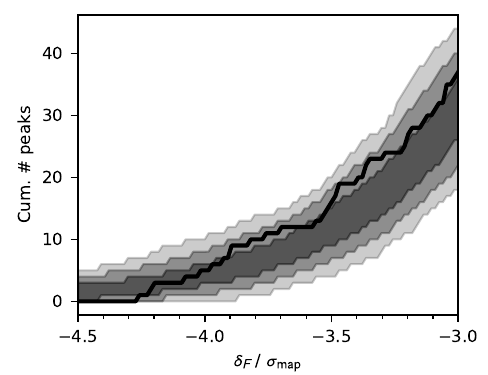} \hspace{0.5in}
    \includegraphics[width=3.25in]{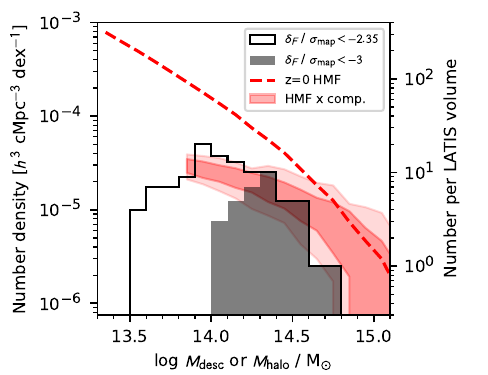}
    \caption{\emph{Left panel:} The cumulative number of LATIS absorption peaks as a function of $\delta_F$ (solid line) is compared to the envelope of curves seen in the mock surveys (shaded regions show the central 68\%, 95\%, and 99.7\%). \emph{Right panel:} The number density of Ly$\alpha$ absorption peaks as a function of the estimated descendant mass $M_{\rm desc}$. The open histogram shows all peaks with $\dFsm < -2.35$, the threshold used in the \citet{Qezlou22} prescription, while the filled histogram shows the subset with $\dFsm < -3$ that are the focus of this paper. The red dashed curve shows the $z=0$ halo mass function from MDPL2. After weighting halos by the mass-dependent completeness computed by \citet[][Fig.~13]{Qezlou22} and computing the number density in many LATIS-sized subvolumes to roughly estimate cosmic variance, we obtain the red bands (enclosing the central 68\% and 95\% of subvolumes) which can be compared to the open histogram.}
    \label{fig:hmf}
\end{figure*}

\begin{figure}
    \centering
    \includegraphics[width=3.2in]{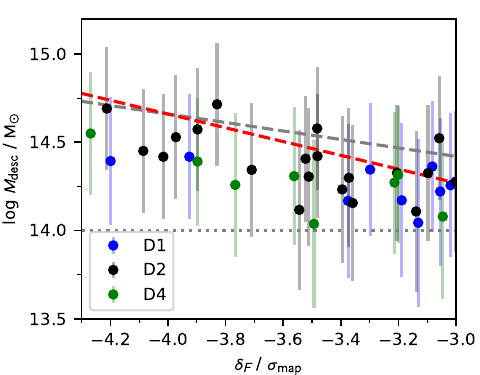}
    \caption{The Ly$\alpha$ flux contrast $\dFsm$ of the 37 IGM-selected overdensities is compared to the estimated mass $M_{\rm desc}$ of their most massive descendant ($z=0$) halo. Protoclusters are often defined as objects with $M_{\rm desc} > 10^{14} \msol$ (above dotted line). Colors denote the three survey fields. Dashed lines show the expectation value of $\log M_{\rm desc}$ conditional on $\delta_F$ from \citet[][red]{Qezlou22} and \citet[][gray]{Lee16}. \label{fig:signif_mdesc}}
\end{figure}

\subsection{IGM-selected Overdensities in LATIS}
\label{sec:odprops}

We find 37 IGM-selected overdensities satisfying $\dFsm < -3$ across the three LATIS fields (20, 9, and 8 in COSMOS, D1, and D4, respectively).\footnote{We made one modification to the \citet{Qezlou22} prescription: we considered absorption peaks separated by less than $\sigma_{\rm sm} = 4$ $h^{-1}$ cMpc to be totally blended, and we deleted the weaker peak to retain one structure in the catalog. This removed only one peak within the $\delta_F$ from our catalog.} Their properties are listed in Table~\ref{tab:catalog}, and their Ly$\alpha$ absorption and mass estimates (see below) are summarized in Figs.~\ref{fig:hmf} and \ref{fig:signif_mdesc}. These structures show non-zero Ly$\alpha$ absorption with a detection significance ranging from 4.0$\sigma$--7.1$\sigma$, defined as the ratio of $|\delta_F|$ to the noise in the map at the position of the absorption peak (Section~\ref{sec:mocks}). The detection significance differs from $\dFsm$, because $\sigma_{\rm map}$ includes large-scale structure and is not a measure of noise.

We use a naming convention similar to \citet{Newman22}: a prefix LATIS2 denoting the second  catalog described in this paper (the first being \citealt{Newman22}), the name of the field (note COSMOS $=$ D2), and an integer denoting the rank order of the absorption peak within its field, sorted by $\delta_F$ and beginning with zero. Thus the strongest peak in the COSMOS field in the current maps is LATIS2-D2-00.

\citet{Newman20} tabulated only the strongest absorption peaks with $\dFsm < -3.8$ in an earlier version of the maps. All of these structures, whose names begin with LATIS1, have counterparts in the current catalog, which are are listed in Table~\ref{tab:catalog}. However, the absorption strength and thus the rank order of these structures have shifted in some cases. These shifts are due to the introduction of new observations combined with refinements to the reductions and analysis methods. These changes are detailed, including a peak-by-peak comparison, in the companion paper (N25). Here we note simply that the $\dFsm$ of the LATIS1 structures has increased by 0.3, on average, and that the standard deviation of the earlier and present $\dFsm$ measurements is 0.4; this can be compared to the random error of 0.63. 

By adding noise realizations (Section~\ref{sec:mocks}) to the observed maps, we estimate uncertainties in the coordinates of the absorption peaks to be 2.7~\cMpch~per sky coordinate and 1.5~\cMpch~along the line of sight ($1\sigma$). These correspond to 2.3 arcmin and $\Delta z = 0.0018$ (150 km~s${}^{-1}$), respectively, at a redshift $z=2.5$.

\subsection{Densities and Masses}
\label{sec:identifying}

The smoothed $\delta_F$ is closely related to the matter overdensity $\delta_m$ \citep[e.g.,][]{Lee14A,Lee16,Qezlou22}. Using the MDPL2 mocks, we find that the expectation value of $\delta_m$ conditional on the observed $\delta_F^{\rm rec}$ is described by $\langle \delta_m | \delta_F^{\rm rec} \rangle = 10.6 (\delta_F^{\rm rec})^2 - 4.03 \delta_F^{\rm rec} - 0.04$, where both $\delta_m$ and $\delta_F$ are smoothed with a $\sigma = 4$~\cMpch~kernel.\footnote{This differs from the formula given by \citet{Qezlou22}, Equation 5, which refers to the matter overdensity evaluated in redshift space. The redshift-space overdensity is appropriate for the purpose of estimating masses from the maps, but here we provide $\delta_m$ in real space for easier comparison to other studies. We link voxels in the real-space $\delta_m$ and redshift-space $\delta_F$ maps using the mean line-of-sight peculiar velocity, although this correction ultimately makes little difference.} The variance about this estimator is approximately $\sigma_{\delta_m} = 0.2 (1 + \delta_m)$. We find that our 37 overdensities typically have $\delta_m \approx 1$, comparable to published estimates of other protoclusters when the overdensity is defined within a similar volume \citep[e.g.,][]{Steidel98,Shi19}. For comparison with other overdensity estimates, we note that our Gaussian kernel with $\sigma = 4$~\cMpch~has the same volume as a top-hat cube with a side length of 14.7 cMpc, and so is comparable to the $(15~{\rm cMpc})^3$ volume often used to measure protocluster overdensities \citep[e.g.,][]{Chiang14}. The dispersion in the smoothed density field at $z=2.5$ is $({\rm Var}~\delta_m)^{1/2} = 0.285$; thus, our sample consists of $\approx 3$-5$\sigma$ density fluctuations.

In addition to this point estimate of the density, the tomographic maps also contain information about the extent of structures. \citet{Qezlou22} developed a method to associate empirical volumes and masses to structures in Ly$\alpha$ tomographic maps and to relate these to physical quantities of interest. The method defines volumes associated with absorption peaks as isocontours of $\dFsm = -2$. When such a contour contains multiple absorption peaks that have $\dFsm < -2.35$, the contour volume is partitioned into ``watersheds'' associated with each such absorption peak using a simple algorithm borrowed from image segmentation. By integrating the  $\delta_m(\delta_F^{\rm rec})$ relation over a watershed, a raw tomographic mass $M_{\rm tomo, raw}$ can be estimated for each structure. \citet{Lee16} first introduced a similar method, although the masses are not directly comparable since they used a different contour level. After applying a small 0.14~dex offset (Equation 7) to $M_{\rm tomo, raw}$, \citet{Qezlou22} found that the resulting $M_{\rm tomo}$ is an excellent estimator ($\sigma = 0.12$~dex) of the dark matter mass $M_{\rm DM}$ contained within the volume. However, the volume itself (i.e., the $\dFsm = -2$ contour) is noisy, resulting in uncertainties of $\simeq$0.3~dex compared to an ideal observation.  We do not separately estimate $M_{\rm DM}$ from $M_{\rm tomo}$ following Equation 8 of \citet{Qezlou22}, because that relation corrects a bias that becomes significant only at lower masses. For our sample, $M_{\rm DM} \approx M_{\rm tomo}$.

We report $M_{\rm tomo}$ (Table~\ref{tab:catalog}) as an indicator of the mass of an IGM-selected overdensity at the observed redshift, and we estimate its uncertainty as the standard deviation of masses obtained when noise realizations are added to the observed maps. We emphasize that $M_{\rm tomo}$ is defined entirely empirically as the dark matter mass enclosed within a particular contour level in a smoothed $\delta_F$ map. It can be used to estimate other masses (see below), but it does not have a direct numerical equality with any theoretical mass (e.g., a bound mass, the most massive halo formed by the observed epoch or by $z=0$) or any mass that would be obtained by another observational technique (e.g., a dynamical mass, or a mass estimated from a galaxy overdensity and a known bias). The masses of the IGM-selected structure span a decade, from $\log M_{\rm tomo} / \msol = 14.2$-15.2. We also report the volume of the watershed, i.e., the volume within a $\dFsm < -2$ contour that is associated with a given absorption peak.

We use $M_{\rm tomo}$ to estimate $M_{\rm desc}$, the mass of the most massive descendant halo at $z=0$, using the prescription of \citet[][Equation 9]{Qezlou22} that follows on the work of \citet{Stark15} and \citet{Lee16}.\footnote{We correct an error in \citet{Qezlou22}, in which calculated $M_{\rm tomo}$ and $M_{\rm DM}$ values were too high by approximately a factor of $h^{-1}$.} $M_{\rm tomo}$ is larger than $M_{\rm desc}$, because not all material within our selected contour level will collapse into the main halo. We estimate uncertainties in $M_{\rm desc}$ of $\simeq 0.4$~dex following \citet{Qezlou22}; see their Figure 11. Qezlou et al.~showed that a noiseless tomography survey reduces the error in $M_{\rm desc}$ by only $\sim0.1$~dex, and therefore much of this uncertainty is irreducible within the context of our map resolution and analysis methods. Fig.~\ref{fig:signif_mdesc} shows the estimated descendant masses $M_{\rm desc}$ for each structure. The expectation value of $M_{\rm desc}$ would place every overdensity within the protocluster regime, but the uncertainties allow an increasing probability of $M_{\rm desc} < 10^{14}~\msol$ as $\delta_F$ increases (weaker absorption). These mass estimates and their uncertainties are consistent with our expectation that 85\% of our IGM-selected overdensities are protoclusters. 

Alternatively we could estimate $M_{\rm desc}$ directly from the peak absorption $\dFsm$. Using the results of \citet{Qezlou22}, we built the estimator $\log M_{\rm desc} = 13.10 - 0.38 \dFsm$. Fig.~\ref{fig:signif_mdesc} shows that the resulting $M_{\rm desc}$ estimates would be slightly higher, by an average of 0.13~dex. This can be considered a systematic uncertainy in $M_{\rm desc}$, which is much smaller than the random errors for individual overdensities.

The number density of absorption peaks in the LATIS maps as a function of their strength is consistent with expectations from the MDPL2 mock surveys (Fig.~\ref{fig:hmf}, left panel). Considering the distribution of $M_{\rm desc}$ (right panel), we further find good agreement with the $z=0$ halo mass function after multiplying it by the mass-dependent completeness determined by \citet[][see their Fig.~13]{Qezlou22}. This can be seen by comparing the red bands to the open histogram, which includes all LATIS absorption peaks with $\dFsm < -2.35$, the threshold used by \citet{Qezlou22}. The number of LATIS peaks at the highest masses $\log M_{\rm desc} / \msol > 14.6$ is slightly lower than the average mock survey, but still within $2\sigma$. We therefore confirm that the abundance and masses of the IGM-selected overdensities in LATIS are consistent with cosmological expectations.

\begin{deluxetable*}{lcccccccccl}
\tabletypesize{\scriptsize}
\tablecolumns{11}
\tablecaption{LATIS IGM-selected Overdensities with $\dFsm < -3$}
\tablehead{\colhead{Name} & \colhead{$\dFsm$} & \colhead{R.A.} & \colhead{Decl.} & \colhead{$z$} & 
\colhead{Signif.} & \colhead{$\delta_m$} &
\colhead{$\log M_{\rm tomo}$} & \colhead{$\log M_{\rm desc}$} & \colhead{Volume} & \colhead{Cross-IDs}}
\startdata
LATIS2-D4-00 & -4.27 & 333.938 & -17.467 & 2.533 & 6.2 & 1.2 $\pm$ 0.4 & 14.92 $\pm$ 0.22 & 14.55 $\pm$ 0.35 & 2947 & \\
LATIS2-D2-00 & -4.21 & 150.033 & 2.175 & 2.685 & 6.5 & 1.2 $\pm$ 0.4 & 15.11 $\pm$ 0.22 & 14.69 $\pm$ 0.35 & 4427 & LATIS1-D2-4, Antu \\
LATIS2-D1-00 & -4.20 & 36.202 & -4.224 & 2.452 & 6.2 & 1.3 $\pm$ 0.5 & 14.70 $\pm$ 0.22 & 14.39 $\pm$ 0.36 & 1633 &  \\
LATIS2-D2-01 & -4.09 & 149.532 & 1.974 & 2.462 & 5.0 & 1.1 $\pm$ 0.4 & 14.78 $\pm$ 0.24 & 14.45 $\pm$ 0.35 & 2098 & LATIS1-D2-1 \\
LATIS2-D2-02 & -4.02 & 149.978 & 1.970 & 2.683 & 6.3 & 1.1 $\pm$ 0.4 & 14.73 $\pm$ 0.25 & 14.42 $\pm$ 0.35 & 1933 & Antu  \\
LATIS2-D2-03 & -3.97 & 150.502 & 2.173 & 2.322 & 6.3 & 1.1 $\pm$ 0.4 & 14.89 $\pm$ 0.25 & 14.53 $\pm$ 0.35 & 2719 & \\
LATIS2-D1-01 & -3.93 & 36.203 & -4.353 & 2.455 & 6.4 & 1.1 $\pm$ 0.4 & 14.73 $\pm$ 0.26 & 14.42 $\pm$ 0.35 & 1871 & LATIS1-D1-1 \\
LATIS2-D2-04 & -3.90 & 150.332 & 2.274 & 2.457 & 6.1 & 1.1 $\pm$ 0.4 & 14.95 $\pm$ 0.26 & 14.57 $\pm$ 0.35 & 3136 & LATIS1-D2-3, Hyp \\
LATIS2-D4-01 & -3.90 & 334.100 & -17.705 & 2.595 & 6.8 & 1.0 $\pm$ 0.4 & 14.70 $\pm$ 0.26 & 14.39 $\pm$ 0.36 & 1817 & LATIS1-D4-0 \\
LATIS2-D2-05 & -3.83 & 150.474 & 2.445 & 2.473 & 5.9 & 1.0 $\pm$ 0.4 & 15.15 $\pm$ 0.27 & 14.72 $\pm$ 0.35 & 4992 & Hyp \\
LATIS2-D4-02 & -3.77 & 333.963 & -17.797 & 2.381 & 5.3 & 1.0 $\pm$ 0.4 & 14.51 $\pm$ 0.28 & 14.26 $\pm$ 0.41 & 1191 & \\
LATIS2-D2-06 & -3.71 & 149.539 & 2.146 & 2.529 & 5.6 & 1.0 $\pm$ 0.4 & 14.63 $\pm$ 0.29 & 14.34 $\pm$ 0.38 & 1527 & \\
LATIS2-D4-03 & -3.56 & 334.233 & -17.678 & 2.538 & 4.9 & 0.9 $\pm$ 0.4 & 14.58 $\pm$ 0.31 & 14.31 $\pm$ 0.39 & 1413 & \\
LATIS2-D2-07 & -3.54 & 150.089 & 2.175 & 2.560 & 5.5 & 0.9 $\pm$ 0.4 & 14.32 $\pm$ 0.31 & 14.12 $\pm$ 0.45 & 750 & LATIS1-D2-0, D13 19 \\
LATIS2-D2-08 & -3.52 & 150.161 & 2.289 & 2.442 & 5.9 & 0.9 $\pm$ 0.4 & 14.72 $\pm$ 0.31 & 14.41 $\pm$ 0.36 & 1895 & L16, Hyp, CCPC* \\
LATIS2-D2-09 & -3.51 & 149.643 & 2.102 & 2.420 & 5.2 & 0.9 $\pm$ 0.4 & 14.58 $\pm$ 0.31 & 14.31 $\pm$ 0.39 & 1391 & \\
LATIS2-D4-04 & -3.49 & 334.056 & -17.372 & 2.617 & 5.1 & 0.9 $\pm$ 0.4 & 14.21 $\pm$ 0.32 & 14.04 $\pm$ 0.48 & 596 & \\
LATIS2-D2-10 & -3.48 & 150.330 & 2.245 & 2.497 & 5.0 & 0.9 $\pm$ 0.4 & 14.96 $\pm$ 0.32 & 14.58 $\pm$ 0.35 & 3279 & Hyp, H6 \\
LATIS2-D2-11 & -3.48 & 149.704 & 2.230 & 2.679 & 5.3 & 0.9 $\pm$ 0.4 & 14.74 $\pm$ 0.32 & 14.42 $\pm$ 0.35 & 1971 & LATIS1-D2-2, Antu \\
LATIS2-D2-12 & -3.40 & 150.046 & 2.004 & 2.500 & 4.6 & 0.9 $\pm$ 0.4 & 14.48 $\pm$ 0.33 & 14.23 $\pm$ 0.42 & 1095 & COSTCO-V \\
LATIS2-D1-02 & -3.37 & 36.138 & -4.548 & 2.568 & 5.6 & 0.9 $\pm$ 0.4 & 14.39 $\pm$ 0.33 & 14.17 $\pm$ 0.44 & 866 & LATIS1-D1-0 \\
LATIS2-D2-13 & -3.37 & 150.104 & 2.188 & 2.435 & 4.4 & 0.9 $\pm$ 0.4 & 14.57 $\pm$ 0.33 & 14.30 $\pm$ 0.39 & 1342 & L16, Hyp \\
LATIS2-D2-14 & -3.36 & 149.883 & 2.312 & 2.683 & 5.4 & 0.9 $\pm$ 0.4 & 14.37 $\pm$ 0.33 & 14.16 $\pm$ 0.44 & 831 & D13 42, Antu \\
LATIS2-D1-03 & -3.30 & 36.157 & -4.297 & 2.650 & 5.4 & 0.9 $\pm$ 0.4 & 14.63 $\pm$ 0.34 & 14.35 $\pm$ 0.38 & 1536 &  \\
LATIS2-D4-05 & -3.21 & 334.070 & -17.735 & 2.512 & 5.1 & 0.8 $\pm$ 0.4 & 14.53 $\pm$ 0.35 & 14.27 $\pm$ 0.40 & 1295 &  \\
LATIS2-D2-15 & -3.20 & 150.458 & 2.409 & 2.320 & 4.9 & 0.8 $\pm$ 0.4 & 14.61 $\pm$ 0.35 & 14.33 $\pm$ 0.38 & 1501 & D13 34 \\
LATIS2-D4-06 & -3.20 & 334.174 & -17.326 & 2.530 & 5.1 & 0.8 $\pm$ 0.4 & 14.59 $\pm$ 0.35 & 14.32 $\pm$ 0.39 & 1478 & \\
LATIS2-D1-04 & -3.19 & 36.644 & -4.156 & 2.545 & 5.5 & 0.9 $\pm$ 0.4 & 14.39 $\pm$ 0.36 & 14.17 $\pm$ 0.44 & 892 & \\
LATIS2-D2-16 & -3.14 & 149.677 & 2.394 & 2.681 & 4.4 & 0.8 $\pm$ 0.4 & 14.30 $\pm$ 0.36 & 14.11 $\pm$ 0.46 & 747 & Antu  \\
LATIS2-D1-05 & -3.13 & 36.662 & -4.553 & 2.440 & 4.5 & 0.8 $\pm$ 0.4 & 14.22 $\pm$ 0.36 & 14.04 $\pm$ 0.48 & 599 &  \\
LATIS2-D2-17 & -3.10 & 149.934 & 1.992 & 2.549 & 4.9 & 0.8 $\pm$ 0.4 & 14.61 $\pm$ 0.37 & 14.33 $\pm$ 0.38 & 1526 & \\
LATIS2-D1-06 & -3.08 & 36.280 & -4.234 & 2.255 & 5.7 & 0.8 $\pm$ 0.4 & 14.66 $\pm$ 0.37 & 14.36 $\pm$ 0.37 & 1690 &  \\
LATIS2-D2-18 & -3.06 & 150.559 & 2.174 & 2.479 & 5.0 & 0.8 $\pm$ 0.4 & 14.88 $\pm$ 0.37 & 14.52 $\pm$ 0.35 & 2810 & Hyp, H8 \\
LATIS2-D1-07 & -3.05 & 36.166 & -4.282 & 2.564 & 4.0 & 0.8 $\pm$ 0.4 & 14.46 $\pm$ 0.37 & 14.22 $\pm$ 0.42 & 1057 &  \\
LATIS2-D4-07 & -3.05 & 334.158 & -17.594 & 2.585 & 4.5 & 0.7 $\pm$ 0.3 & 14.26 $\pm$ 0.37 & 14.08 $\pm$ 0.47 & 699 &  \\
LATIS2-D1-08 & -3.02 & 36.490 & -4.499 & 2.360 & 5.2 & 0.8 $\pm$ 0.4 & 14.51 $\pm$ 0.38 & 14.26 $\pm$ 0.41 & 1198 & \\
LATIS2-D2-19 & -3.00 & 150.421 & 2.001 & 2.411 & 4.6 & 0.8 $\pm$ 0.4 & 14.54 $\pm$ 0.38 & 14.28 $\pm$ 0.40 & 1305 & 
\enddata
\tablecomments{The $1\sigma$ uncertainty in $\dFsm$ is 0.63. Coordinate uncertainties are 2.3 arcmin in each of R.A.~and Decl.~and 0.002 in redshift. Mass units are $\msol$. The volume reported in $h^{-3}$ cMpc${}^{3}$ is that of the watershed associated with the absorption peak \citep{Qezlou22}. Notes on Cross-IDs: LATIS1 denotes strong Ly$\alpha$ absorption peaks discussed by \citet{Newman22} using a previous version of the LATIS IGM maps. Antu denotes components of the extended structure described in the companion paper (N25). Hyp denotes structures within the Hyperion region, and H6/H8 denote individual peaks discussed by \citet{Cucciati18}. L16 = \citet{Lee16}, COSTCO = \citet{Ata22}, D13 = \citet[][with ID number given]{Diener13}.\label{tab:catalog}}
\end{deluxetable*}

Our identification of IGM absorption-selected structures with large-scale matter overdensities relies on the Ly$\alpha$ absorption tracing the diffuse IGM. Strong absorption can also be produced by HCD absorption lines that are produced by dense gas close to typical galaxies, which do not necessarily indicate a 4~\cMpch-scale overdensity. Thus it is important to investigate whether rare map features could be produced by HCD lines or any other features associated with an individual sight line, including data reduction artifacts. In Appendix~\ref{sec:appendix_robustness}, we describe quantitative tests verifying that the strongest absorption peaks in LATIS are not notably sensitive to individual sight lines. In Section~\ref{sec:discussion} we will discuss the possibility that locally enhanced ionization could affect the detectability of protoclusters or our estimates of their masses.

\subsection{Visualizing IGM-selected Overdensities}
\label{sec:visualizing}

\begin{figure*}[!p]
    \centering
    \begin{interactive}{animation}{anc/fig4.mp4}
    \includegraphics[width=0.95\linewidth]{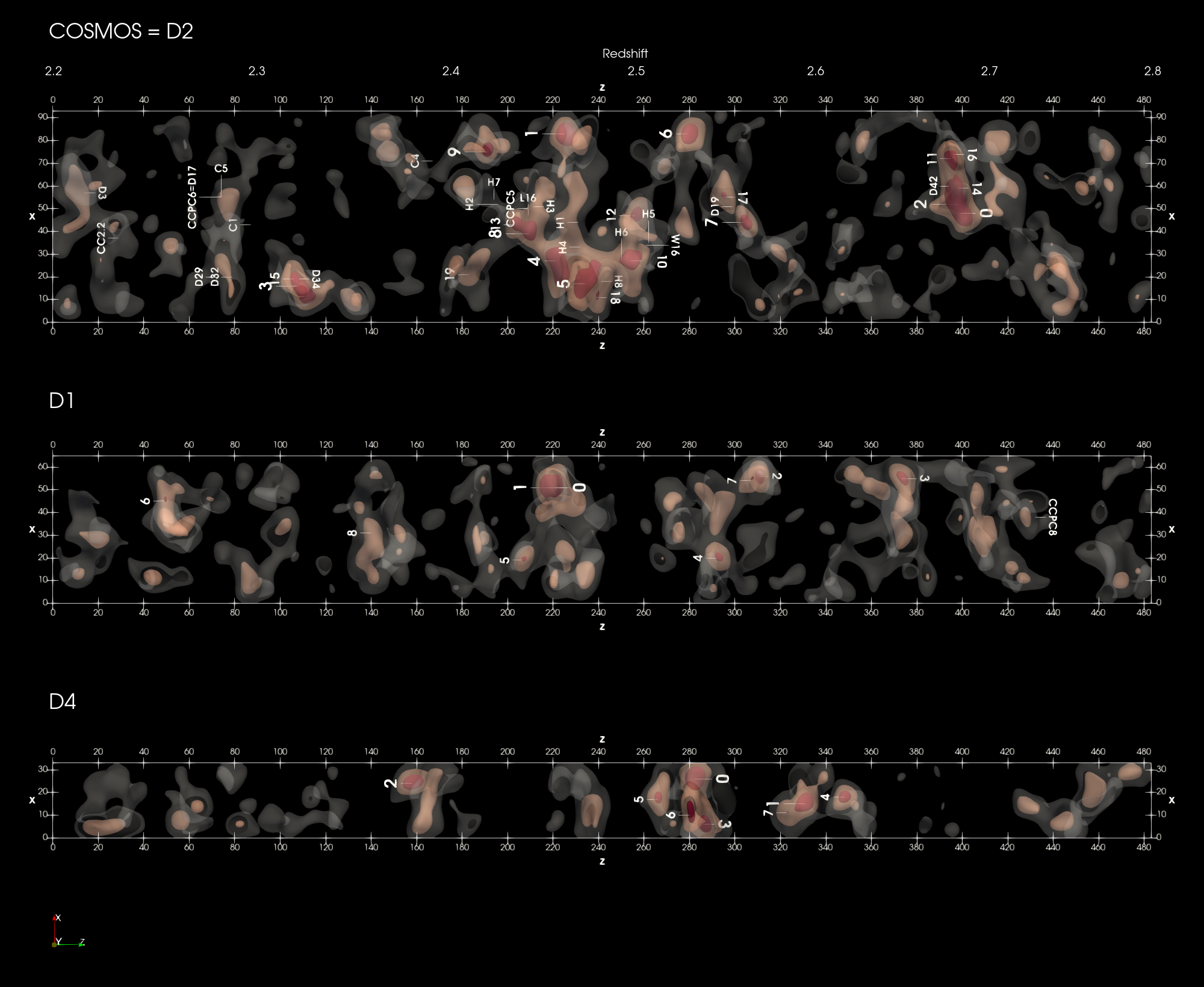}
    \end{interactive}
    \caption{Visualizations of the IGM tomographic maps in the three LATIS survey fields, focusing on the high-density regions. Three isocontours of Ly$\alpha$ absorption are shown at $\dFsm = -1$, $-2$, and $-3$ (from most to least transparent, and from gray to red). Axis coordinates are in units of \cMpch; a conversion to redshift is shown at top. The still image shows a projection along the $y$ axis. The animated version shows the maps rotating about their $z$ axes. LATIS overdensities listed in Table~\ref{tab:catalog} are indicated by lone numbers, with larger and bolder numbers indicating the stronger absorption peaks ($\dFsm < -3.5$). Other labels beginning with letters show structures identified in other studies: D \citep{Diener13}, CCPC \citep{Franck16}, CC2.2 \citep{Darvish20}, C \citep[][COSTCO]{Ata22}, H \citep[][Hyperion]{Cucciati18} and H8 (new; Section~\ref{sec:hyperion}), W16 \citep{Wang16}, L16 \citep{Lee16}. See Section~\ref{sec:litcompare} for further discussion of individual structures. We note that the $x$, $y$, and $z$ axes are aligned with -R.A., Decl., and redshift, respectively, and therefore define a left-handed coordinate system. Coordinates are displayed correctly, but the parity is inverted by ParaView, the software used to create this visualization \citep{ParaView}.\\
    {\bf The animation is available at \url{https://vimeo.com/1096973540/3138527060} and as ancillary data on arXiv.}}
    \label{fig:dFonly}
\end{figure*}

\begin{figure*}[!p]
    \centering
    \begin{interactive}{animation}{anc/fig5.mp4}
    \includegraphics[width=0.95\linewidth]{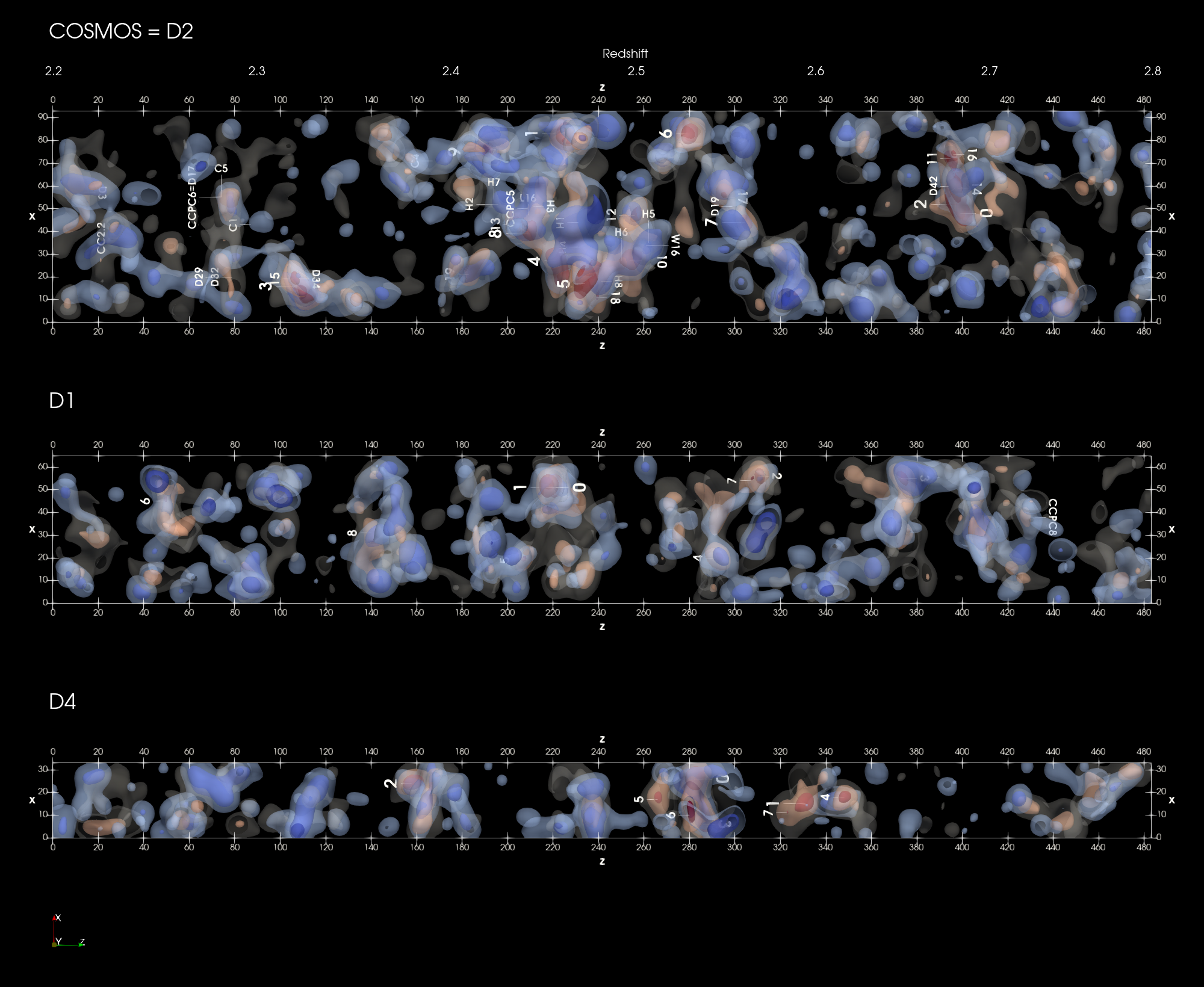}
    \end{interactive}
    \caption{Visualizations of the IGM tomography and galaxy density maps in the three LATIS survey fields. The Ly$\alpha$ absorption contours from Fig.~\ref{fig:dFonly} are repeated here. Blue/white contours show smoothed galaxy overdensity levels of $\delta_{\rm LBG} = 1$ and 3. The animated version shows the maps rotating about their $z$ axes. See the caption of Fig.~\ref{fig:dFonly} for information on the labels and axes.\\
    {\bf The animation is available at \url{https://vimeo.com/1096989868/758514d0e8} and as ancillary data on arXiv.}}
    \label{fig:dFgals}
\end{figure*}

Fig.~\ref{fig:dFonly} depicts the LATIS IGM maps and the positions of the 37 IGM-selected overdensities within them. To better visualize the 3D large-scale structure, we provide an animation of the maps rotating about the $z$ axis. In addition to the IGM-selected structures identified with numbers (e.g., 1 in the COSMOS map indicates LATIS2-D2-01 in Table~\ref{tab:catalog}), we also show the positions of previously identified protoclusters that we will discuss in Section~\ref{sec:litcompare}. It is visually apparent that the overdensities are clustered, as expected in hierarchical structure formation. The most striking region lines in the COSMOS (D2) field around $z = 2.47$, where many absorption peaks are concentrated, especially in the eastern half (low $x$ coordinates). This is the proto-supercluster Hyperion (see Section~\ref{sec:hyperion}). The second-largest region of contiguous strong Ly$\alpha$ absorption lies in the COSMOS field at $z = 2.68$. This complex structure, which we call Antu, is discussed in the companion paper (N25). It contains five of our IGM-selected overdensities within a volume of $\approx 10^4$ $h^{-3}$ cMpc${}^3$. In the D4 field, we note a concentration of strong Ly$\alpha$ absorption peaks around $z = 2.53$ as an interesting region for further study, which may extend beyond the region mapped by LATIS.

Fig.~\ref{fig:dFgals} shows a similar visualization focusing on the correlation between Ly$\alpha$ absorption and the positions of LATIS galaxies. The contours now focus on the stronger Ly$\alpha$ absorption regions with $\dFsm < -2$, while points denote LATIS galaxies and blue isocontours indicate a galaxy overdensity of $\delta_{\rm LBG} = 3$. It is visually clear that galaxies are correlated with large-scale IGM absorption, a relationship that N25 examine in detail. We find that spatial correlation between galaxies and Ly$\alpha$ absorption is clearer in the animation, which helps to mitigate projection effects.

\begin{figure*}
    \centering
    \begin{interactive}{animation}{anc/fig6.mp4}
    \includegraphics[width=0.8\linewidth]{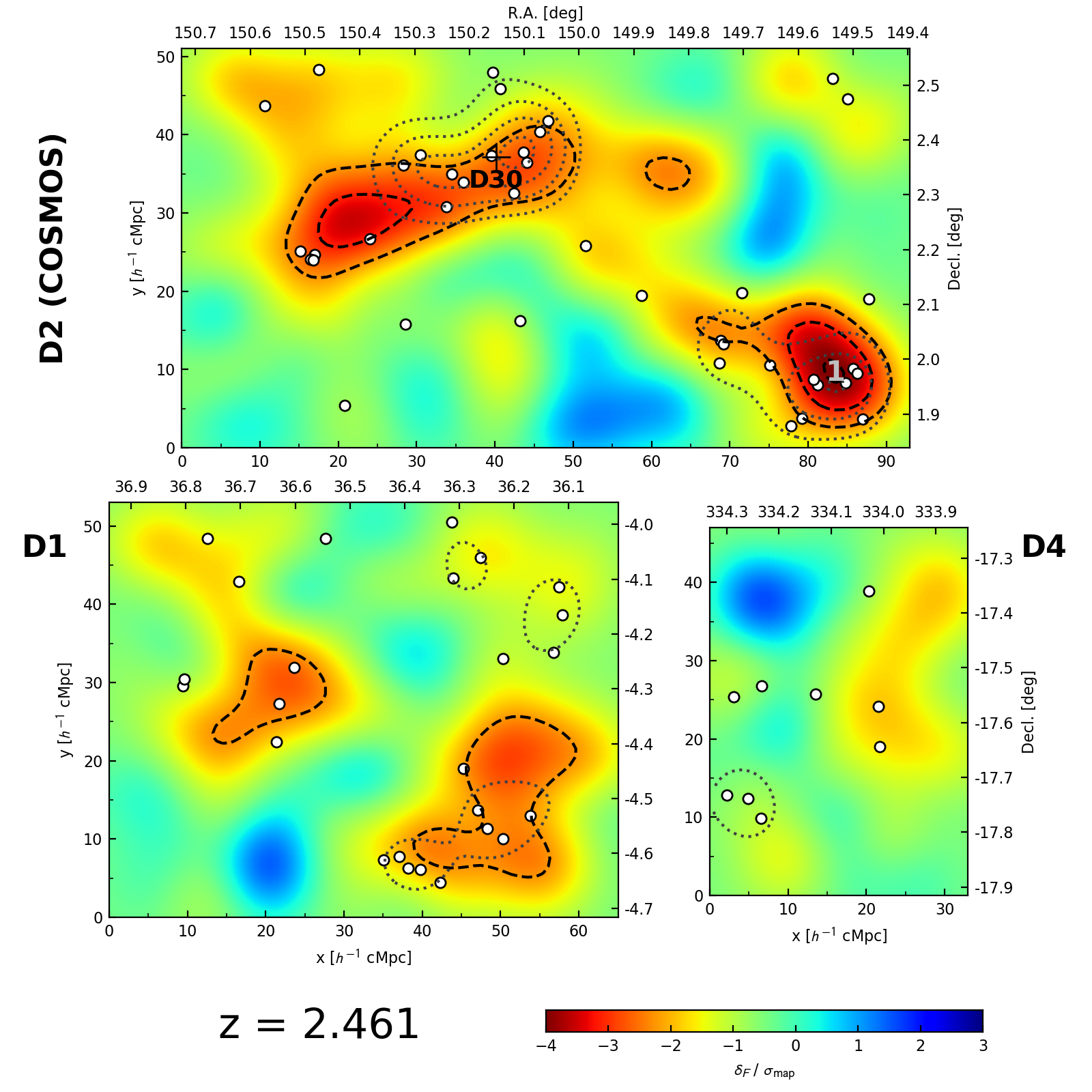}
    \end{interactive}
    \caption{Cross sections of the three LATIS maps at fixed redshift. The animated figure scans through the full redshift range. The smoothed Ly$\alpha$ flux contrast $\delta_F$ is indicated by colors and using the dashed ($\dFsm = -2, -3, \ldots$) and solid ($\dFsm = +2, +3, \ldots$) contours. Points show the positions of LATIS LBGs and QSOs, and dotted contours show the smoothed galaxy overdensity $\delta_{\rm LBG} / \sigma = 2, 4, 6, \ldots$, where $\sigma$ is the standard deviation of the galaxy density map. Gray numbers indicate the IGM-selected overdensities cataloged in this paper (e.g., 1 in the COSMOS field is LATIS2-D2-01). Black labels and crosses indicate the positions of protocluters and protogroups in the literature. See the caption of Fig.~\ref{fig:dFgals} for references. In the animation, we show all of the CCPC \citep{Franck16} and \citet{Diener13} structures covered by LATIS, not only the subset shown in Fig.~\ref{fig:dFgals}. Both $\delta_F$ and $\delta_{\rm LBG}$ are smoothed with a Gaussian 4~\cMpch~kernel. The left and bottom axes show the map coordinates in units of \cMpch. The top and right axes give celestial coordinates R.A.~and decl., respectively.\\
    {\bf The animation is available at \url{https://vimeo.com/1096990028/30503e878d} and as ancillary data on arXiv.}}
    \label{fig:scananim}
\end{figure*}

We complement these side-on views of the LATIS maps with an animation in Fig.~\ref{fig:scananim} that scans along the redshift axis. We find that this is often the most informative visualization for examining the LATIS maps in the regions around individual structures (Section~\ref{sec:litcompare}), whereas the global visualizations in Figs.~\ref{fig:dFonly} and \ref{fig:dFgals} can provide more insight into the large-scale environment.

Finally, in Appendix~\ref{sec:appendix_maps} we provide a detailed visualization of the 3D structure of LBGs and IGM absorption surrounding the 16 strongest Ly$\alpha$ absorption peaks in the LATIS maps, those having $\dFsm < -3.5$. 

Although many IGM-selected overdensities are essentially pointlike in the LATIS maps, there are also a number of cases in which the 3D structure is resolved. Often this structure is evident in both the galaxy and IGM maps and aligns beautifully well (e.g., LATIS2-D4-00, D2-04, D1-01, and D2-08; see Appendix~\ref{sec:appendix_maps}). The maps also often trace connections in the cosmic web between neighboring overdensities (e.g., LATIS2-D2-00, D2-04, D2-05, and D2-08). We note a wide variety in the galaxy content of the IGM-selected overdensities, with some that are quite rich (e.g., LATIS2-D4-00, D2-03, and D2-08) and others that are remarkably poor (e.g., LATIS-D2-00, D2-05, D2-07, and D4-01). The galaxy content of the IGM-selected overdensities is the subject of the companion paper (N25).

\section{Comparisons to Galaxy-Selected Samples of Protoclusters}
\label{sec:litcompare}

LATIS overlaps many previous searches for protoclusters and protogroups, especially within the COSMOS field, which employed a variety of different techniques. Here we briefly review several structures identified by these searches and examine their counterparts, or lack thereof, in the LATIS IGM and galaxy maps. 

\subsection{Hyperion}
\label{sec:hyperion}

The richest region in the LATIS maps lies in the COSMOS field at $z \sim 2.45$. Individual structures within this region have previously been identified as overdensities of Lyman-break galaxies \citep{Diener13}, dusty star-forming galaxies \citep{Casey15}, a radio galaxy \citep{Castignani14}, distant red galaxies (DRGs; \citealt{Wang16}), sources with photometric redshifts \citep{Chiang15}, Ly$\alpha$ emitters \citep{Huang22}, and in one case, as a region of excess IGM absorption \citep{Lee16}. \citet{Cucciati18} presented a unified view of the region based on LBG spectra from the VIMOS Ultra-Deep Survey (VUDS; \citealt{LeFevre15}) and zCOSMOS \citep{Lilly07} and labeled the region as a proto-supercluster named Hyperion. Although a complete comparison of the rich structure of Hyperion as seen in galaxy and IGM maps is beyond the scope of this paper, we highlight some salient features.

\citet{Cucciati18} identified seven galaxy density peaks within Hyperion labeled H1 through H7 in Figs.~\ref{fig:dFgals} and \ref{fig:scananim}. We find an excellent correspondence with the LATIS galaxy density map, locating a peak near the \citet{Cucciati18} coordinates for H1--H6.\footnote{The rms positional difference is only 1~\cMpch~per coordinate, and even that may be limited by the voxel scale.} In the case of H7, to which \citet{Cucciati18} ascribed the lowest mass, we do not find a distinctly separate peak from the nearby H2. Furthermore, we locate an additional galaxy density peak at coordinates (R.A., Decl., redshift) = $(150.459, 2.259, 2.481)$, close to H6, with a similar overdensity to the other Hyperion peaks ($\delta_{\rm LBG} = 8 \pm 2$). This structure expands the previously known extent of Hyperion to the east, likely because it lies beyond the VUDS survey area. We label this new galaxy density peak as H8 and identify it with the Ly$\alpha$ absorption peak LATIS2-D2-18.\footnote{Although the galaxy and absorption peak positions are separated by 9~\cMpch~in the sky plane, the region of strong absorption is broad and encompasses the galaxy structure.}

\begin{figure*}
    \centering
    \includegraphics[width=\linewidth]{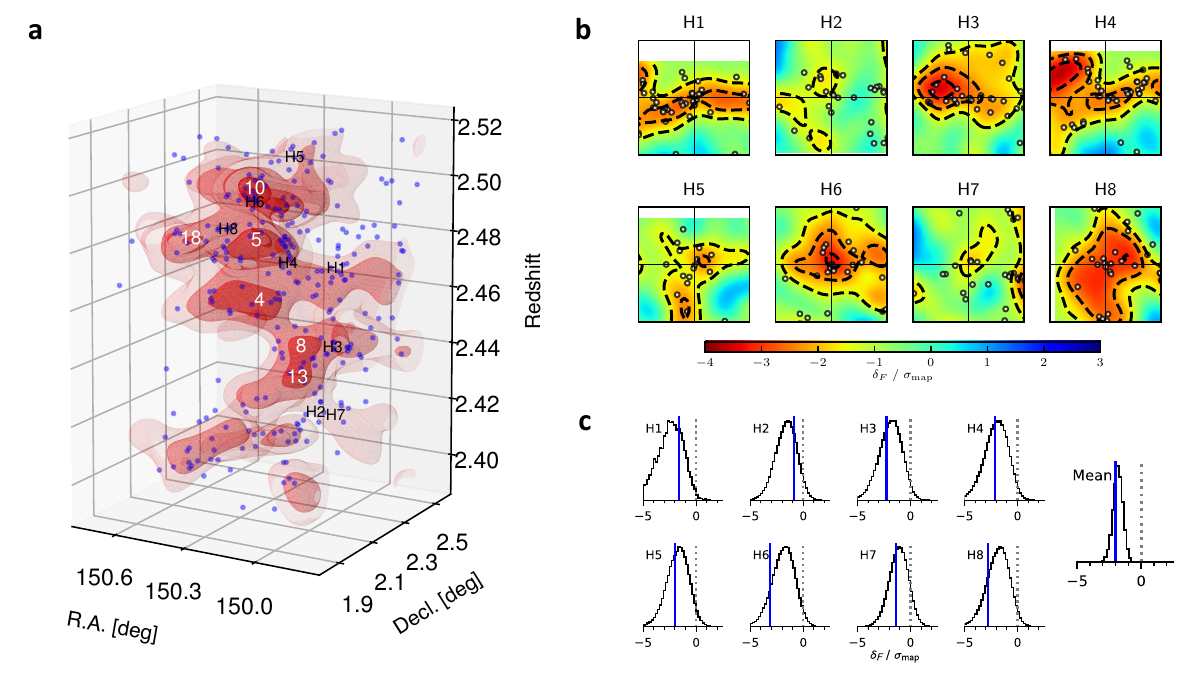}
    \caption{\emph{Panel (a):} The Hyperion proto-supercluster region within the COSMOS field as traced by IGM absorption and galaxies. The three red contour levels indicate $\dFsm = -1$, $-2$, and $-3$, from lightest to darkest. Blue points show the positions of LATIS galaxies. The Ly$\alpha$ absorption peaks are labeled with numbers, i.e., 4 indicates LATIS2-D2-04. The galaxy density peaks identified by \citet{Cucciati18} are labeled H1--H7, and H8 indicates the new galaxy density peak identified with LATIS. \emph{Panel (b):} For each of the eight galaxy density peaks, a cross section through the IGM map is shown. Each panel spans 40~\cMpch~on a side, with R.A.~increasing to the left and decl. toward the top. Contours indicate the $\dFsm = -1$, $-2$, and $-3$ levels. Points show LATIS galaxies within $\pm 6$~\cMpch~along the line of sight. \emph{Panel (c):} For each structure H1--H8 in panel (b), the observed IGM absorption $\dFsm$ at the galaxy density peak position (blue vertical lines) is compared to the conditional distribution $p(\delta_F | \delta_{\rm LBG})$ from the mock surveys (black histograms). The panel labeled ``Mean'' shows the average $\langle \dFsm \rangle$ over the eight peaks and the corresponding distribution in the mock surveys.}
    \label{fig:hyperion}
\end{figure*}

In Fig.~\ref{fig:hyperion} we compare the distribution of galaxies and Ly$\alpha$ absorption (panel a) in Hyperion. Galaxies broadly follow the IGM map, yet the positions of the \citet{Cucciati18} galaxy density peaks do not correspond very closely to the Ly$\alpha$ absorption peaks. The highest galaxy density peak, H1 \citep{Casey15,Cucciati18}, is the heart of Hyperion in galaxy maps, but in the IGM map, H1 lies within an elongated region of absorption that is only moderately strong. It is not the strongest in the Hyperion region, and there is no clear absorption peak at its location. Among the six IGM-selected overdensities within Hyperion (see Table~\ref{tab:catalog}), the clearest association with the \citet{Cucciati18} galaxy density peaks is of H6 with LATIS2-D2-10, but in other cases, the absorption peaks are often separated by $\sim$10~\cMpch~from the nearest galaxy peak. We expect that this is, at least in part, a result of observational uncertainty in the position of a peak within a very broad region of significant Ly$\alpha$ absorption. In addition, there is one IGM peak that is markedly poor in LATIS galaxies: LATIS2-D2-05, which is central to Hyperion in the IGM map but contains no significant overdensity of the bright LBGs traced by LATIS at its peak.

We now consider the compatibility of the galaxy and Ly$\alpha$ absorption distributions in a more quantitative way. Specifically, we examine the Ly$\alpha$ absorption observed at the positions of the \citet{Cucciati18} galaxy density peaks along with the new H8. Panel (b) of Fig.~\ref{fig:hyperion} shows cross sections of the IGM maps at each galaxy density peak, while panel (c) compares the observed $\dFsm$ at each peak to the conditional probability density function $p(\delta_F^{\rm rec} | \delta_{\rm LBG}^{\rm rec})$ built from the mock surveys. We observe absorption ($\delta_F < 0$) at every galaxy peak, and the amount of absorption is always consistent with the FGPA-based mock surveys, evaluated in regions with matched $\delta_{\rm LBG}^{\rm rec}$. The mean $\langle \dFsm \rangle$ over H1--H8 is better constrained than the individual peaks, with an observed value of $-2.0$ consistent with $-1.8 \pm 0.4$ in the mocks.\footnote{We also considered maps in which $\delta_F$ and $\delta_{\rm LBG}$ were each smoothed with a Gaussian kernel of $\sigma_{\rm sm} = 8$~\cMpch~(following \citet{Dong23}; see Section~\ref{sec:discussion}), larger than the $\sigma_{\rm sm} = 4$~\cMpch~used throughout the rest of this paper. With this larger kernel, we similarly find an observed $\langle \dFsm \rangle = -2.2$ that is consistent with the distribution $-1.9 \pm 0.3$ in the mock surveys.} The FGPA assumes that the IGM gas is on a tight temperature--density relation and that the ionizing radiation field is spatially uniform; it therefore neglects any local enhancements in gas ionization due to feedback and radiation from galaxies. The concordance that we observed thus implies that gas in the Hyperion density peaks, on average and evaluated on large scales ($\sigma = 4$~\cMpch), has not been strongly affected by galaxy formation (see Section~\ref{sec:discussion}).

The H5 peak was also identified by \citet{Wang16} as a DRG overdensity, confirmed by H$\alpha$ and CO spectroscopy. They detected extended X-ray emission and, along with other evidence, inferred the existence of a mature, collapsed cluster-sized halo of mass $M_{200} = 10^{13.9} M_{\odot}$. \citet{Champagne21} argued that this X-ray emission, if it is real, does not necessarily indicate the presence of a hot diffuse medium, and is more likely to be an inverse Compton ghost, which results from the upscattering of microwave background photons by electrons in an extinguished active galactic nucleus (AGN) jet. We find a substantial galaxy overdensity ($\delta_{\rm LBG} = 7 \pm 2$) and moderate Ly$\alpha$ absorption ($\delta_F = -1.8 \pm 0.6$) at the position of H5 that is well within the conditional distribution shown in Fig.~\ref{fig:hyperion}. This implies that cool gas is present on~4 \cMpch~scales. However, the putative X-ray emission is detected at radii $< 0.4$~\cMpch. Our observations cannot exclude a compact and hot gas phase, nor an extended multiphase structure containing both cool and hot gas.

Part of Hyperion was examined by \citet{Lee16} using the CLAMATO maps. They reported a system of strong absorption around $z=2.442$, whose coordinates lie between LATIS2-D2-08 and LATIS2-D2-13, and thus we have recovered the same structure. We measure $\dFsm = -3.5 \pm 0.6$ and $-3.4 \pm 0.7$ for these two peaks, respectively, consistent with $\dFsm = -4.0$ reported by \citet{Lee16}. They estimated that the system will form at least one cluster with mass $\log M_{\rm desc} / \msol = 14.6 \pm 0.2$, consistent with our estimates (Table~\ref{tab:catalog}). This IGM structure overlaps H3 as well as CCPC5-z24-005 (see below).

\subsection{CC2.2}

\citet{Darvish20} located a protocluster in the COSMOS field at $z=2.232$ using narrowband H$\alpha$ imaging, confirmed via Keck/MOSFIRE spectroscopy. We find a nearby LATIS galaxy overdensity ($\delta_{\rm LBG} = 6 \pm 2$), consistent with the $\delta_{\rm gal} = 6.6 \pm 0.3$ reported by \citet{Darvish20}. We detect only moderate IGM absorption of $\dFsm = -1.0 \pm 0.6$ at the position of the LATIS galaxy peak ($\dFsm = -1.7 \pm 0.6$ at the nearby absorption peak). Considering LATIS data only, the observed absorption lies easily within the conditional distribution $p(\delta_F | \delta_{\rm LBG})$ in the mock surveys, which has mean and standard deviation $-1.6 \pm 1.0$. We explored whether the observed absorption would still be consistent with the more constrained estimate by Darvish et al., which was based on 35 H$\alpha$ emitters and was evaluated over a comparable volume to the smoothing kernel that we use. Lacking more detailed information, we assume that the H$\alpha$ emitter bias is similar to that of the LATIS LBGs. Even if we consider their $\delta_{\rm gal} = 6.6$ to be a noiseless measure, we still find that the modest observed $\delta_F$ is unsurprising and consistent with noise in the IGM maps. 

\subsection{COSTCO}
\label{sec:costco}

\citet{Ata22} used constrained cosmological simulations to reconstruct the matter density field and its evolution within COSMOS. The simulations were constrained by the galaxy density field as measured using the zCOSMOS, VUDS, MOSDEF, and ZFIRE surveys \citep{Lilly07,LeFevre15,Kriek15,Nanayakkara16}, and protoclusters were identified based on the distribution of descendant halo masses over a suite of realizations. Five new protoclusters were identified, and among these, COSTCO-I, -IV and -V (see their Table~1) lie with the redshift range of our IGM maps. \citet{Ata22} noted that their method is able to identify extended regions exhibiting mild overdensities, which may be at an earlier stage of collapse than the protoclusters typically identified by other means.

COSTCO-I was investigated by \citet{Lee16} and \citet{Dong23} who found that the Ly$\alpha$ transmission in the CLAMATO maps was surprisingly high. \citet{Dong23} considered this to be evidence of large-scale gas heating in this protocluster, resulting in increased ionization (see Section~\ref{sec:discussion}). At the location of COSTCO-I, we also find no excess Ly$\alpha$ absorption ($\dFsm = +0.1 \pm 0.7$), confirming the CLAMATO result, but we do not find a significant overdensity of LATIS galaxies.\footnote{There is a galaxy peak with associated IGM absorption at a separation of 12 \cMpch, but it is a closer match to CCPC-z22-006.} All of the members of the protocluster ``core'' were identified using MOSDEF redshifts \citep{Lee16}, and we find that all are fainter that the LATIS flux limit of $r = 24.8$. 

We find similar results for COSTCO-IV: mean transmission ($\delta_F = +0.1$) and no LATIS galaxy overdensity at the \citet{Ata22} position. But in this case, we also do not see any clear overdensity of VUDS or zCOSMOS galaxies, and the structure is not covered by the MOSDEF and ZFIRE surveys. The putative protocluster appears very diffuse in the Ata et al.~maps (see their Fig.~4) and may be an example that can be identified only by using a dynamical approach.

COSTCO-V, on the other hand, is likely associated with the absorption peak LATIS2-D2-12 ($\dFsm = -3.4$, $\delta_{\rm LBG} = 5.2 \pm 1.8$).

\subsection{Diener et al.}

\citet{Diener13} located a sample of protogroups and protoclusters in the COSMOS field using the zCOSMOS-deep spectroscopic survey. There are 21 structures in their catalog within the redshift range of the LATIS IGM maps ($z = 2.2$-2.8). As already shown by \citet{Newman20}, these structures, as an ensemble, are clearly overdense in the LATIS IGM and LBG maps. We find that 19 of 21 have higher than average galaxy density ($\delta_{\rm LBG} > 0$) and lower than average Ly$\alpha$ flux ($\delta_F < 0$), with means of $\langle \delta_{\rm LBG} \rangle = 3.7$ and $\langle \dFsm \rangle = -1.3$. 

\citet{Diener13} evaluated the likely descendant masses of their structures and found that it extends to halo masses as low as $10^{12} \msol$. Thus we do not expect to detect most of their sample in our IGM maps. We do find a correspondences to an IGM-selected overdensities in three cases: the Diener et al.~IDs 19, 34, and 42 are LATIS2-D2-07, -15, and -14, respectively. In many other cases, a Diener et al.~structure corresponds to weaker but still significant absorption in the LATIS maps, and these cases are shown in Figures~\ref{fig:dFonly} and \ref{fig:dFgals}. The animation in Figure~\ref{fig:scananim} includes all of the Diener et al.~structures within the LATIS footprint.

\subsection{CCPC}

\citet{Franck16} identified protoclusters from a compilation of public spectroscopic redshift catalogs. Within the LATIS map volume, they found five structures. CCPC-z22-006 was considered by \citet{Ata22} who did not find a convincing protocluster candidate. We confirm a galaxy overdensity ($\delta_{\rm LBG} = 3.0 \pm 1.5$) separated by 7~\cMpch~from the CCPC coordinates (in 3D), which is accompanied by fairly strong IGM absorption ($\dFsm = -2.9 \pm 0.6$) just below the threshold for inclusion in our sample. CCPC-z24-005 is the highest confidence and richest of the five CCPC structures. Its reported coordinates are within 3~\cMpch~of a strong LATIS galaxy density peak ($\delta_{\rm LBG} = 7.6 \pm 2.2$) that overlaps LATIS2-D2-08, one of the strongest absorption peaks in LATIS  ($\dFsm = -3.5 \pm 0.7$). Next, we find a possible counterpart of CCPC-z27-008 in the D1 field at a separation of 12~\cMpch, a moderate galaxy overdensity ($\delta_{\rm LBG} = 3.3 \pm 1.5$) that overlaps a moderate IGM absorption peak ($\dFsm = -2.5 \pm 0.7$). We see no counterpart in either our galaxy or IGM maps for the other two structures, CCPC-z27-007 (COSMOS) and CCPC-z27-013 (D1), but we note that these were uncertain detections based on two to three galaxies.

\section{Discussion}
\label{sec:discussion}

LATIS was designed to provide a significant samples of over- and underdense large-scale environments at $z\sim2.5$ that are traced in a galaxy-independent manner. Its comoving volume contains, on average, $\sim$30 massive galaxy clusters with $M_{\rm vir} > 10^{14.5} \msol$ at $z=0$. The catalog presented in this paper shows that LATIS was successful in delivering a novel and sizable sample of IGM-selected overdensities. We identified  37 such overdensities by applying a threshold of $\dFsm < -3$ to the smoothed Ly$\alpha$ transmission maps (Section~\ref{sec:catalog}). Using the prescriptions developed by \citet[][and see \citealt{Lee16}]{Qezlou22}, we estimated the matter overdensity and tomographic mass $M_{\rm tomo}$ of each IGM-selected overdensity at the observed redshift, along with the mass $M_{\rm desc}$ of the largest descendant halo at $z=0$. 

We find that our sample comprises overdensities of $\delta_m \approx 1$, representing $\approx 3$-5$\sigma$ density fluctuations, with tomographic masses $M_{\rm tomo} = 10^{14.2}$-$10^{15.2} \msol$. They collapse into $z=0$ halos with estimated masses from $M_{\rm desc} = 10^{14.0}$-$10^{14.7} \msol$. The space density of IGM-selected overdensities as a function of $\delta_F$ or $M_{\rm desc}$ agrees well with expectations from mock surveys  (Fig.~\ref{fig:hmf}), supporting the reliability of the mass estimates. We expect that 85\% of our sample consists of protoclusters (i.e., $M_{\rm desc} > 10^{14} \msol$), a fraction that increases to 93\% for the 16 strongest absorption peaks ($\dFsm < -3.5$), while nearly all of the remainder are still the progenitors of massive galaxy groups ($M_{\rm desc} = 10^{13.5-14.0} \msol$; Section~\ref{sec:catalog}). In common with other protocluster searches, the completeness  of our catalog is strong function of mass, increasing from 10\%-90\% as $M_{\rm desc}$ increases from $10^{14.2}~\msol$ to $10^{15.1}~\msol$. 

The LATIS catalog presented in this paper represents a roughly order-of-magnitude increase in the number of IGM-selected overdensities known. Previously, a portion of Hyperion was identified in the CLAMATO maps \citep{Lee16,Horowitz22}. In addition, three protoclusters have been detected in IGM absorption using quasar surveys and then confirmed as high galaxy overdensities (BOSS1441, BOSS1244, BOSS1542; \citealt{Cai17B,Shi21,Zheng21}). It is difficult to compare the Ly$\alpha$ absorption observed in these systems to the LATIS sample due to the different techniques used. However, their descendant masses are estimated to be $M_{\rm desc} \geq 1.4 \times 10^{15} \msol$, a range that is not accessible with a LATIS-sized volume (the expected number of such halos is 0.7). As described in Section~\ref{sec:intro}, this selection technique is highly incomplete but can be deployed over enormous volumes, providing a complementary way to study the most extreme overdensities.

Our catalog opens new opportunities for studying large-scale structures at cosmic noon. An important caveat is that the catalog is a highly compressed peak-based representation of a continuous absorption field that can have complex structure. In many cases, absorption peaks are relatively isolated, and the quantities listed in Table~\ref{tab:catalog} are a nearly complete summary. But in rich regions like Hyperion, which contain extended regions of absorption that connect the peaks, considering the peaks alone could be misleading. Only two of eight galaxy density peaks in Hyperion have a close counterpart in our IGM-selected catalog, which could be superficially interpreted as a gross mismatch between galaxy and IGM tracers. Yet we see that galaxies and Ly$\alpha$ absorption do broadly trace one another in this region, and we find statistical consistency between the expected and observed IGM absorption conditioned on the galaxy density in the Hyperion peaks. Particularly in such complex regions, it may be necessary to rely on the full IGM maps to study, for example, the variation of galaxy properties with local density.

Another important caution is that the absence of an individual protocluster from our catalog does not straightforwardly contradict its reality or its estimated mass, nor does it necessarily imply that its gas is unusually transparent to Ly$\alpha$ photons. Incompleteness is significant (Section~\ref{sec:identifying}), and a rather broad distribution of absorption is expected for a given measured galaxy overdensity due to noise in both measures. Reviewing the literature, we found that some previously known protoclusters have counterparts in our catalog, while a number of others showed Ly$\alpha$ absorption that does not meet our threshold ($\dFsm < -3$) but was still significantly detected and compatible with the observed $\delta_{\rm LBG}$ according to our mock surveys.

\subsection{Possible Effects of Locally Enhanced Ionization on Protocluster Detection and Masses}

Protocluster mass estimation using IGM tomography is complementary to other methods. One strength is that the IGM maps trace the full extended region of diffuse material that will collapse into a cluster-mass halo. Another is that unlike dynamical estimates derived from galaxy velocities, IGM-derived protocluster masses do not rely on any assumption of virialized equilibrium. Finally, unlike mass estimates based on galaxy overdensities, IGM-based estimates do not require knowledge of galaxy bias, and they are immune from selection effects in galaxy spectroscopic surveys that could potentially lead to environment-dependent biases in the galaxy overdensity (see Section~\ref{sec:intro}; \citealt{Newman22}). 

On the other hand, Ly$\alpha$ tomography maps use neutral gas as a tracer, which represents a small fraction of the baryons. A concern is that locally enhanced ionization within protoclusters could complicate their detection and estimates of mass. Such enhanced ionization could arise from shock heated gas, induced by gravitational infall or galaxy outflows, or from an elevated intensity of ionizing radiation.

Using the IllustrisTNG300 simulation \citep{Nelson19}, \citet{Qezlou22} found that the transmitted Ly$\alpha$ flux in overdense regions, when both are smoothed to LATIS resolution, is \emph{on average} very close to an FGPA-based calculation, which does not model feedback or hydrodynamics at all and assumes a spatially uniform ionizing background. In the companion paper (N25) we show that our FGPA-based mocks accurately model the mean IGM transmission as a function of density. This suggests that hydrodynamic effects in \emph{most} protoclusters will minimally affect their appearance in our IGM maps.

\citet{Miller21} accounted for local ionization effects using a simple model of AGN proximity zones within the IllustrisTNG100 simulation. They placed AGN in the most massive halos assuming a 100\% duty cycle, a choice that will maximize the impact of AGN radiation on protocluster gas. They found that the ionization is indeed enhanced near massive halos, but this enhancement is confined to distances of $\lesssim 1$~\cMpch. When the Ly$\alpha$ transmission is smoothed on $\sigma = 4$~\cMpch~scales like the LATIS maps, it was little affected by the inclusion of collisional ionization or AGN photoionization in the model. The typical sensitivity of $\dFsm$ to these choices is $\lesssim 0.2$ (see their Fig.~4), $3\times$ smaller than the measurement uncertainty in LATIS. Only in a few individual protoclusters did the models lead to differences of $\Delta \dFsm \approx 0.6$, roughly equal to our measurement uncertainty. \citet{Miller21} concluded that at the resolution of LATIS and similar Ly$\alpha$ tomography maps like CLAMATO, the uncertain ionization in protoclusters will minimally affect the completeness of protocluster discovery or mass estimates for individual systems. Systematic shifts in tomographic mass estimates were also small, $\Delta \log M_{\rm desc} <0.1$ (\citealt{Miller21}; see their Fig.~4).\footnote{\citet{Miller21} contrasted their linear fit connecting $\dFsm$ to $M_{z=0}$ ($M_{\rm desc}$, in our terminology) to the one derived by \citet{Lee16} using the FGPA. The rather large difference suggests a large sensitivity to the astrophysics. However, we suggest that the different fitting formulae arose mostly from different choices of the independent variable ($\dFsm$ or $\log M_{z=0}$).}

\citet{Dong23} presented COSTCO-I as a case study of a protocluster with unusually transparent gas, $\delta_F \approx 0$. Since, as discussed above, a large-scale (several cMpc) increase in gas ionization is not easily explained by AGN feedback as implemented in the IllustrisTNG simulations, which distribute the heating isotropically, nor by an enhanced ionizing radiation field near AGN, Dong et al.~suggested that AGN jets, implemented in some cosmological simulations \citep{Dave19}, could be a plausible means to heat gas over larger (several cMpc) scales. \citet{Dong24} considered a broader suite of simulations and concluded that some form of AGN feedback may be needed to explain COSTCO-I and possibly several other protoclusters showing a much more modestly enhanced Ly$\alpha$ transmission. 

Locations like COSTCO-I are important and interesting puzzles, but they appear to represent an extreme phenomenon. In most protoclusters, the gas ionization is probably not much affected by hydrodynamical effects or feedback from galaxies. As we discuss in the companion paper (N25), removing even 8\% of the excess Ly$\alpha$ absorption from all IGM-selected overdensities would lead to a 5$\sigma$ deficit in the number density of such structures. Most protoclusters thus cannot be much more transparent than we have estimated using the FGPA or IllustrisTNG.

This finding supports the validity of selecting and estimating the masses of overdensities via Ly$\alpha$ absorption. However, it is not inconsistent with the erasure of IGM absorption in a minority of protoclusters like COSTCO-I. Interestingly, in addition to the unusual IGM transparency, the galaxy population in COSTCO-I may also be atypical. We find that the members of its core, which were identified using MOSDEF near-infrared spectra, are all fainter than the LATIS flux limit (Section~\ref{sec:costco}), which is uncommon for a randomly selected sample of MOSDEF galaxies at the same redshift. This points to the importance of near-infrared galaxy spectroscopy covering much larger portions of LATIS to assemble more complete samples of both galaxies and protoclusters.

\subsection{Data Availability and User's Guide}

With this paper, we provide a public release of the LATIS IGM tomography maps on Zenodo\footnote{\dataset[doi: 10.5281/zenodo.15200365]{\doi{10.5281/zenodo.15200365}}}. The release of the LATIS spectra, redshifts, and related data products will occur in an imminent separate publication, to be submitted by the end of August 2025.

Our catalog and maps can support a variety of studies on protoclusters at cosmic noon. Several features are important to bear in mind in future applications: (1) The protocluster catalog has significant mass-dependent incompleteness (Section~\ref{sec:identifying}). Therefore the absence of a protocluster from the catalog, which is defined by a $\dFsm < -3$ threshold, is itself not very informative, although the $\delta_F$ value at the corresponding map position may be. (2) The positions of the IGM-selected overdensities have significant uncertainties (Table~\ref{tab:catalog}) that need to be taken into account when comparing to other tracers, e.g., a galaxy overdensity. In many cases, the strongest Ly$\alpha$ absorption peaks are embedded within a complex and extended region of absorption, and we encourage users to inspect the maps rather than relying on the peak catalog alone. (3) Comparing IGM tomography and galaxy density maps requires accounting for noise in both tracers and, ideally, reference to mock surveys.

\subsection{Outlook}

We have focused primarily on the IGM-derived properties of our sample. In the companion paper (N25), we consider the LBG content of these structures. The IGM maps and the catalog presented in this paper will enable a variety of other studies of environment-dependent galaxy evolution in the cosmic noon era, which we are currently pursuing (B. Chartab et al. 2025, in preparation). These include the influence of the large-scale environment on star formation, accretion and outflows as traced by galaxy metallicity, and quenching. Such work will benefit from the rich ancillary data available within the LATIS fields. In particular, we note that half of LATIS lies within COSMOS and enjoys substantial overlap with the large JWST imaging survey COSMOS Web \citep{Casey23} and approved spectroscopic surveys covering the same region.

\begin{acknowledgments}
This paper includes data gathered with the 6.5~m Magellan Telescopes located at Las Campanas Observatory, Chile. We gratefully acknowledge the support of the Observatory staff. A.B.N.~and S.B.~acknowledge support from the National Science Foundation under grant Nos.~2108014 and 2107821, respectively. B.C.L.~and D.H.~acknowledge support from NSF grant No.~1908422. S.B.~acknowledges funding from NASA ATP 80NSSC22K1897. B.C.L.~is supported by the international Gemini Observatory, a program of NSF NOIRLab, which is managed by the Association of Universities for Research in Astronomy (AURA) under a cooperative agreement with the U.S. National Science Foundation, on behalf of the Gemini partnership of Argentina, Brazil, Canada, Chile, the Republic of Korea, and the United States of America. 
\end{acknowledgments}

\bibliography{main}{}

\begin{thebibliography}{}
\expandafter\ifx\csname natexlab\endcsname\relax\def\natexlab#1{#1}\fi
\providecommand{\url}[1]{\href{#1}{#1}}
\providecommand{\dodoi}[1]{doi:~\href{http://doi.org/#1}{\nolinkurl{#1}}}
\providecommand{\doeprint}[1]{\href{http://ascl.net/#1}{\nolinkurl{http://ascl.net/#1}}}
\providecommand{\doarXiv}[1]{\href{https://arxiv.org/abs/#1}{\nolinkurl{https://arxiv.org/abs/#1}}}

\bibitem[{J. Ahrens {et~al.}(2005)Ahrens, Geveci, \& Law}]{ParaView}
Ahrens, J., Geveci, B., \& Law, C. 2005, Visualization Handbook, ed. C.~D. Hansen \& C.~R. Johnson (Burlington, MA, USA: Elsevier Inc.), 717--731.
\newblock \url{https://www.sciencedirect.com/book/9780123875822/visualization-handbook}

\bibitem[{S. {Andreon} {et~al.}(2014){Andreon}, {Newman}, {Trinchieri}, {Raichoor}, {Ellis}, \& {Treu}}]{Andreon14}
{Andreon}, S., {Newman}, A.~B., {Trinchieri}, G., {et~al.} 2014, \bibinfo{title}{{JKCS 041: a Coma cluster progenitor at z = 1.803},} \aap, 565, A120, \dodoi{10.1051/0004-6361/201323077}

\bibitem[{S. {Andreon} {et~al.}(2023){Andreon}, {Romero}, {Aussel}, {Bhandarkar}, {Devlin}, {Dicker}, {Ladjelate}, {Lowe}, {Mason}, {Mroczkowski}, {Raichoor}, {Sarazin}, \& {Trinchieri}}]{Andreon23}
{Andreon}, S., {Romero}, C., {Aussel}, H., {et~al.} 2023, \bibinfo{title}{{Witnessing the intracluster medium assembly at the cosmic noon in JKCS 041},} \mnras, 522, 4301, \dodoi{10.1093/mnras/stad1270}

\bibitem[{F. {Arrigoni Battaia} {et~al.}(2018){Arrigoni Battaia}, {Chen}, {Fumagalli}, {Cai}, {Calistro Rivera}, {Xu}, {Smail}, {Prochaska}, {Yang}, \& {De Breuck}}]{ArrigoniBattaia18}
{Arrigoni Battaia}, F., {Chen}, C.-C., {Fumagalli}, M., {et~al.} 2018, \bibinfo{title}{{Overdensity of submillimeter galaxies around the z ≃ 2.3 MAMMOTH-1 nebula. The environment and powering of an enormous Lyman-{\ensuremath{\alpha}} nebula},} \aap, 620, A202, \dodoi{10.1051/0004-6361/201834195}

\bibitem[{M. {Ata} {et~al.}(2021){Ata}, {Kitaura}, {Lee}, {Lemaux}, {Kashino}, {Cucciati}, {Hern{\'a}ndez-S{\'a}nchez}, \& {Le F{\`e}vre}}]{Ata21}
{Ata}, M., {Kitaura}, F.-S., {Lee}, K.-G., {et~al.} 2021, \bibinfo{title}{{BIRTH of the COSMOS field: primordial and evolved density reconstructions during cosmic high noon},} \mnras, 500, 3194, \dodoi{10.1093/mnras/staa3318}

\bibitem[{M. {Ata} {et~al.}(2022){Ata}, {Lee}, {Vecchia}, {Kitaura}, {Cucciati}, {Lemaux}, {Kashino}, \& {M{\"u}ller}}]{Ata22}
{Ata}, M., {Lee}, K.-G., {Vecchia}, C.~D., {et~al.} 2022, \bibinfo{title}{{Predicted future fate of COSMOS galaxy protoclusters over 11 Gyr with constrained simulations},} Nature Astronomy, 6, 857, \dodoi{10.1038/s41550-022-01693-0}

\bibitem[{M. {Brodwin} {et~al.}(2011){Brodwin}, {Stern}, {Vikhlinin}, {Stanford}, {Gonzalez}, {Eisenhardt}, {Ashby}, {Bautz}, {Dey}, {Forman}, {Gettings}, {Hickox}, {Jannuzi}, {Jones}, {Mancone}, {Miller}, {Moustakas}, {Ruel}, {Snyder}, \& {Zeimann}}]{Brodwin11}
{Brodwin}, M., {Stern}, D., {Vikhlinin}, A., {et~al.} 2011, \bibinfo{title}{{X-ray Emission from Two Infrared-selected Galaxy Clusters at z > 1.4 in the IRAC Shallow Cluster Survey},} \apj, 732, 33, \dodoi{10.1088/0004-637X/732/1/33}

\bibitem[{M. {Brodwin} {et~al.}(2012){Brodwin}, {Gonzalez}, {Stanford}, {Plagge}, {Marrone}, {Carlstrom}, {Dey}, {Eisenhardt}, {Fedeli}, {Gettings}, {Jannuzi}, {Joy}, {Leitch}, {Mancone}, {Snyder}, {Stern}, \& {Zeimann}}]{Brodwin12}
{Brodwin}, M., {Gonzalez}, A.~H., {Stanford}, S.~A., {et~al.} 2012, \bibinfo{title}{{IDCS J1426.5+3508: Sunyaev-Zel'dovich Measurement of a Massive Infrared-selected Cluster at z = 1.75},} \apj, 753, 162, \dodoi{10.1088/0004-637X/753/2/162}

\bibitem[{Z. {Cai} {et~al.}(2016){Cai}, {Fan}, {Peirani}, {Bian}, {Frye}, {McGreer}, {Prochaska}, {Lau}, {Tejos}, {Ho}, \& {Schneider}}]{Cai16}
{Cai}, Z., {Fan}, X., {Peirani}, S., {et~al.} 2016, \bibinfo{title}{{Mapping the Most Massive Overdensity Through Hydrogen (MAMMOTH) I: Methodology},} \apj, 833, 135, \dodoi{10.3847/1538-4357/833/2/135}

\bibitem[{Z. {Cai} {et~al.}(2017){Cai}, {Fan}, {Bian}, {Zabludoff}, {Yang}, {Prochaska}, {McGreer}, {Zheng}, {Kashikawa}, {Wang}, {Frye}, {Green}, \& {Jiang}}]{Cai17B}
{Cai}, Z., {Fan}, X., {Bian}, F., {et~al.} 2017, \bibinfo{title}{{Mapping the Most Massive Overdensities through Hydrogen (MAMMOTH). II. Discovery of the Extremely Massive Overdensity BOSS1441 at z = 2.32},} \apj, 839, 131, \dodoi{10.3847/1538-4357/aa6a1a}

\bibitem[{C.~M. {Casey} {et~al.}(2015){Casey}, {Cooray}, {Capak}, {Fu}, {Kovac}, {Lilly}, {Sanders}, {Scoville}, \& {Treister}}]{Casey15}
{Casey}, C.~M., {Cooray}, A., {Capak}, P., {et~al.} 2015, \bibinfo{title}{{A Massive, Distant Proto-cluster at z = 2.47 Caught in a Phase of Rapid Formation?},} \apjl, 808, L33, \dodoi{10.1088/2041-8205/808/2/L33}

\bibitem[{C.~M. {Casey} {et~al.}(2023){Casey}, {Kartaltepe}, {Drakos}, {Franco}, {Harish}, {Paquereau}, {Ilbert}, {Rose}, {Cox}, {Nightingale}, {Robertson}, {Silverman}, {Koekemoer}, {Massey}, {McCracken}, {Rhodes}, {Akins}, {Allen}, {Amvrosiadis}, {Arango-Toro}, {Bagley}, {Bongiorno}, {Capak}, {Champagne}, {Chartab}, {Ch{\'a}vez Ortiz}, {Chworowsky}, {Cooke}, {Cooper}, {Darvish}, {Ding}, {Faisst}, {Finkelstein}, {Fujimoto}, {Gentile}, {Gillman}, {Gould}, {Gozaliasl}, {Hayward}, {He}, {Hemmati}, {Hirschmann}, {Jahnke}, {Jin}, {Khostovan}, {Kokorev}, {Lambrides}, {Laigle}, {Larson}, {Leung}, {Liu}, {Liaudat}, {Long}, {Magdis}, {Mahler}, {Mainieri}, {Manning}, {Maraston}, {Martin}, {McCleary}, {McKinney}, {McPartland}, {Mobasher}, {Pattnaik}, {Renzini}, {Rich}, {Sanders}, {Sattari}, {Scognamiglio}, {Scoville}, {Sheth}, {Shuntov}, {Sparre}, {Suzuki}, {Talia}, {Toft}, {Trakhtenbrot}, {Urry}, {Valentino}, {Vanderhoof}, {Vardoulaki}, {Weaver}, {Whitaker}, {Wilkins}, {Yang}, \& {Zavala}}]{Casey23}
{Casey}, C.~M., {Kartaltepe}, J.~S., {Drakos}, N.~E., {et~al.} 2023, \bibinfo{title}{{COSMOS-Web: An Overview of the JWST Cosmic Origins Survey},} \apj, 954, 31, \dodoi{10.3847/1538-4357/acc2bc}

\bibitem[{G. {Castignani} {et~al.}(2014){Castignani}, {Chiaberge}, {Celotti}, {Norman}, \& {De Zotti}}]{Castignani14}
{Castignani}, G., {Chiaberge}, M., {Celotti}, A., {Norman}, C., \& {De Zotti}, G. 2014, \bibinfo{title}{{Cluster Candidates around Low-power Radio Galaxies at z \raisebox{-0.5ex}\textasciitilde 1-2 in COSMOS},} \apj, 792, 114, \dodoi{10.1088/0004-637X/792/2/114}

\bibitem[{S. {Caucci} {et~al.}(2008){Caucci}, {Colombi}, {Pichon}, {Rollinde}, {Petitjean}, \& {Sousbie}}]{Caucci08}
{Caucci}, S., {Colombi}, S., {Pichon}, C., {et~al.} 2008, \bibinfo{title}{{Recovering the topology of the intergalactic medium at z \raisebox{-0.5ex}\textasciitilde 2},} \mnras, 386, 211, \dodoi{10.1111/j.1365-2966.2008.13016.x}

\bibitem[{J.~B. {Champagne} {et~al.}(2021){Champagne}, {Casey}, {Zavala}, {Cooray}, {Dannerbauer}, {Fabian}, {Hayward}, {Long}, \& {Spilker}}]{Champagne21}
{Champagne}, J.~B., {Casey}, C.~M., {Zavala}, J.~A., {et~al.} 2021, \bibinfo{title}{{Comprehensive Gas Characterization of a z = 2.5 Protocluster: A Cluster Core Caught in the Beginning of Virialization?},} \apj, 913, 110, \dodoi{10.3847/1538-4357/abf4e6}

\bibitem[{Y.-K. {Chiang} {et~al.}(2014){Chiang}, {Overzier}, \& {Gebhardt}}]{Chiang14}
{Chiang}, Y.-K., {Overzier}, R., \& {Gebhardt}, K. 2014, \bibinfo{title}{{Discovery of a Large Number of Candidate Protoclusters Traced by \raisebox{-0.5ex}\textasciitilde15 Mpc-scale Galaxy Overdensities in COSMOS},} \apjl, 782, L3, \dodoi{10.1088/2041-8205/782/1/L3}

\bibitem[{Y.-K. {Chiang} {et~al.}(2017){Chiang}, {Overzier}, {Gebhardt}, \& {Henriques}}]{Chiang17}
{Chiang}, Y.-K., {Overzier}, R.~A., {Gebhardt}, K., \& {Henriques}, B. 2017, \bibinfo{title}{{Galaxy Protoclusters as Drivers of Cosmic Star Formation History in the First 2 Gyr},} \apjl, 844, L23, \dodoi{10.3847/2041-8213/aa7e7b}

\bibitem[{Y.-K. {Chiang} {et~al.}(2015){Chiang}, {Overzier}, {Gebhardt}, {Finkelstein}, {Chiang}, {Hill}, {Blanc}, {Drory}, {Chonis}, {Zeimann}, {Hagen}, {Schneider}, {Jogee}, {Ciardullo}, \& {Gronwall}}]{Chiang15}
{Chiang}, Y.-K., {Overzier}, R.~A., {Gebhardt}, K., {et~al.} 2015, \bibinfo{title}{{Surveying Galaxy Proto-clusters in Emission: A Large-scale Structure at z = 2.44 and the Outlook for HETDEX},} \apj, 808, 37, \dodoi{10.1088/0004-637X/808/1/37}

\bibitem[{R.~A.~C. {Croft} {et~al.}(1998){Croft}, {Weinberg}, {Katz}, \& {Hernquist}}]{Croft98}
{Croft}, R. A.~C., {Weinberg}, D.~H., {Katz}, N., \& {Hernquist}, L. 1998, \bibinfo{title}{{Recovery of the Power Spectrum of Mass Fluctuations from Observations of the Ly{\ensuremath{\alpha}} Forest},} \apj, 495, 44, \dodoi{10.1086/305289}

\bibitem[{O. {Cucciati} {et~al.}(2018){Cucciati}, {Lemaux}, {Zamorani}, {Le F{\`e}vre}, {Tasca}, {Hathi}, {Lee}, {Bardelli}, {Cassata}, {Garilli}, {Le Brun}, {Maccagni}, {Pentericci}, {Thomas}, {Vanzella}, {Zucca}, {Lubin}, {Amorin}, {Cassar{\`a}}, {Cimatti}, {Talia}, {Vergani}, {Koekemoer}, {Pforr}, \& {Salvato}}]{Cucciati18}
{Cucciati}, O., {Lemaux}, B.~C., {Zamorani}, G., {et~al.} 2018, \bibinfo{title}{{The progeny of a cosmic titan: a massive multi-component proto-supercluster in formation at z = 2.45 in VUDS},} \aap, 619, A49, \dodoi{10.1051/0004-6361/201833655}

\bibitem[{B. {Darvish} {et~al.}(2020){Darvish}, {Scoville}, {Martin}, {Sobral}, {Mobasher}, {Rettura}, {Matthee}, {Capak}, {Chartab}, {Hemmati}, {Masters}, {Nayyeri}, {O'Sullivan}, {Paulino-Afonso}, {Sattari}, {Shahidi}, {Salvato}, {Lemaux}, {F{\`e}vre}, \& {Cucciati}}]{Darvish20}
{Darvish}, B., {Scoville}, N.~Z., {Martin}, C., {et~al.} 2020, \bibinfo{title}{{Spectroscopic Confirmation of a Coma Cluster Progenitor at z {\ensuremath{\sim}} 2.2},} \apj, 892, 8, \dodoi{10.3847/1538-4357/ab75c3}

\bibitem[{R. {Dav{\'e}} {et~al.}(2019){Dav{\'e}}, {Angl{\'e}s-Alc{\'a}zar}, {Narayanan}, {Li}, {Rafieferantsoa}, \& {Appleby}}]{Dave19}
{Dav{\'e}}, R., {Angl{\'e}s-Alc{\'a}zar}, D., {Narayanan}, D., {et~al.} 2019, \bibinfo{title}{{SIMBA: Cosmological simulations with black hole growth and feedback},} \mnras, 486, 2827, \dodoi{10.1093/mnras/stz937}

\bibitem[{L. {Di Mascolo} {et~al.}(2023){Di Mascolo}, {Saro}, {Mroczkowski}, {Borgani}, {Churazov}, {Rasia}, {Tozzi}, {Dannerbauer}, {Basu}, {Carilli}, {Ginolfi}, {Miley}, {Nonino}, {Pannella}, {Pentericci}, \& {Rizzo}}]{DiMascolo23}
{Di Mascolo}, L., {Saro}, A., {Mroczkowski}, T., {et~al.} 2023, \bibinfo{title}{{Forming intracluster gas in a galaxy protocluster at a redshift of 2.16},} \nat, 615, 809, \dodoi{10.1038/s41586-023-05761-x}

\bibitem[{C. {Diener} {et~al.}(2013){Diener}, {Lilly}, {Knobel}, {Zamorani}, {Lemson}, {Kampczyk}, {Scoville}, {Carollo}, {Contini}, {Kneib}, {Le Fevre}, {Mainieri}, {Renzini}, {Scodeggio}, {Bardelli}, {Bolzonella}, {Bongiorno}, {Caputi}, {Cucciati}, {de la Torre}, {de Ravel}, {Franzetti}, {Garilli}, {Iovino}, {Kova{\v{c}}}, {Lamareille}, {Le Borgne}, {Le Brun}, {Maier}, {Mignoli}, {Pello}, {Peng}, {Perez Montero}, {Presotto}, {Silverman}, {Tanaka}, {Tasca}, {Tresse}, {Vergani}, {Zucca}, {Bordoloi}, {Cappi}, {Cimatti}, {Coppa}, {Koekemoer}, {L{\'o}pez-Sanjuan}, {McCracken}, {Moresco}, {Nair}, {Pozzetti}, \& {Welikala}}]{Diener13}
{Diener}, C., {Lilly}, S.~J., {Knobel}, C., {et~al.} 2013, \bibinfo{title}{{Proto-groups at 1.8 \&lt; z \&lt; 3 in the zCOSMOS-deep Sample},} \apj, 765, 109, \dodoi{10.1088/0004-637X/765/2/109}

\bibitem[{C. {Dong} {et~al.}(2023){Dong}, {Lee}, {Ata}, {Horowitz}, \& {Momose}}]{Dong23}
{Dong}, C., {Lee}, K.-G., {Ata}, M., {Horowitz}, B., \& {Momose}, R. 2023, \bibinfo{title}{{Observational Evidence for Large-scale Gas Heating in a Galaxy Protocluster at z = 2.30},} \apjl, 945, L28, \dodoi{10.3847/2041-8213/acba89}

\bibitem[{C. {Dong} {et~al.}(2024){Dong}, {Lee}, {Cui}, {Dav{\'e}}, \& {Sorini}}]{Dong24}
{Dong}, C., {Lee}, K.-G., {Cui}, W., {Dav{\'e}}, R., \& {Sorini}, D. 2024, \bibinfo{title}{{The effect of AGN feedback on the Lyman-{\ensuremath{\alpha}} forest signature of galaxy protoclusters at z 2.3},} \mnras, 532, 4876, \dodoi{10.1093/mnras/stae1830}

\bibitem[{J.~R. {Franck} \& S.~S. {McGaugh}(2016){Franck} \& {McGaugh}}]{Franck16}
{Franck}, J.~R., \& {McGaugh}, S.~S. 2016, \bibinfo{title}{{The Candidate Cluster and Protocluster Catalog (CCPC) II. Spectroscopically Identified Structures Spanning 2< z < 6.6},} \apj, 833, 15, \dodoi{10.3847/0004-637X/833/1/15}

\bibitem[{R. {Gobat} {et~al.}(2011){Gobat}, {Daddi}, {Onodera}, {Finoguenov}, {Renzini}, {Arimoto}, {Bouwens}, {Brusa}, {Chary}, {Cimatti}, {Dickinson}, {Kong}, \& {Mignoli}}]{Gobat11}
{Gobat}, R., {Daddi}, E., {Onodera}, M., {et~al.} 2011, \bibinfo{title}{{A mature cluster with X-ray emission at z = 2.07},} \aap, 526, A133, \dodoi{10.1051/0004-6361/201016084}

\bibitem[{J.~E. {Gunn} \& B.~A. {Peterson}(1965){Gunn} \& {Peterson}}]{Gunn65}
{Gunn}, J.~E., \& {Peterson}, B.~A. 1965, \bibinfo{title}{{On the Density of Neutral Hydrogen in Intergalactic Space.},} \apj, 142, 1633, \dodoi{10.1086/148444}

\bibitem[{B. {Horowitz} {et~al.}(2019){Horowitz}, {Lee}, {White}, {Krolewski}, \& {Ata}}]{Horowitz19}
{Horowitz}, B., {Lee}, K.-G., {White}, M., {Krolewski}, A., \& {Ata}, M. 2019, \bibinfo{title}{{TARDIS. I. A Constrained Reconstruction Approach to Modeling the z {\ensuremath{\sim}} 2.5 Cosmic Web Probed by Ly{\ensuremath{\alpha}} Forest Tomography},} \apj, 887, 61, \dodoi{10.3847/1538-4357/ab4d4c}

\bibitem[{B. {Horowitz} {et~al.}(2022){Horowitz}, {Lee}, {Ata}, {M{\"u}ller}, {Krolewski}, {Prochaska}, {Hennawi}, {White}, {Schlegel}, {Rich}, {Nugent}, {Suzuki}, {Kashino}, {Koekemoer}, \& {Lemaux}}]{Horowitz22}
{Horowitz}, B., {Lee}, K.-G., {Ata}, M., {et~al.} 2022, \bibinfo{title}{{Second Data Release of the COSMOS Ly{\ensuremath{\alpha}} Mapping and Tomography Observations: The First 3D Maps of the Detailed Cosmic Web at 2.05 < z < 2.55},} \apjs, 263, 27, \dodoi{10.3847/1538-4365/ac982d}

\bibitem[{W. {Hu} {et~al.}(2021){Hu}, {Wang}, {Infante}, {Rhoads}, {Zheng}, {Yang}, {Malhotra}, {Barrientos}, {Jiang}, {Gonz{\'a}lez-L{\'o}pez}, {Prieto}, {Perez}, {Hibon}, {Galaz}, {Coughlin}, {Harish}, {Kong}, {Kang}, {Khostovan}, {Pharo}, {Valdes}, {Wold}, {Walker}, \& {Zheng}}]{Hu21}
{Hu}, W., {Wang}, J., {Infante}, L., {et~al.} 2021, \bibinfo{title}{{A Lyman-{\ensuremath{\alpha}} protocluster at redshift 6.9},} Nature Astronomy, 5, 485, \dodoi{10.1038/s41550-020-01291-y}

\bibitem[{Y. {Huang} {et~al.}(2022){Huang}, {Lee}, {Cucciati}, {Lemaux}, {Sawicki}, {Malavasi}, {Ramakrishnan}, {Xue}, {Cassara}, {Chiang}, {Dey}, {Gwyn}, {Hathi}, {Pentericci}, {Prescott}, \& {Zamorani}}]{Huang22}
{Huang}, Y., {Lee}, K.-S., {Cucciati}, O., {et~al.} 2022, \bibinfo{title}{{Evaluating Ly{\ensuremath{\alpha}} Emission as a Tracer of the Largest Cosmic Structure at z 2.47},} \apj, 941, 134, \dodoi{10.3847/1538-4357/ac9ea4}

\bibitem[{K. {Ito} {et~al.}(2023){Ito}, {Tanaka}, {Valentino}, {Toft}, {Brammer}, {Gould}, {Ilbert}, {Kashikawa}, {Kubo}, {Liang}, {McCracken}, \& {Weaver}}]{Ito23}
{Ito}, K., {Tanaka}, M., {Valentino}, F., {et~al.} 2023, \bibinfo{title}{{COSMOS2020: Discovery of a Protocluster of Massive Quiescent Galaxies at z = 2.77},} \apjl, 945, L9, \dodoi{10.3847/2041-8213/acb49b}

\bibitem[{A. {Klypin} {et~al.}(2016){Klypin}, {Yepes}, {Gottl{\"o}ber}, {Prada}, \& {He{\ss}}}]{Klypin16}
{Klypin}, A., {Yepes}, G., {Gottl{\"o}ber}, S., {Prada}, F., \& {He{\ss}}, S. 2016, \bibinfo{title}{{MultiDark simulations: the story of dark matter halo concentrations and density profiles},} \mnras, 457, 4340, \dodoi{10.1093/mnras/stw248}

\bibitem[{M. {Kriek} {et~al.}(2015){Kriek}, {Shapley}, {Reddy}, {Siana}, {Coil}, {Mobasher}, {Freeman}, {de Groot}, {Price}, {Sanders}, {Shivaei}, {Brammer}, {Momcheva}, {Skelton}, {van Dokkum}, {Whitaker}, {Aird}, {Azadi}, {Kassis}, {Bullock}, {Conroy}, {Dav{\'e}}, {Kere{\v s}}, \& {Krumholz}}]{Kriek15}
{Kriek}, M., {Shapley}, A.~E., {Reddy}, N.~A., {et~al.} 2015, \bibinfo{title}{{The MOSFIRE Deep Evolution Field (MOSDEF) Survey: Rest-frame Optical Spectroscopy for \~{}1500 H-selected Galaxies at 1.37 $\lt$ z $\lt$ 3.8},} \apjs, 218, 15, \dodoi{10.1088/0067-0049/218/2/15}

\bibitem[{A. {Krolewski} {et~al.}(2018){Krolewski}, {Lee}, {White}, {Hennawi}, {Schlegel}, {Nugent}, {Luki{\'c}}, {Stark}, {Koekemoer}, {Le F{\`e}vre}, {Lemaux}, {Maier}, {Rich}, {Salvato}, \& {Tasca}}]{Krolewski18}
{Krolewski}, A., {Lee}, K.-G., {White}, M., {et~al.} 2018, \bibinfo{title}{{Detection of z {\ensuremath{\sim}} 2.3 Cosmic Voids from 3D Ly{\ensuremath{\alpha}} Forest Tomography in the COSMOS Field},} \apj, 861, 60, \dodoi{10.3847/1538-4357/aac829}

\bibitem[{O. {Le F{\`e}vre} {et~al.}(2015){Le F{\`e}vre}, {Tasca}, {Cassata}, {Garilli}, {Le Brun}, {Maccagni}, {Pentericci}, {Thomas}, {Vanzella}, {Zamorani}, {Zucca}, {Amorin}, {Bardelli}, {Capak}, {Cassar{\`a}}, {Castellano}, {Cimatti}, {Cuby}, {Cucciati}, {de la Torre}, {Durkalec}, {Fontana}, {Giavalisco}, {Grazian}, {Hathi}, {Ilbert}, {Lemaux}, {Moreau}, {Paltani}, {Ribeiro}, {Salvato}, {Schaerer}, {Scodeggio}, {Sommariva}, {Talia}, {Taniguchi}, {Tresse}, {Vergani}, {Wang}, {Charlot}, {Contini}, {Fotopoulou}, {L{\'o}pez-Sanjuan}, {Mellier}, \& {Scoville}}]{LeFevre15}
{Le F{\`e}vre}, O., {Tasca}, L.~A.~M., {Cassata}, P., {et~al.} 2015, \bibinfo{title}{{The VIMOS Ultra-Deep Survey: \raisebox{-0.5ex}\textasciitilde10 000 galaxies with spectroscopic redshifts to study galaxy assembly at early epochs 2 < z ≃ 6},} \aap, 576, A79, \dodoi{10.1051/0004-6361/201423829}

\bibitem[{K.-G. {Lee} {et~al.}(2014{\natexlab{a}}){Lee}, {Hennawi}, {White}, {Croft}, \& {Ozbek}}]{Lee14A}
{Lee}, K.-G., {Hennawi}, J.~F., {White}, M., {Croft}, R. A.~C., \& {Ozbek}, M. 2014{\natexlab{a}}, \bibinfo{title}{{Observational Requirements for Ly{\ensuremath{\alpha}} Forest Tomographic Mapping of Large-scale Structure at z \raisebox{-0.5ex}\textasciitilde 2},} \apj, 788, 49, \dodoi{10.1088/0004-637X/788/1/49}

\bibitem[{K.-G. {Lee} {et~al.}(2014{\natexlab{b}}){Lee}, {Hennawi}, {Stark}, {Prochaska}, {White}, {Schlegel}, {Eilers}, {Arinyo-i-Prats}, {Suzuki}, {Croft}, {Caputi}, {Cassata}, {Ilbert}, {Garilli}, {Koekemoer}, {Le Brun}, {Le F{\`e}vre}, {Maccagni}, {Nugent}, {Taniguchi}, {Tasca}, {Tresse}, {Zamorani}, \& {Zucca}}]{Lee14B}
{Lee}, K.-G., {Hennawi}, J.~F., {Stark}, C., {et~al.} 2014{\natexlab{b}}, \bibinfo{title}{{Ly{\ensuremath{\alpha}} Forest Tomography from Background Galaxies: The First Megaparsec-resolution Large-scale Structure Map at z > 2},} \apjl, 795, L12, \dodoi{10.1088/2041-8205/795/1/L12}

\bibitem[{K.-G. {Lee} {et~al.}(2016){Lee}, {Hennawi}, {White}, {Prochaska}, {Font-Ribera}, {Schlegel}, {Rich}, {Suzuki}, {Stark}, {Le F{\`e}vre}, {Nugent}, {Salvato}, \& {Zamorani}}]{Lee16}
{Lee}, K.-G., {Hennawi}, J.~F., {White}, M., {et~al.} 2016, \bibinfo{title}{{Shadow of a Colossus: A z = 2.44 Galaxy Protocluster Detected in 3D Ly{\ensuremath{\alpha}} Forest Tomographic Mapping of the COSMOS Field},} \apj, 817, 160, \dodoi{10.3847/0004-637X/817/2/160}

\bibitem[{K.-G. {Lee} {et~al.}(2018){Lee}, {Krolewski}, {White}, {Schlegel}, {Nugent}, {Hennawi}, {M{\"u}ller}, {Pan}, {Prochaska}, {Font-Ribera}, {Suzuki}, {Glazebrook}, {Kacprzak}, {Kartaltepe}, {Koekemoer}, {Le F{\`e}vre}, {Lemaux}, {Maier}, {Nanayakkara}, {Rich}, {Sanders}, {Salvato}, {Tasca}, \& {Tran}}]{Lee18}
{Lee}, K.-G., {Krolewski}, A., {White}, M., {et~al.} 2018, \bibinfo{title}{{First Data Release of the COSMOS Ly{\ensuremath{\alpha}} Mapping and Tomography Observations: 3D Ly{\ensuremath{\alpha}} Forest Tomography at 2.05 < z < 2.55},} \apjs, 237, 31, \dodoi{10.3847/1538-4365/aace58}

\bibitem[{Y. {Liang} {et~al.}(2021){Liang}, {Kashikawa}, {Cai}, {Fan}, {Prochaska}, {Shimasaku}, {Tanaka}, {Uchiyama}, {Ito}, {Shimakawa}, {Nagamine}, {Shimizu}, {Onoue}, \& {Toshikawa}}]{Liang21}
{Liang}, Y., {Kashikawa}, N., {Cai}, Z., {et~al.} 2021, \bibinfo{title}{{Statistical Correlation between the Distribution of Ly{\ensuremath{\alpha}} Emitters and Intergalactic Medium H I at z {\ensuremath{\sim}} 2.2 Mapped by the Subaru/Hyper Suprime-Cam},} \apj, 907, 3, \dodoi{10.3847/1538-4357/abcd93}

\bibitem[{S.~J. {Lilly} {et~al.}(2007){Lilly}, {Le F{\`e}vre}, {Renzini}, {Zamorani}, {Scodeggio}, {Contini}, {Carollo}, {Hasinger}, {Kneib}, {Iovino}, {Le Brun}, {Maier}, {Mainieri}, {Mignoli}, {Silverman}, {Tasca}, {Bolzonella}, {Bongiorno}, {Bottini}, {Capak}, {Caputi}, {Cimatti}, {Cucciati}, {Daddi}, {Feldmann}, {Franzetti}, {Garilli}, {Guzzo}, {Ilbert}, {Kampczyk}, {Kovac}, {Lamareille}, {Leauthaud}, {Le Borgne}, {McCracken}, {Marinoni}, {Pello}, {Ricciardelli}, {Scarlata}, {Vergani}, {Sanders}, {Schinnerer}, {Scoville}, {Taniguchi}, {Arnouts}, {Aussel}, {Bardelli}, {Brusa}, {Cappi}, {Ciliegi}, {Finoguenov}, {Foucaud}, {Franceschini}, {Halliday}, {Impey}, {Knobel}, {Koekemoer}, {Kurk}, {Maccagni}, {Maddox}, {Marano}, {Marconi}, {Meneux}, {Mobasher}, {Moreau}, {Peacock}, {Porciani}, {Pozzetti}, {Scaramella}, {Schiminovich}, {Shopbell}, {Smail}, {Thompson}, {Tresse}, {Vettolani}, {Zanichelli}, \& {Zucca}}]{Lilly07}
{Lilly}, S.~J., {Le F{\`e}vre}, O., {Renzini}, A., {et~al.} 2007, \bibinfo{title}{{zCOSMOS: A Large VLT/VIMOS Redshift Survey Covering 0 < z < 3 in the COSMOS Field},} \apjs, 172, 70, \dodoi{10.1086/516589}

\bibitem[{A.~B. {Mantz} {et~al.}(2020){Mantz}, {Allen}, {Morris}, {Canning}, {Bayliss}, {Bleem}, {Floyd}, \& {McDonald}}]{Mantz20}
{Mantz}, A.~B., {Allen}, S.~W., {Morris}, R.~G., {et~al.} 2020, \bibinfo{title}{{Deep XMM-Newton observations of the most distant SPT-SZ galaxy cluster},} \mnras, 496, 1554, \dodoi{10.1093/mnras/staa1581}

\bibitem[{A.~B. {Mantz} {et~al.}(2014){Mantz}, {Abdulla}, {Carlstrom}, {Greer}, {Leitch}, {Marrone}, {Muchovej}, {Adami}, {Birkinshaw}, {Bremer}, {Clerc}, {Giles}, {Horellou}, {Maughan}, {Pacaud}, {Pierre}, \& {Willis}}]{Mantz14}
{Mantz}, A.~B., {Abdulla}, Z., {Carlstrom}, J.~E., {et~al.} 2014, \bibinfo{title}{{The XXL Survey. V. Detection of the Sunyaev-Zel'dovich Effect of the Redshift 1.9 Galaxy Cluster XLSSU J021744.1-034536 with CARMA},} \apj, 794, 157, \dodoi{10.1088/0004-637X/794/2/157}

\bibitem[{A.~B. {Mantz} {et~al.}(2018){Mantz}, {Abdulla}, {Allen}, {Carlstrom}, {Logan}, {Marrone}, {Maughan}, {Willis}, {Pacaud}, \& {Pierre}}]{Mantz18}
{Mantz}, A.~B., {Abdulla}, Z., {Allen}, S.~W., {et~al.} 2018, \bibinfo{title}{{The XXL Survey. XVII. X-ray and Sunyaev-Zel'dovich properties of the redshift 2.0 galaxy cluster XLSSC 122},} \aap, 620, A2, \dodoi{10.1051/0004-6361/201630096}

\bibitem[{I. {McConachie} {et~al.}(2022){McConachie}, {Wilson}, {Forrest}, {Marsan}, {Muzzin}, {Cooper}, {Annunziatella}, {Marchesini}, {Chan}, {Gomez}, {Abdullah}, {Saracco}, \& {Nantais}}]{McConachie22}
{McConachie}, I., {Wilson}, G., {Forrest}, B., {et~al.} 2022, \bibinfo{title}{{Spectroscopic Confirmation of a Protocluster at z = 3.37 with a High Fraction of Quiescent Galaxies},} \apj, 926, 37, \dodoi{10.3847/1538-4357/ac2b9f}

\bibitem[{J.~S.~A. {Miller} {et~al.}(2019){Miller}, {Bolton}, \& {Hatch}}]{Miller19}
{Miller}, J. S.~A., {Bolton}, J.~S., \& {Hatch}, N. 2019, \bibinfo{title}{{Searching for the shadows of giants: characterizing protoclusters with line of sight Lyman-{\ensuremath{\alpha}} absorption},} \mnras, 489, 5381, \dodoi{10.1093/mnras/stz2504}

\bibitem[{J.~S.~A. {Miller} {et~al.}(2021){Miller}, {Bolton}, \& {Hatch}}]{Miller21}
{Miller}, J. S.~A., {Bolton}, J.~S., \& {Hatch}, N.~A. 2021, \bibinfo{title}{{Searching for the shadows of giants - II. The effect of local ionization on the Ly {\ensuremath{\alpha}} absorption signatures of protoclusters at redshift z 2.4},} \mnras, 506, 6001, \dodoi{10.1093/mnras/stab2083}

\bibitem[{S.~I. {Muldrew} {et~al.}(2018){Muldrew}, {Hatch}, \& {Cooke}}]{Muldrew18}
{Muldrew}, S.~I., {Hatch}, N.~A., \& {Cooke}, E.~A. 2018, \bibinfo{title}{{Galaxy evolution in protoclusters},} \mnras, 473, 2335, \dodoi{10.1093/mnras/stx2454}

\bibitem[{T. {Nanayakkara} {et~al.}(2016){Nanayakkara}, {Glazebrook}, {Kacprzak}, {Yuan}, {Tran}, {Spitler}, {Kewley}, {Straatman}, {Cowley}, {Fisher}, {Labbe}, {Tomczak}, {Allen}, \& {Alcorn}}]{Nanayakkara16}
{Nanayakkara}, T., {Glazebrook}, K., {Kacprzak}, G.~G., {et~al.} 2016, \bibinfo{title}{{ZFIRE: A KECK/MOSFIRE Spectroscopic Survey of Galaxies in Rich Environments at z {\tilde} 2},} \apj, 828, 21, \dodoi{10.3847/0004-637X/828/1/21}

\bibitem[{D. {Nelson} {et~al.}(2019){Nelson}, {Springel}, {Pillepich}, {Rodriguez-Gomez}, {Torrey}, {Genel}, {Vogelsberger}, {Pakmor}, {Marinacci}, {Weinberger}, {Kelley}, {Lovell}, {Diemer}, \& {Hernquist}}]{Nelson19}
{Nelson}, D., {Springel}, V., {Pillepich}, A., {et~al.} 2019, \bibinfo{title}{{The IllustrisTNG simulations: public data release},} Computational Astrophysics and Cosmology, 6, 2, \dodoi{10.1186/s40668-019-0028-x}

\bibitem[{A.~B. {Newman} {et~al.}(2014){Newman}, {Ellis}, {Andreon}, {Treu}, {Raichoor}, \& {Trinchieri}}]{Newman14}
{Newman}, A.~B., {Ellis}, R.~S., {Andreon}, S., {et~al.} 2014, \bibinfo{title}{{Spectroscopic Confirmation of the Rich z = 1.80 Galaxy Cluster JKCS 041 using the WFC3 Grism: Environmental Trends in the Ages and Structure of Quiescent Galaxies},} \apj, 788, 51, \dodoi{10.1088/0004-637X/788/1/51}

\bibitem[{A.~B. {Newman} {et~al.}(2020){Newman}, {Rudie}, {Blanc}, {Kelson}, {Rhoades}, {Hare}, {P{\'e}rez}, {Benson}, {Dressler}, {Gonzalez}, {Kollmeier}, {Konidaris}, {Mulchaey}, {Rauch}, {Le F{\`e}vre}, {Lemaux}, {Cucciati}, \& {Lilly}}]{Newman20}
{Newman}, A.~B., {Rudie}, G.~C., {Blanc}, G.~A., {et~al.} 2020, \bibinfo{title}{{LATIS: The Ly{\ensuremath{\alpha}} Tomography IMACS Survey},} \apj, 891, 147, \dodoi{10.3847/1538-4357/ab75ee}

\bibitem[{A.~B. {Newman} {et~al.}(2022){Newman}, {Rudie}, {Blanc}, {Qezlou}, {Bird}, {Kelson}, {P{\'e}rez}, {Congiu}, {Lemaux}, {Dressler}, \& {Mulchaey}}]{Newman22}
{Newman}, A.~B., {Rudie}, G.~C., {Blanc}, G.~A., {et~al.} 2022, \bibinfo{title}{{A population of ultraviolet-dim protoclusters detected in absorption},} \nat, 606, 475, \dodoi{10.1038/s41586-022-04681-6}

\bibitem[{A.~B. {Newman} {et~al.}(2024){Newman}, {Qezlou}, {Chartab}, {Rudie}, {Blanc}, {Bird}, {Benson}, {Kelson}, \& {Lemaux}}]{Newman24}
{Newman}, A.~B., {Qezlou}, M., {Chartab}, N., {et~al.} 2024, \bibinfo{title}{{LATIS: Constraints on the Galaxy-Halo Connection at z 2.5 from Galaxy-Galaxy and Galaxy-Ly{\ensuremath{\alpha}} Clustering},} \apj, 961, 27, \dodoi{10.3847/1538-4357/ad0896}

\bibitem[{A.~B. {Newman} {et~al.}(2025){Newman}, {Chartab}, {Qezlou}, {Rudie}, {Blanc}, {Kelson}, {Bird}, {Casey}, {Congiu}, {Cucciati}, {Hung}, {Lemaux}, {Pérez}, \& {Zavala}}]{Newman25}
{Newman}, A.~B., {Chartab}, N., {Qezlou}, M., {et~al.} 2025, \bibinfo{title}{{LATIS: Comparing Galaxy and IGM Tomography Maps as Tracers of Large-Scale Structure and Protoclusters at z ∼ 2.5},} \apj, 988, 48, \dodoi{10.3847/1538-4357/ade3c3}

\bibitem[{R.~A. {Overzier}(2016){Overzier}}]{Overzier16}
{Overzier}, R.~A. 2016, \bibinfo{title}{{The realm of the galaxy protoclusters. A review},} \aapr, 24, 14, \dodoi{10.1007/s00159-016-0100-3}

\bibitem[{C. {Pichon} {et~al.}(2001){Pichon}, {Vergely}, {Rollinde}, {Colombi}, \& {Petitjean}}]{Pichon01}
{Pichon}, C., {Vergely}, J.~L., {Rollinde}, E., {Colombi}, S., \& {Petitjean}, P. 2001, \bibinfo{title}{{Inversion of the Lyman {\ensuremath{\alpha}} forest: three-dimensional investigation of the intergalactic medium},} \mnras, 326, 597, \dodoi{10.1046/j.1365-8711.2001.04595.x}

\bibitem[{ {Planck Collaboration} {et~al.}(2016){Planck Collaboration}, {Ade}, {Aghanim}, {Arnaud}, {Ashdown}, {Aumont}, {Baccigalupi}, {Banday}, {Barreiro}, {Bartlett}, {Bartolo}, {Battaner}, {Battye}, {Benabed}, {Beno{\^\i}t}, {Benoit-L{\'e}vy}, {Bernard}, {Bersanelli}, {Bielewicz}, {Bock}, {Bonaldi}, {Bonavera}, {Bond}, {Borrill}, {Bouchet}, {Boulanger}, {Bucher}, {Burigana}, {Butler}, {Calabrese}, {Cardoso}, {Catalano}, {Challinor}, {Chamballu}, {Chary}, {Chiang}, {Chluba}, {Christensen}, {Church}, {Clements}, {Colombi}, {Colombo}, {Combet}, {Coulais}, {Crill}, {Curto}, {Cuttaia}, {Danese}, {Davies}, {Davis}, {de Bernardis}, {de Rosa}, {de Zotti}, {Delabrouille}, {D{\'e}sert}, {Di Valentino}, {Dickinson}, {Diego}, {Dolag}, {Dole}, {Donzelli}, {Dor{\'e}}, {Douspis}, {Ducout}, {Dunkley}, {Dupac}, {Efstathiou}, {Elsner}, {En{\ss}lin}, {Eriksen}, {Farhang}, {Fergusson}, {Finelli}, {Forni}, {Frailis}, {Fraisse}, {Franceschi}, {Frejsel}, {Galeotta}, {Galli}, {Ganga}, {Gauthier}, {Gerbino}, {Ghosh}, {Giard},
  {Giraud-H{\'e}raud}, {Giusarma}, {Gjerl{\o}w}, {Gonz{\'a}lez-Nuevo}, {G{\'o}rski}, {Gratton}, {Gregorio}, {Gruppuso}, {Gudmundsson}, {Hamann}, {Hansen}, {Hanson}, {Harrison}, {Helou}, {Henrot-Versill{\'e}}, {Hern{\'a}ndez-Monteagudo}, {Herranz}, {Hildebrandt}, {Hivon}, {Hobson}, {Holmes}, {Hornstrup}, {Hovest}, {Huang}, {Huffenberger}, {Hurier}, {Jaffe}, {Jaffe}, {Jones}, {Juvela}, {Keih{\"a}nen}, {Keskitalo}, {Kisner}, {Kneissl}, {Knoche}, {Knox}, {Kunz}, {Kurki-Suonio}, {Lagache}, {L{\"a}hteenm{\"a}ki}, {Lamarre}, {Lasenby}, {Lattanzi}, {Lawrence}, {Leahy}, {Leonardi}, {Lesgourgues}, {Levrier}, {Lewis}, {Liguori}, {Lilje}, {Linden-V{\o}rnle}, {L{\'o}pez-Caniego}, {Lubin}, {Mac{\'\i}as-P{\'e}rez}, {Maggio}, {Maino}, {Mandolesi}, {Mangilli}, {Marchini}, {Maris}, {Martin}, {Martinelli}, {Mart{\'\i}nez-Gonz{\'a}lez}, {Masi}, {Matarrese}, {McGehee}, {Meinhold}, {Melchiorri}, {Melin}, {Mendes}, {Mennella}, {Migliaccio}, {Millea}, {Mitra}, {Miville-Desch{\^e}nes}, {Moneti}, {Montier}, {Morgante}, {Mortlock},
  {Moss}, {Munshi}, {Murphy}, {Naselsky}, {Nati}, {Natoli}, {Netterfield}, {N{\o}rgaard-Nielsen}, {Noviello}, {Novikov}, {Novikov}, {Oxborrow}, {Paci}, {Pagano}, {Pajot}, {Paladini}, {Paoletti}, {Partridge}, {Pasian}, {Patanchon}, {Pearson}, {Perdereau}, {Perotto}, {Perrotta}, {Pettorino}, {Piacentini}, {Piat}, {Pierpaoli}, {Pietrobon}, {Plaszczynski}, {Pointecouteau}, {Polenta}, {Popa}, {Pratt}, {Pr{\'e}zeau}, {Prunet}, {Puget}, {Rachen}, {Reach}, {Rebolo}, {Reinecke}, {Remazeilles}, {Renault}, {Renzi}, {Ristorcelli}, {Rocha}, {Rosset}, {Rossetti}, {Roudier}, {Rouill{\'e} d'Orfeuil}, {Rowan-Robinson}, {Rubi{\~n}o-Mart{\'\i}n}, {Rusholme}, {Said}, {Salvatelli}, {Salvati}, {Sandri}, {Santos}, {Savelainen}, {Savini}, {Scott}, {Seiffert}, {Serra}, {Shellard}, {Spencer}, {Spinelli}, {Stolyarov}, {Stompor}, {Sudiwala}, {Sunyaev}, {Sutton}, {Suur-Uski}, {Sygnet}, {Tauber}, {Terenzi}, {Toffolatti}, {Tomasi}, {Tristram}, {Trombetti}, {Tucci}, {Tuovinen}, {T{\"u}rler}, {Umana}, {Valenziano}, {Valiviita}, {Van Tent},
  {Vielva}, {Villa}, {Wade}, {Wandelt}, {Wehus}, {White}, {White}, {Wilkinson}, {Yvon}, {Zacchei}, \& {Zonca}}]{Planck15}
{Planck Collaboration}, {Ade}, P.~A.~R., {Aghanim}, N., {et~al.} 2016, \bibinfo{title}{{Planck 2015 results. XIII. Cosmological parameters},} \aap, 594, A13, \dodoi{10.1051/0004-6361/201525830}

\bibitem[{M. {Qezlou} {et~al.}(2022){Qezlou}, {Newman}, {Rudie}, \& {Bird}}]{Qezlou22}
{Qezlou}, M., {Newman}, A.~B., {Rudie}, G.~C., \& {Bird}, S. 2022, \bibinfo{title}{{Characterizing Protoclusters and Protogroups at z 2.5 Using Ly{\ensuremath{\alpha}} Tomography},} \apj, 930, 109, \dodoi{10.3847/1538-4357/ac6259}

\bibitem[{D.~D. {Shi} {et~al.}(2021){Shi}, {Cai}, {Fan}, {Zheng}, {Huang}, \& {Xu}}]{Shi21}
{Shi}, D.~D., {Cai}, Z., {Fan}, X., {et~al.} 2021, \bibinfo{title}{{Spectroscopic Confirmation of Two Extremely Massive Protoclusters, BOSS1244 and BOSS1542, at z = 2.24},} \apj, 915, 32, \dodoi{10.3847/1538-4357/abfec0}

\bibitem[{K. {Shi} {et~al.}(2019){Shi}, {Huang}, {Lee}, {Toshikawa}, {Bowen}, {Malavasi}, {Lemaux}, {Cucciati}, {Le Fevre}, \& {Dey}}]{Shi19}
{Shi}, K., {Huang}, Y., {Lee}, K.-S., {et~al.} 2019, \bibinfo{title}{{How Do Galaxies Trace a Large-scale Structure? A Case Study around a Massive Protocluster at Z = 3.13},} \apj, 879, 9, \dodoi{10.3847/1538-4357/ab2118}

\bibitem[{C.~W. {Stark} {et~al.}(2015){Stark}, {White}, {Lee}, \& {Hennawi}}]{Stark15}
{Stark}, C.~W., {White}, M., {Lee}, K.-G., \& {Hennawi}, J.~F. 2015, \bibinfo{title}{{Protocluster discovery in tomographic Ly {\ensuremath{\alpha}} forest flux maps},} \mnras, 453, 311, \dodoi{10.1093/mnras/stv1620}

\bibitem[{C.~C. {Steidel} {et~al.}(1998){Steidel}, {Adelberger}, {Dickinson}, {Giavalisco}, {Pettini}, \& {Kellogg}}]{Steidel98}
{Steidel}, C.~C., {Adelberger}, K.~L., {Dickinson}, M., {et~al.} 1998, \bibinfo{title}{{A Large Structure of Galaxies at Redshift Z approximately 3 and Its Cosmological Implications},} \apj, 492, 428, \dodoi{10.1086/305073}

\bibitem[{T. {Wang} {et~al.}(2016){Wang}, {Elbaz}, {Daddi}, {Finoguenov}, {Liu}, {Schreiber}, {Mart{\'\i}n}, {Strazzullo}, {Valentino}, {van der Burg}, {Zanella}, {Ciesla}, {Gobat}, {Le Brun}, {Pannella}, {Sargent}, {Shu}, {Tan}, {Cappelluti}, \& {Li}}]{Wang16}
{Wang}, T., {Elbaz}, D., {Daddi}, E., {et~al.} 2016, \bibinfo{title}{{Discovery of a Galaxy Cluster with a Violently Starbursting Core at z = 2.506},} \apj, 828, 56, \dodoi{10.3847/0004-637X/828/1/56}

\bibitem[{D. {Weinberg} \&  {et al.}(1999){Weinberg} \& {et al.}}]{Weinberg99}
{Weinberg}, D., \& {et al.} 1999, \bibinfo{title}{{Cosmological tests with the Ly-{\ensuremath{\alpha}} forest (invited review)},} in Evolution of Large Scale Structure : From Recombination to Garching, ed. A.~J. {Banday}, R.~K. {Sheth}, \& L.~N. {da Costa}, 346, \dodoi{10.48550/arXiv.astro-ph/9810142}

\bibitem[{J.~P. {Willis} {et~al.}(2020){Willis}, {Canning}, {Noordeh}, {Allen}, {King}, {Mantz}, {Morris}, {Stanford}, \& {Brammer}}]{Willis20}
{Willis}, J.~P., {Canning}, R.~E.~A., {Noordeh}, E.~S., {et~al.} 2020, \bibinfo{title}{{Spectroscopic confirmation of a mature galaxy cluster at a redshift of 2},} \nat, 577, 39, \dodoi{10.1038/s41586-019-1829-4}

\bibitem[{H. {Yajima} {et~al.}(2022){Yajima}, {Abe}, {Khochfar}, {Nagamine}, {Inoue}, {Kodama}, {Arata}, {Dalla Vecchia}, {Fukushima}, {Hashimoto}, {Kashikawa}, {Kubo}, {Li}, {Matsuda}, {Mawatari}, {Ouchi}, \& {Umehata}}]{Yajima22}
{Yajima}, H., {Abe}, M., {Khochfar}, S., {et~al.} 2022, \bibinfo{title}{{FOREVER22: galaxy formation in protocluster regions},} \mnras, 509, 4037, \dodoi{10.1093/mnras/stab3092}

\bibitem[{X.~Z. {Zheng} {et~al.}(2021){Zheng}, {Cai}, {An}, {Fan}, \& {Shi}}]{Zheng21}
{Zheng}, X.~Z., {Cai}, Z., {An}, F.~X., {Fan}, X., \& {Shi}, D.~D. 2021, \bibinfo{title}{{MAMMOTH: confirmation of two massive galaxy overdensities at z = 2.24 with H{\ensuremath{\alpha}} emitters},} \mnras, 500, 4354, \dodoi{10.1093/mnras/staa2882}

\end{thebibliography}
\bibliographystyle{aasjournalv7}

\clearpage

\begin{appendix}

\section{Robustness Tests}
\label{sec:appendix_robustness}

\begin{figure}
    \centering
    \includegraphics[width=0.5\linewidth]{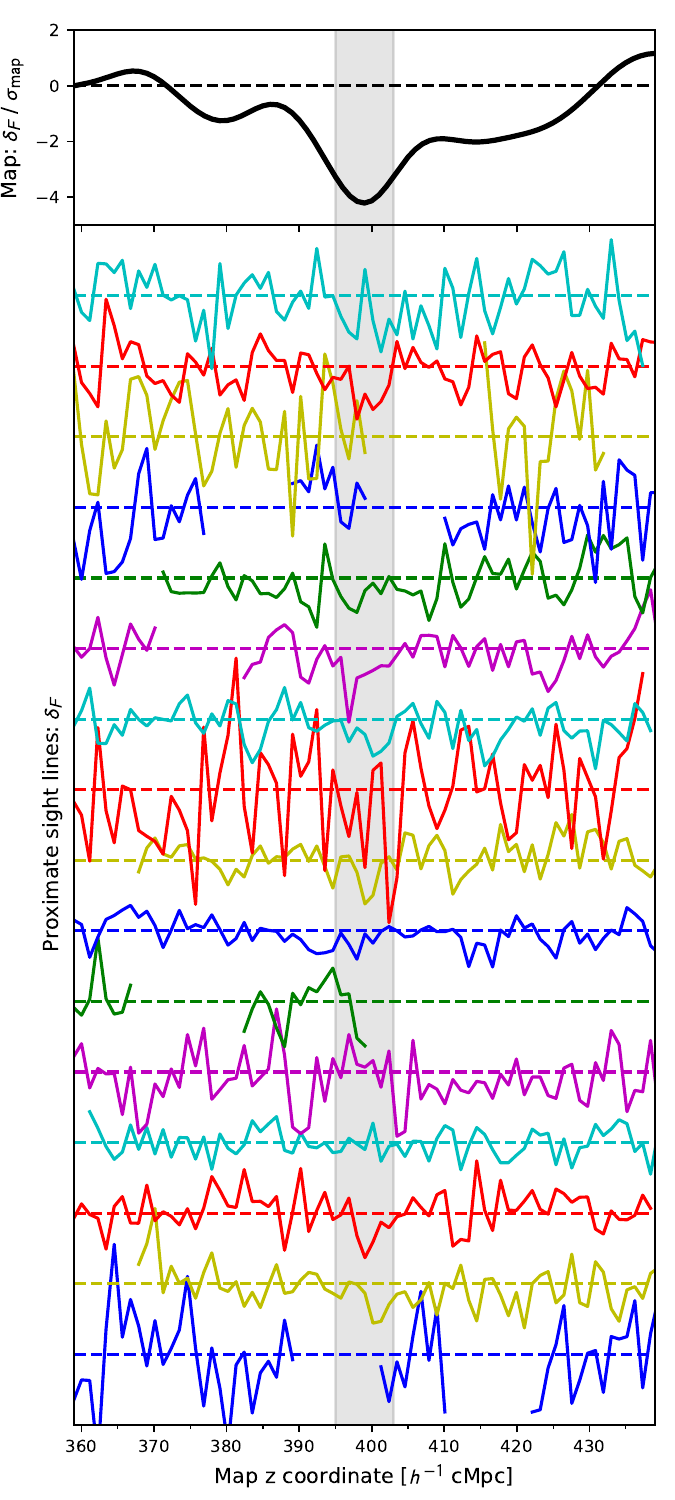}
    \caption{The top panel shows a skewer through the LATIS IGM map intersecting the Ly$\alpha$ absorption peak LATIS2-D2-00. The lower panel shows 16 individual sight lines close to this absorption peak, with an impact parameter $d_\perp < 5$ \cMpch. Each dashed line shows $\delta_F = 0$ for the spectrum with the corresponding color; dashed lines are offset from their neighbors by 2. The vertical band shows the region within $\pm4$~\cMpch~of the absorption peak. Note that 14 of 16 sight lines have greater than average absorption ($\langle \delta_F \rangle < 0$) in this band, demonstrating that the absorption is spatially coherent and not dominated by a single sight line.}
    \label{fig:sightlines}
\end{figure}

\begin{figure}
    \centering
    \includegraphics[width=0.5\linewidth]{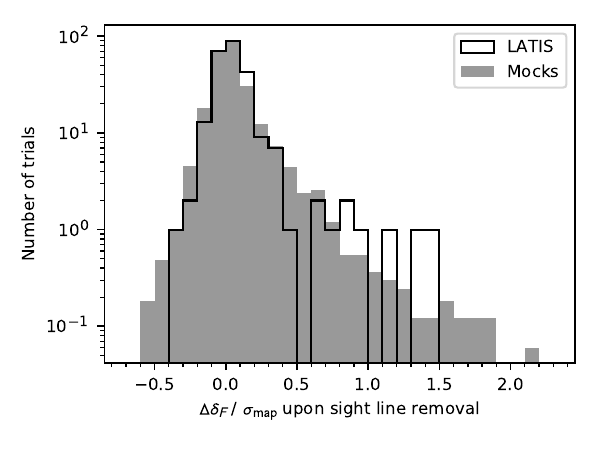}
    \caption{The sensitivity of strong absorption peaks ($\dFsm < -3.5$) to individual nearby sight lines ($d_\perp < 5$ \cMpch). The solid histogram shows the changes to the map absorption strength $\Delta \delta_F$ that result from the removal of a single such sight line. The solid histogram shows the distribution in 20 of the MDPL2-based mock surveys, normalized to match the total number of trials (i.e., close sight line removals) in the observed data set.}
    \label{fig:sightline_sensitivity}
\end{figure}

Previous studies have argued on statistical grounds that HCD lines, which we will consider as those with $N_{\rm HI} \gtrsim 10^{17.2}$~cm${}^{-2}$, are not a major contaminant to IGM tomographic maps in which many sight lines contribute to each resolution element \citep{Lee14A,Stark15,Newman22}. To further mitigate the influence of HCD lines, we identify and mask the strongest absorption lines associated with damped Ly$\alpha$ systems \citep[DLAs,][]{Newman24}. Still, individual lines at slightly lower column densities cannot be identified at the spectral resolution of LATIS, and even a fraction of DLAs may escape detection. Thus it remains important to investigate the sensitivity of Ly$\alpha$ absorption peaks, particularly the rarest and strongest, to individual sight lines.

Fig.~\ref{fig:sightlines} demonstrates, as an example, the sight lines near the strongest Ly$\alpha$ absorption peak in the COSMOS map, LATIS2-D2-00. The absorption is distributed across many sight lines: 14 of 16 close sight lines show greater than average absorption, i.e., an average $\langle \delta_F \rangle < 0$ within $|\Delta z| < 4$~\cMpch~of the peak. Thus the absorption is spatially coherent, as expected when its origin is diffuse IGM gas, and distinct from a map signal produced by an HCD system in a single sight line. (We note that while Fig.~\ref{fig:sightlines} illustrates the distributed nature of the absorption, the amount of absorption in the spectra cannot be straightforwardly compared to the map: the spectra and map are normalized differently, the Wiener filter weights the spectra nontrivially, and significant total weight is given to sight lines more distant than those plotted.)

Turning to a quantitative examination of the full set of strong Ly$\alpha$ absorption peaks, \citet{Newman22} showed that when the sight lines close to such peaks are considered, the observed distribution of the average $\langle \delta_F \rangle$ per sight line (evaluated near the redshift of the absorption peak) matched expectations based on mock surveys constructed within IllustrisTNG300. This analysis showed no population of proximate sight lines exhibiting unexpectedly strong absorption. In addition, explicitly excising the HCD lines from the TNG mocks did not appreciably change the $\langle \delta_F \rangle$ distribution. This indicated that the role of HCDs in producing the strong Ly$\alpha$ absorption peaks in the maps is expected to be negligible.

Here we extend the \citet{Newman22} analysis by considering the sensitivity of each of the strongest observed Ly$\alpha$ absorption peaks to each proximate sight line. We consider all 243 sight lines with a transverse separation $< 5$~\cMpch~from an absorption peak with $\dFsm < -3.5$. The number of such sight lines per absorption peak ranges from 8-26, with an average of 15. Therefore every peak is affected by an appreciable number of sight lines, which mitigates the effect of individual ones. Visual inspection indeed shows coherent widespread absorption across many sight lines, which is illustrated for one example in Fig.~\ref{fig:sightlines}. 

We next reconstruct the tomographic maps by removing each of the 243 close sight lines individually, and we compute the change $\Delta \delta_F$ at the position of the absorption peak. We then repeat this analysis in the MDPL2 mock surveys, which lack HCD lines, to understand the distribution of $\Delta \delta_F$ that is expected from noise alone (i.e., noise in the spectra and finite sampling, but without HCD contamination). Fig.~\ref{fig:sightline_sensitivity} shows that the sensitivity of strong absorption peaks to individual sight lines agrees overall with the mock surveys and is thus well accounted for by known sources of noise. We do find two sight lines whose effect is rather strong, deepening the map absorption by 1.3-1.5~$\sigma_{\rm map}$. Their presence is not very unusual: two or more sight lines have a comparably strong effect in about 10\% of the mock surveys. These two sight lines are near LATIS2-D2-02 and LATIS2-D4-00. Both of these are associated with strong galaxy overdensities, so there is no reason to doubt their reality, although the amount of Ly$\alpha$ absorption may be somewhat more uncertain in these cases.

\section{Maps of the Strongest IGM Absorption Peaks}
\label{sec:appendix_maps}

In Figures~\ref{fig:cubes1}-\ref{fig:cubes4}, we show maps of the 16 strongest Ly$\alpha$ absorption peaks with $\dFsm < -3.5$.

\begin{figure*}
    \centering
    \includegraphics[width=0.48\linewidth]{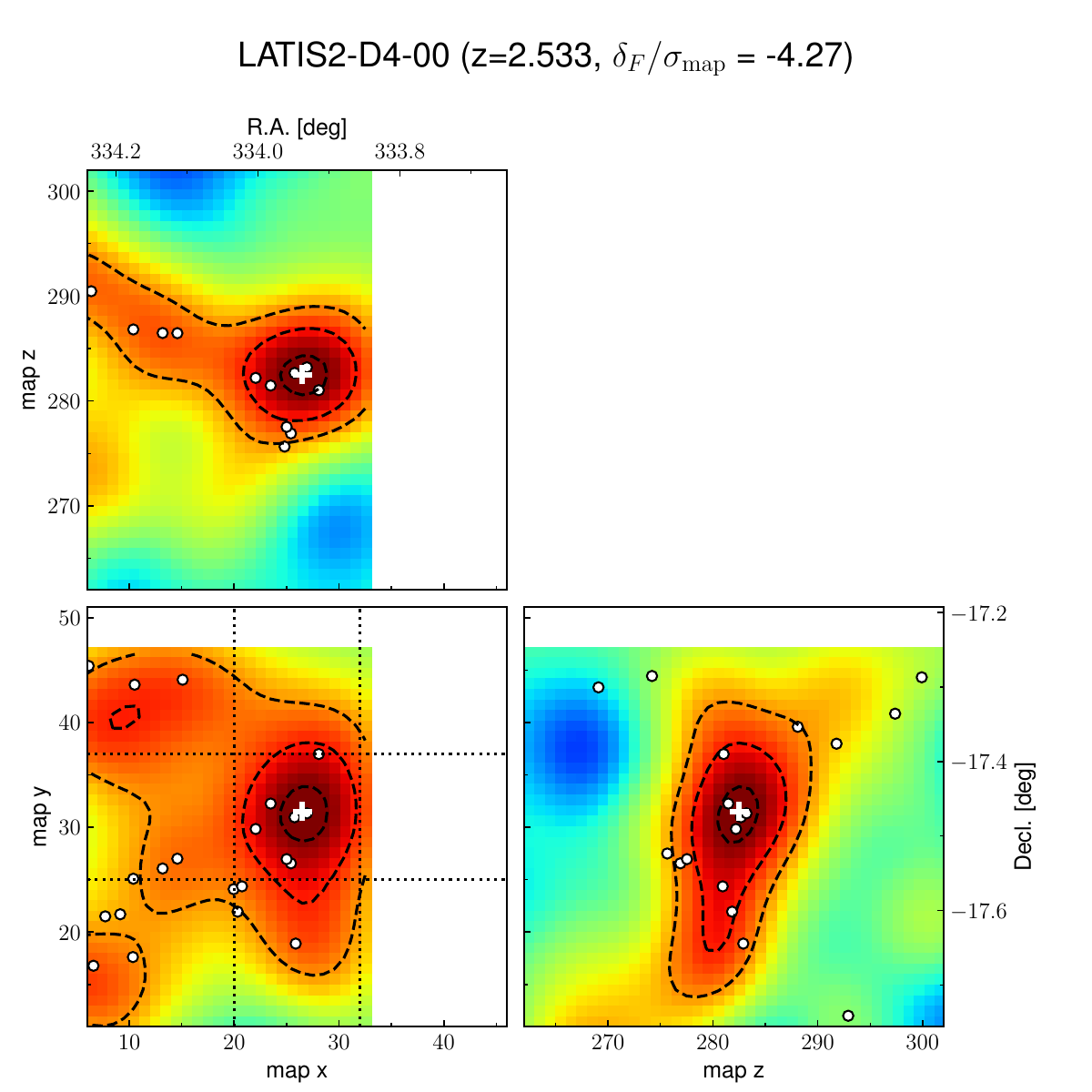} \includegraphics[width=0.48\linewidth]{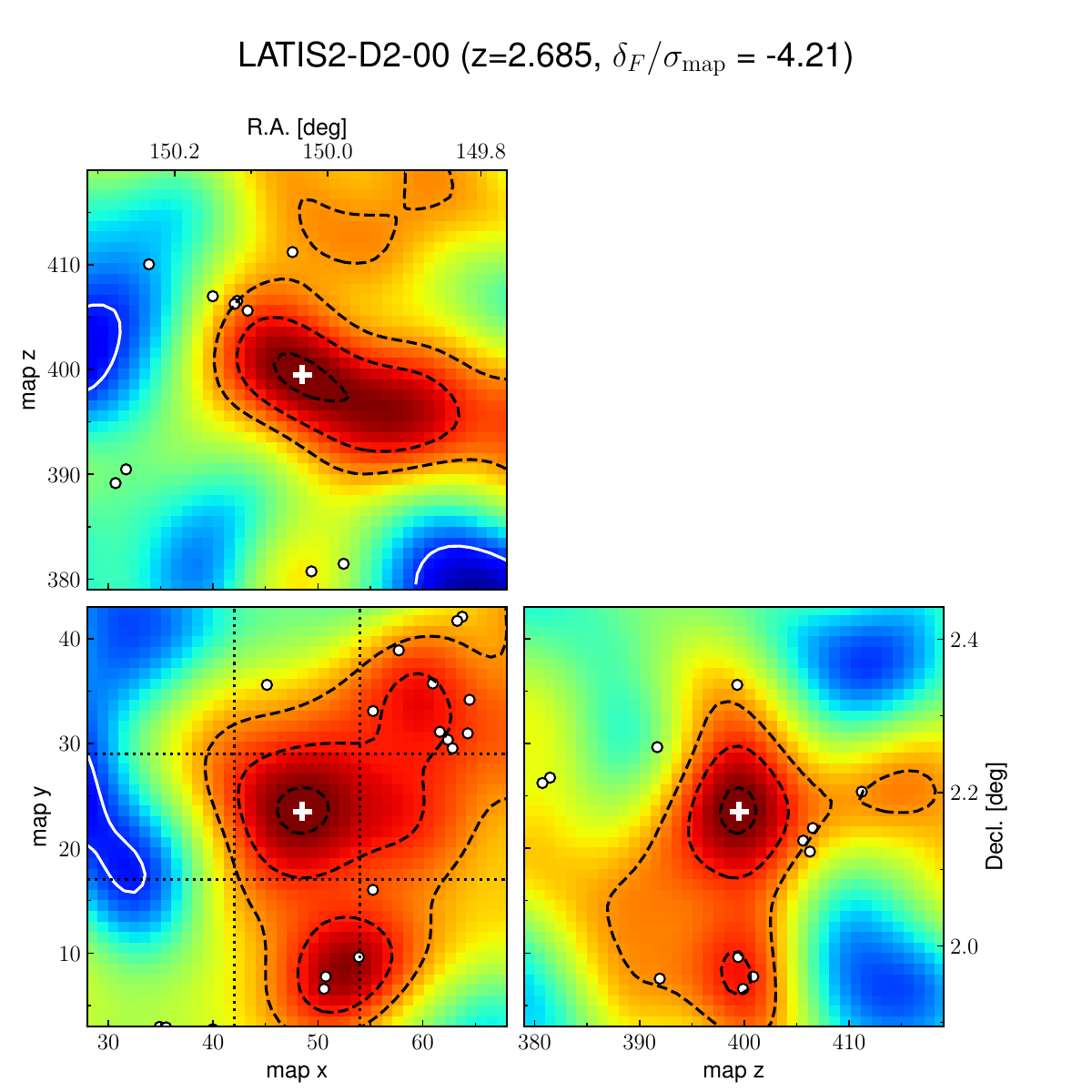} \\
    \includegraphics[width=0.48\linewidth]{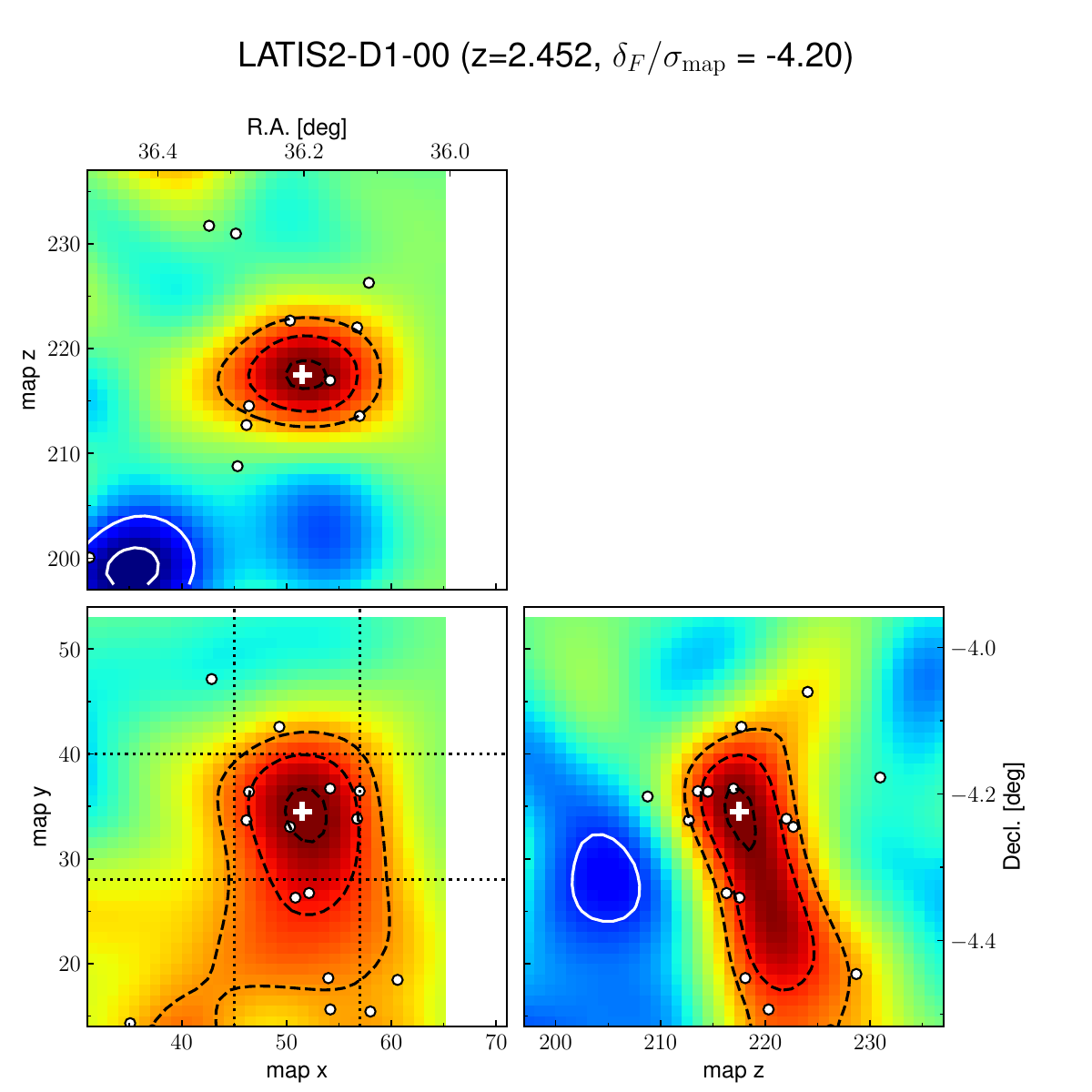} \includegraphics[width=0.48\linewidth]{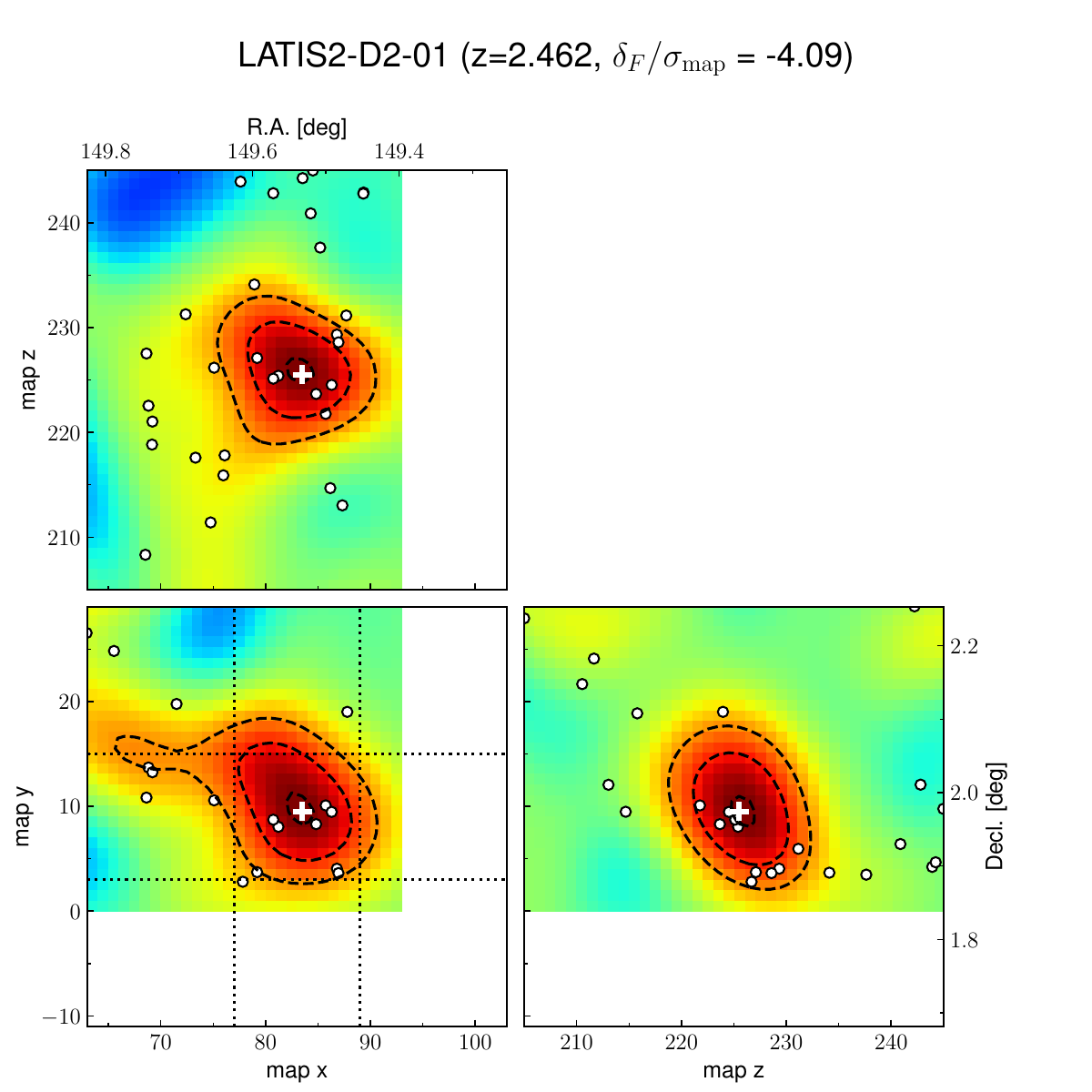} \\
    \includegraphics[width=0.48\linewidth]{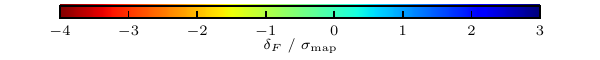} 
    \caption{Visualizations of the IGM maps and galaxy distribution around the four strongest Ly$\alpha$ absorption peaks. For each peak, three cross sections of the IGM map through the peak are shown following the $\dFsm$ colorbar shown at the bottom. Dashed contours show absorption levels $\dFsm = -2$, $-3$, $\ldots$, while solid contours indicate more transparent regions with $\dFsm = +2$, $+3$, $\ldots$. Points show the positions of LATIS LBGs that lie within 6~\cMpch~of each plane; dotted lines in the $xy$ projection indicate this distance from the absorption peak, which is indicated by a white cross. Map coordinates are in \cMpch.}
    \label{fig:cubes1}
\end{figure*}

\begin{figure*}
    \vspace{0.3in}
    \centering
    \includegraphics[width=0.48\linewidth]{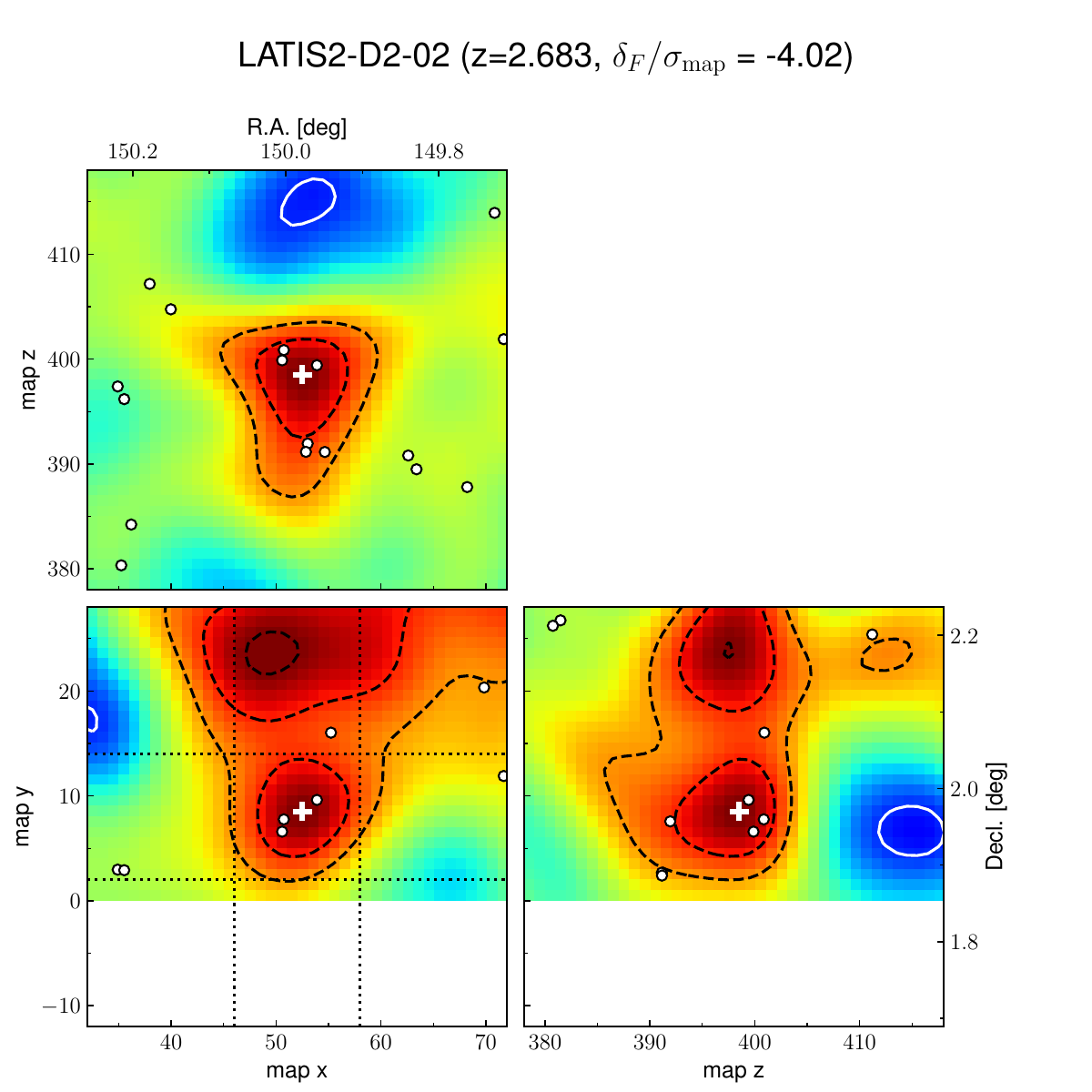} \includegraphics[width=0.48\linewidth]{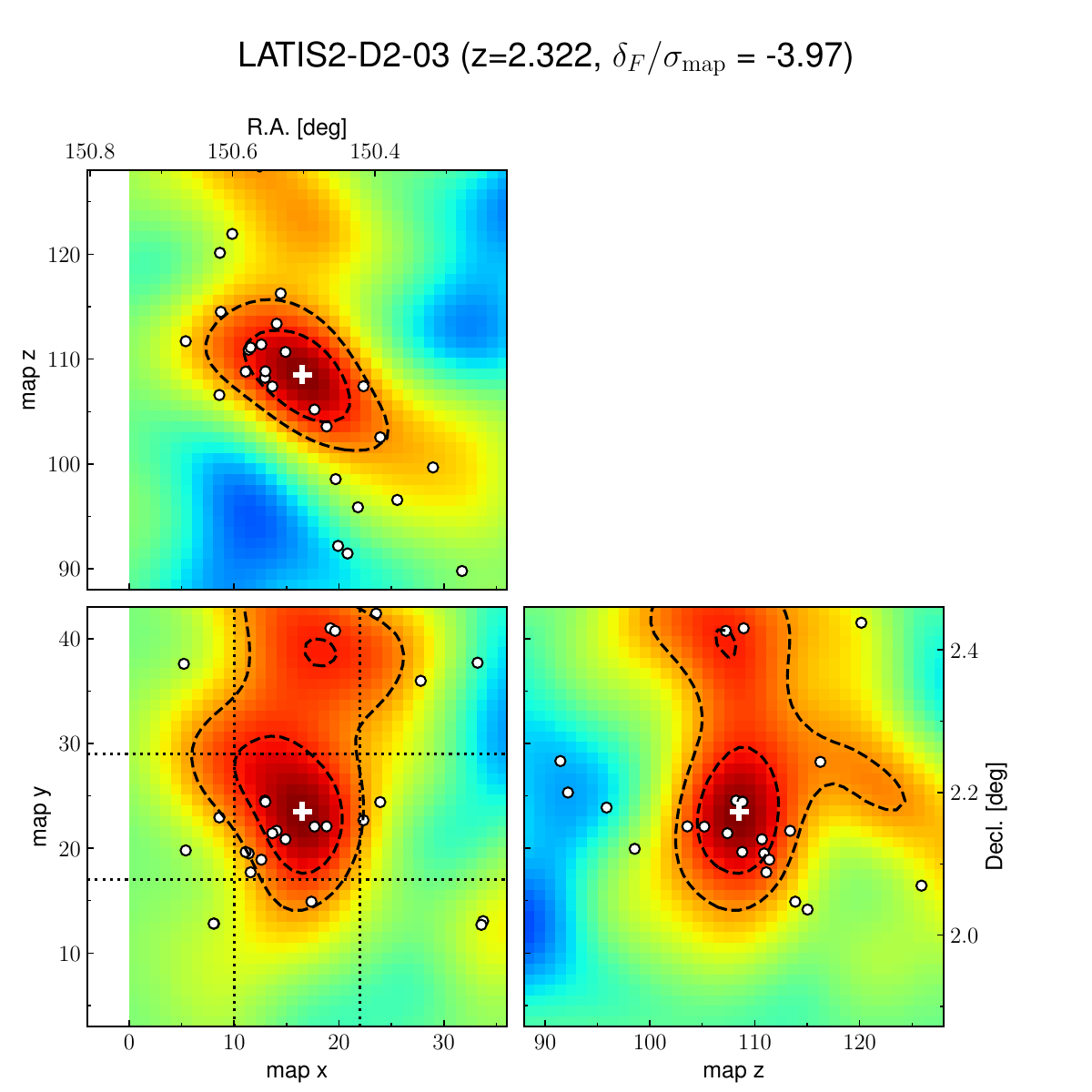} \\
    \includegraphics[width=0.48\linewidth]{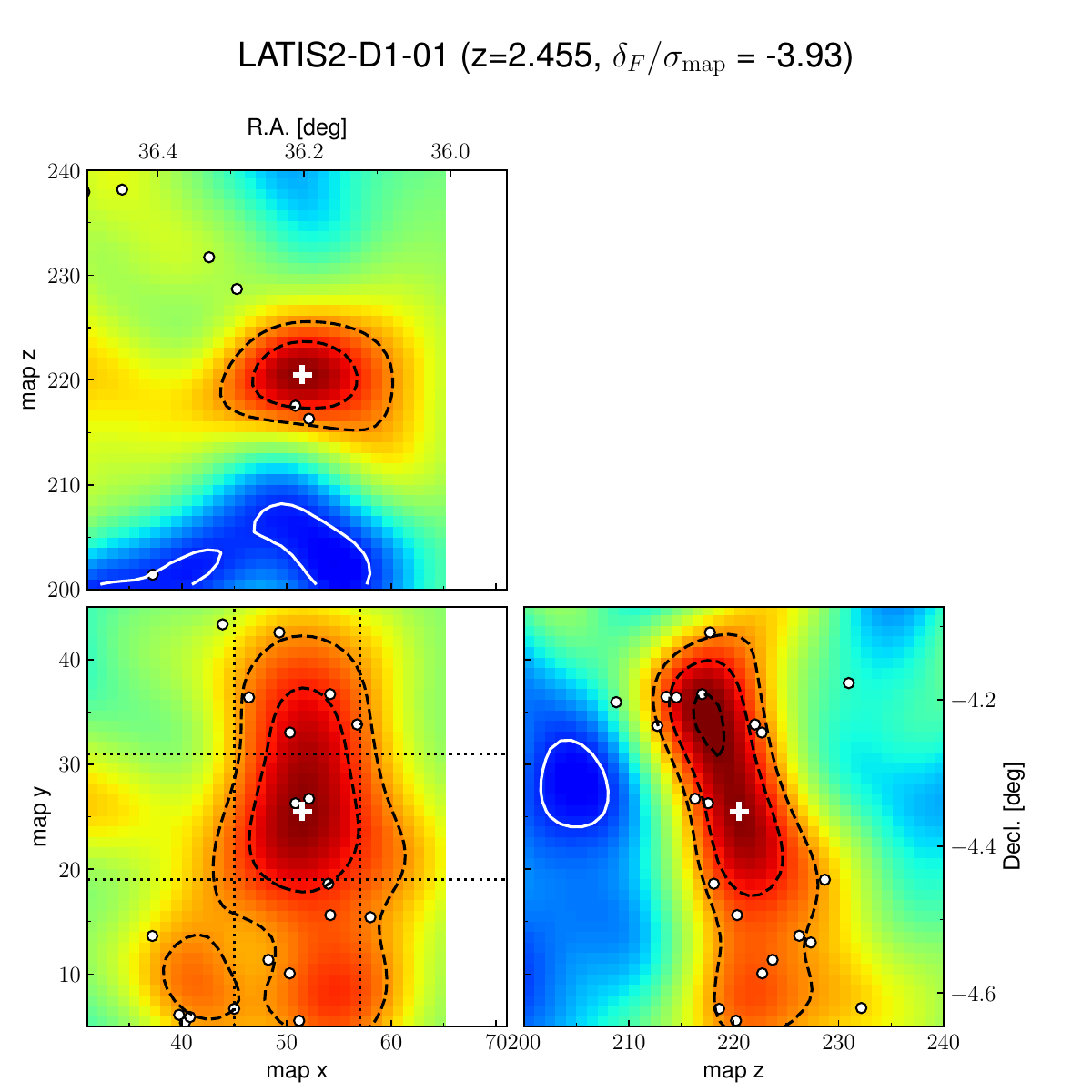} \includegraphics[width=0.48\linewidth]{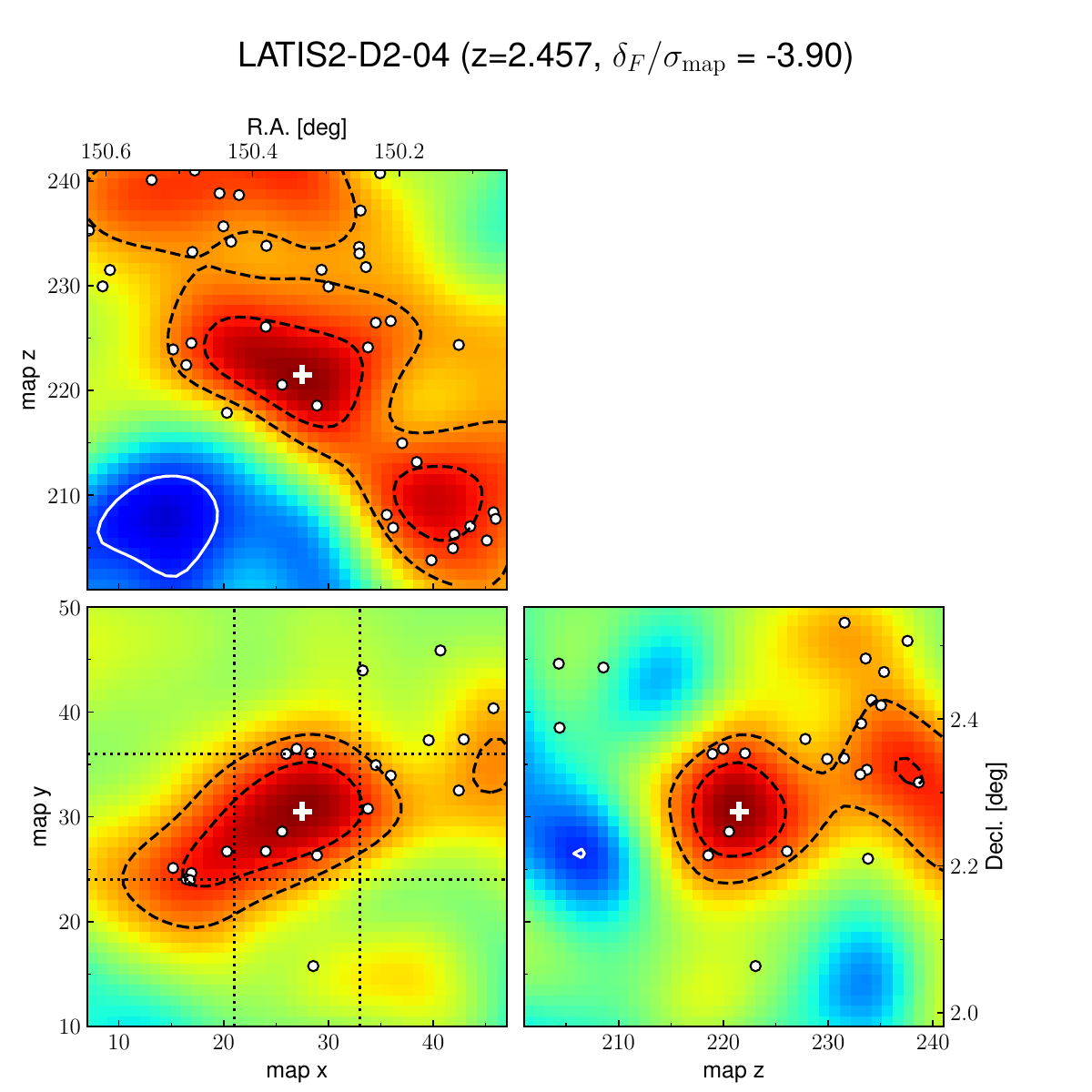} \\
    \includegraphics[width=0.48\linewidth]{minima_dF_colorbar.pdf} 
    \caption{Continuation of Fig.~\ref{fig:cubes1} for the next four strongest Ly$\alpha$ absorption peaks.}
    \label{fig:cubes2}
\end{figure*}

\begin{figure*}
    \vspace{0.3in}
    \centering
    \includegraphics[width=0.48\linewidth]{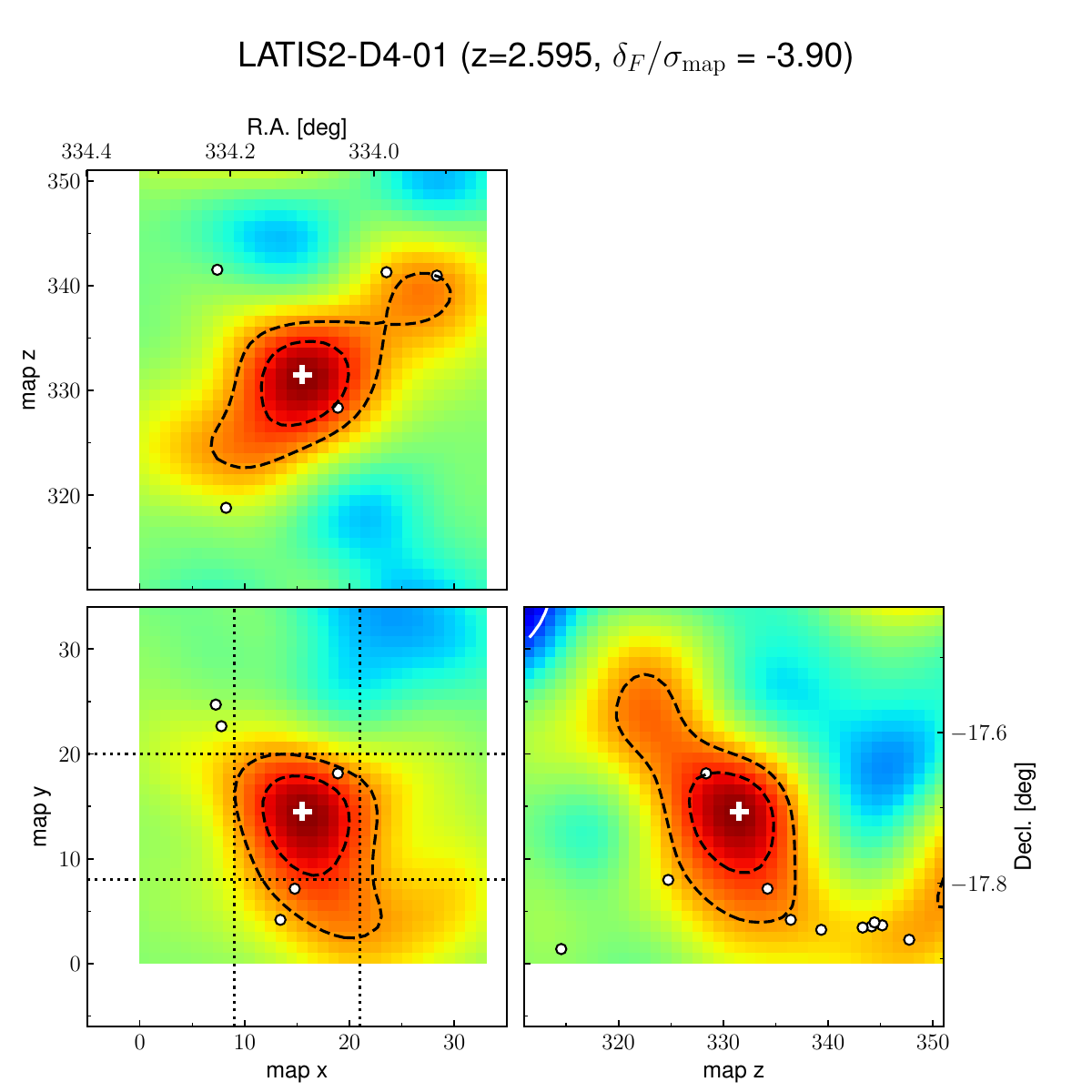} \includegraphics[width=0.48\linewidth]{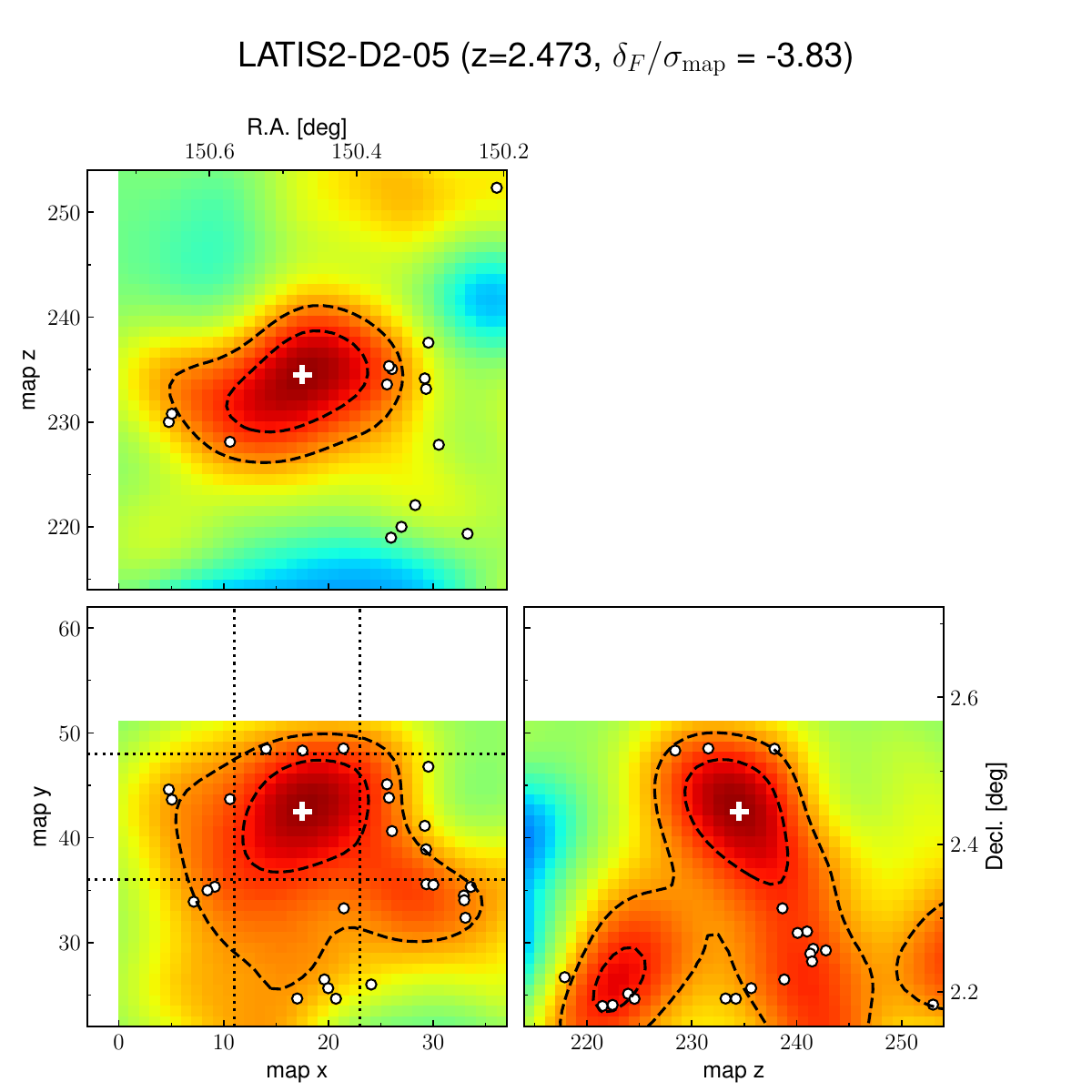} \\
    \includegraphics[width=0.48\linewidth]{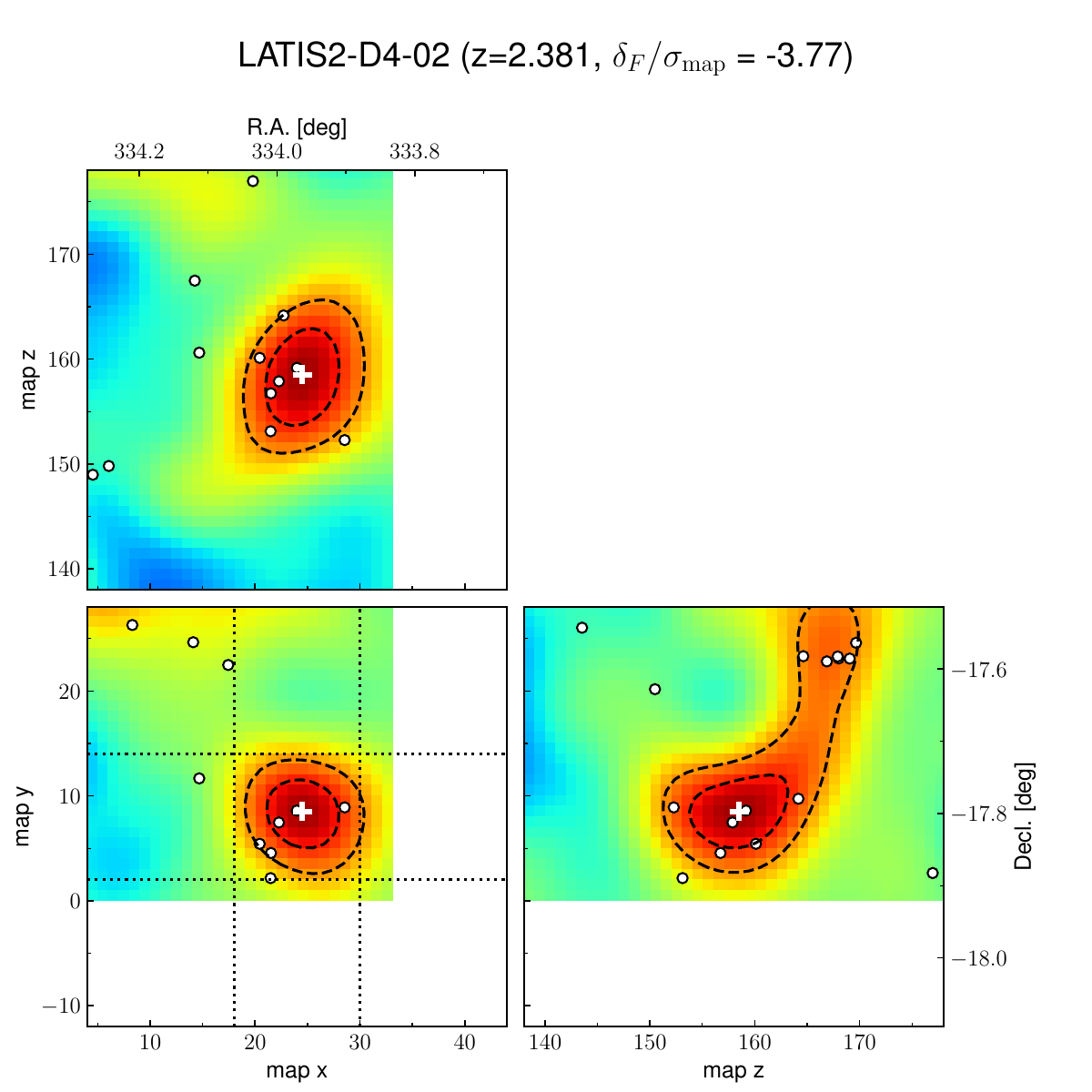} \includegraphics[width=0.48\linewidth]{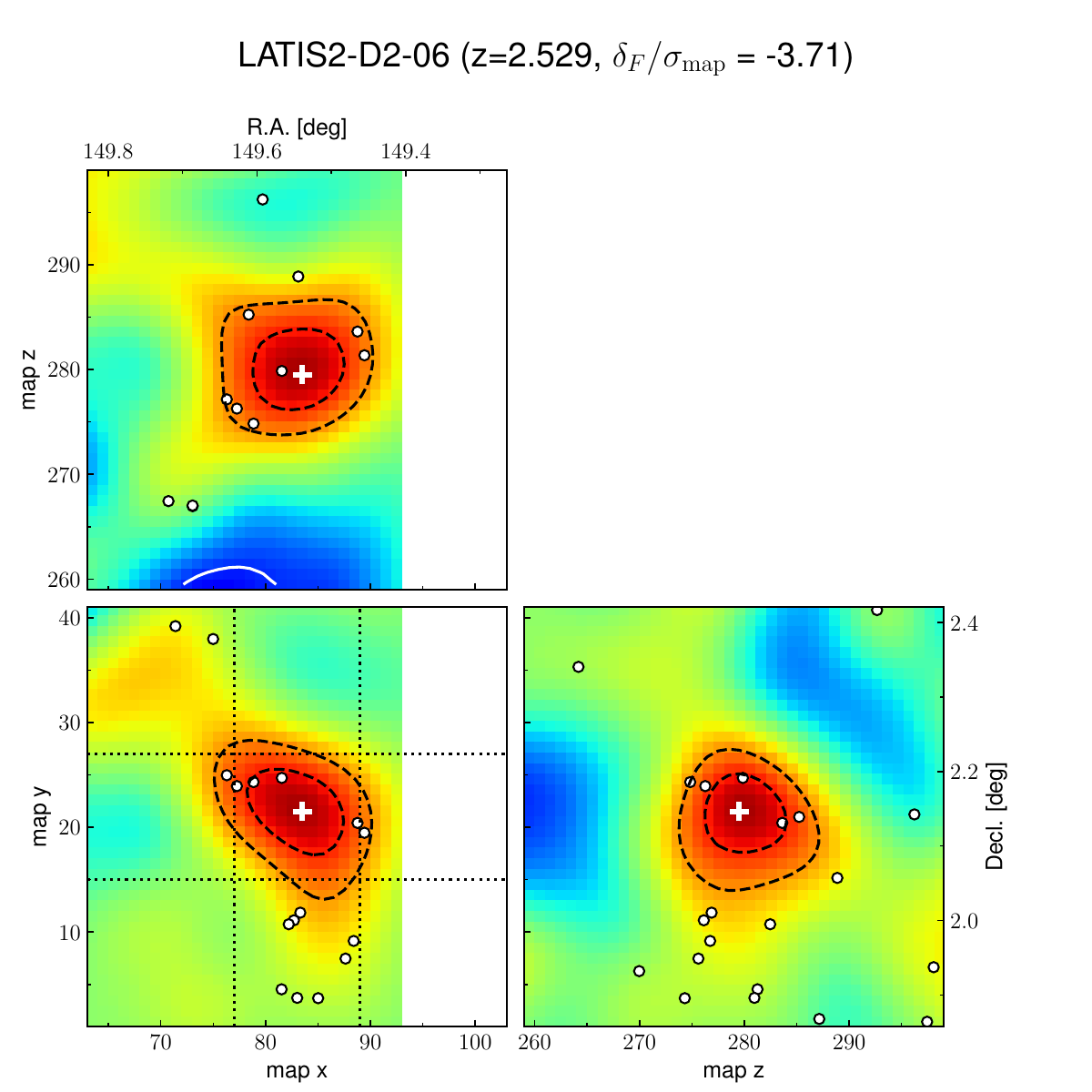} \\
    \includegraphics[width=0.48\linewidth]{minima_dF_colorbar.pdf} 
    \caption{Continuation of Fig.~\ref{fig:cubes2} for the next four strongest Ly$\alpha$ absorption peaks.}
    \label{fig:cubes3}
\end{figure*}

\begin{figure*}
    \vspace{0.3in}
    \centering
    \includegraphics[width=0.48\linewidth]{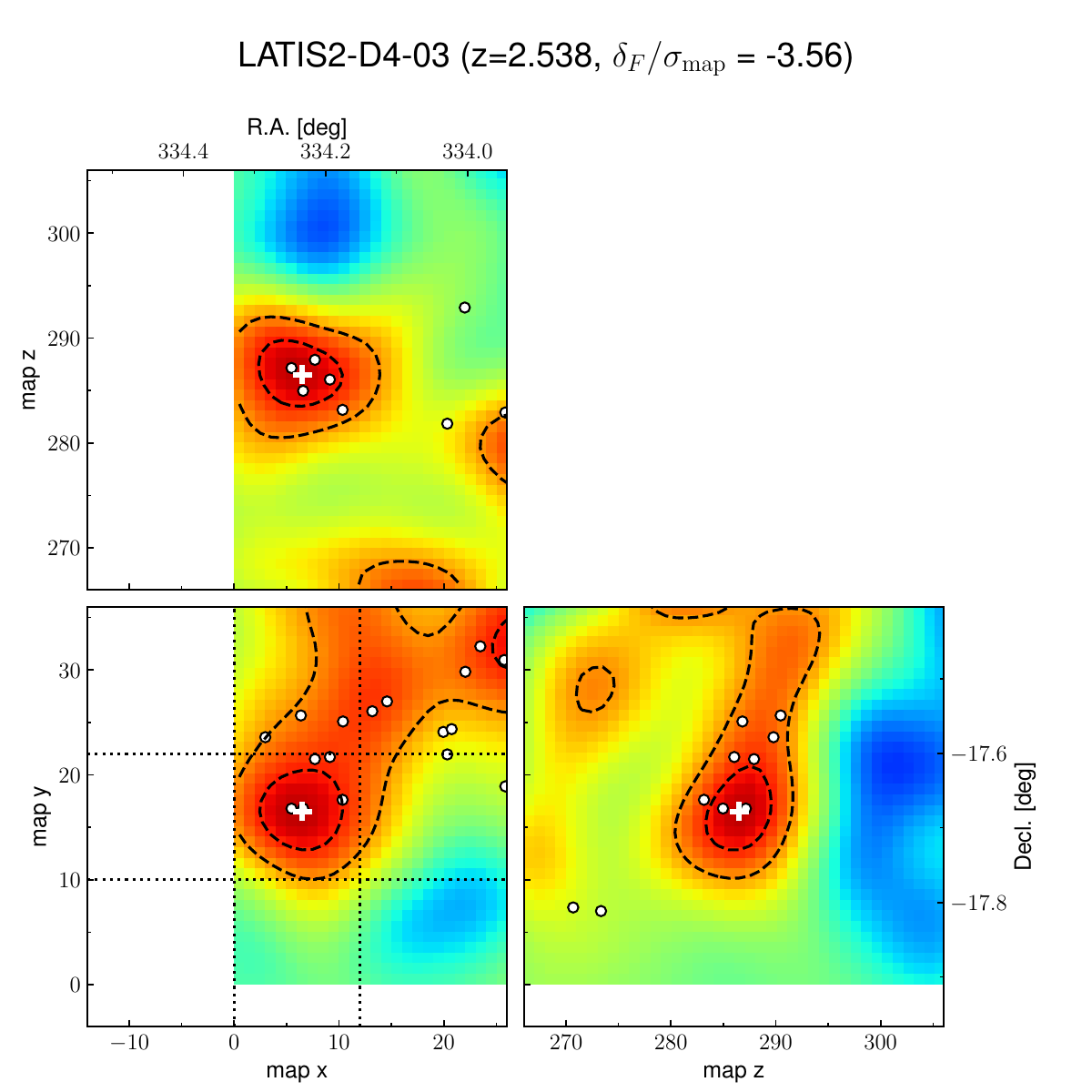} \includegraphics[width=0.48\linewidth]{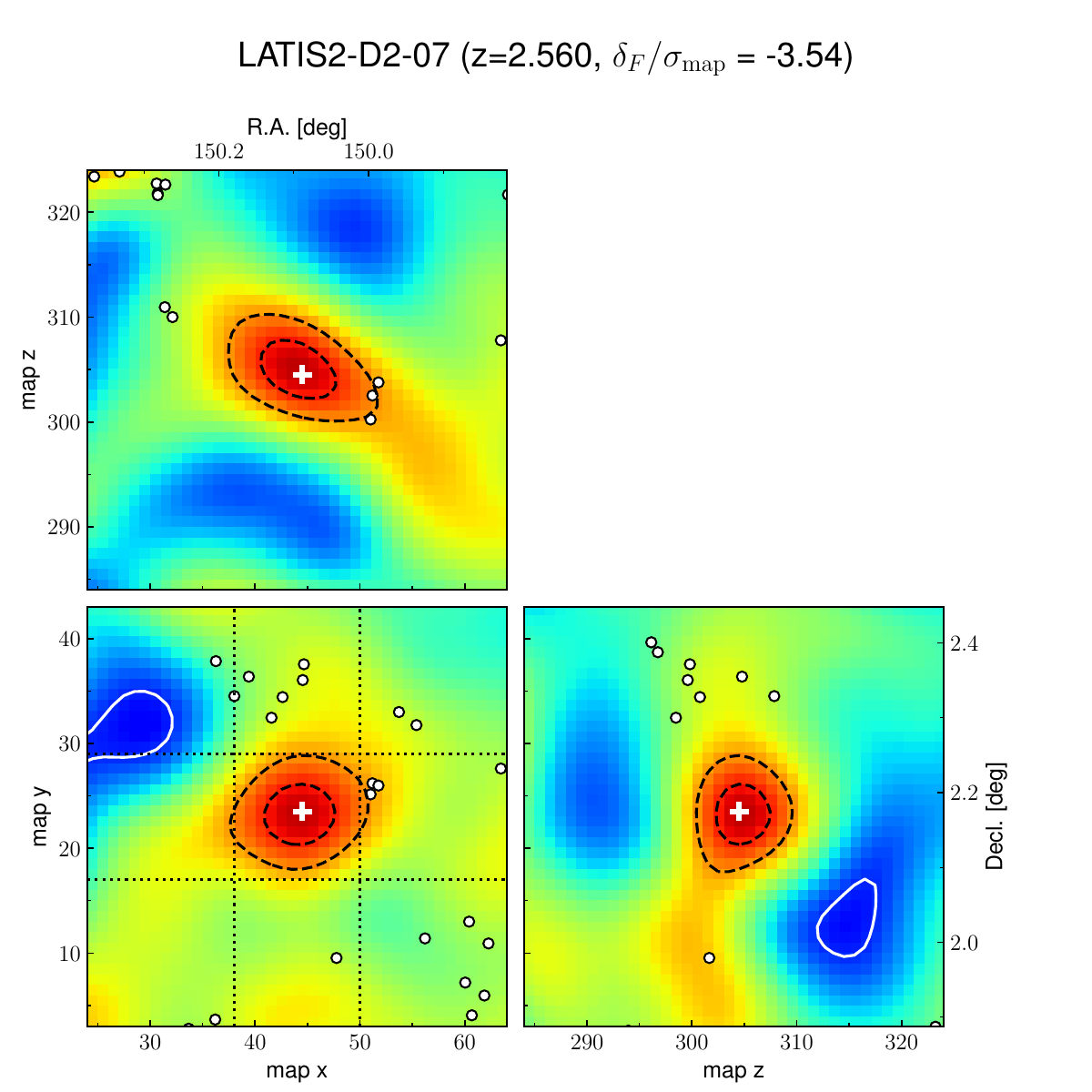} \\
    \includegraphics[width=0.48\linewidth]{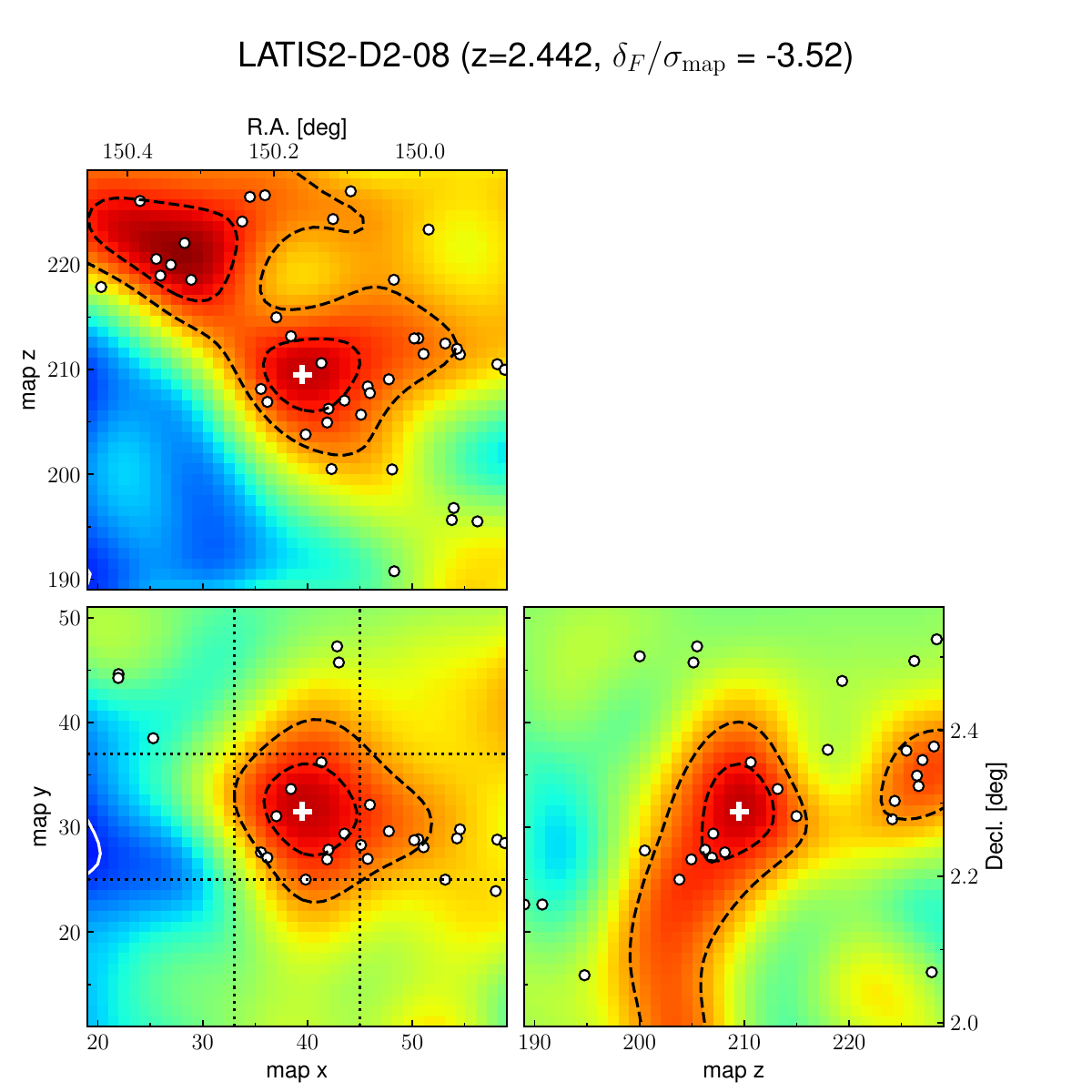} \includegraphics[width=0.48\linewidth]{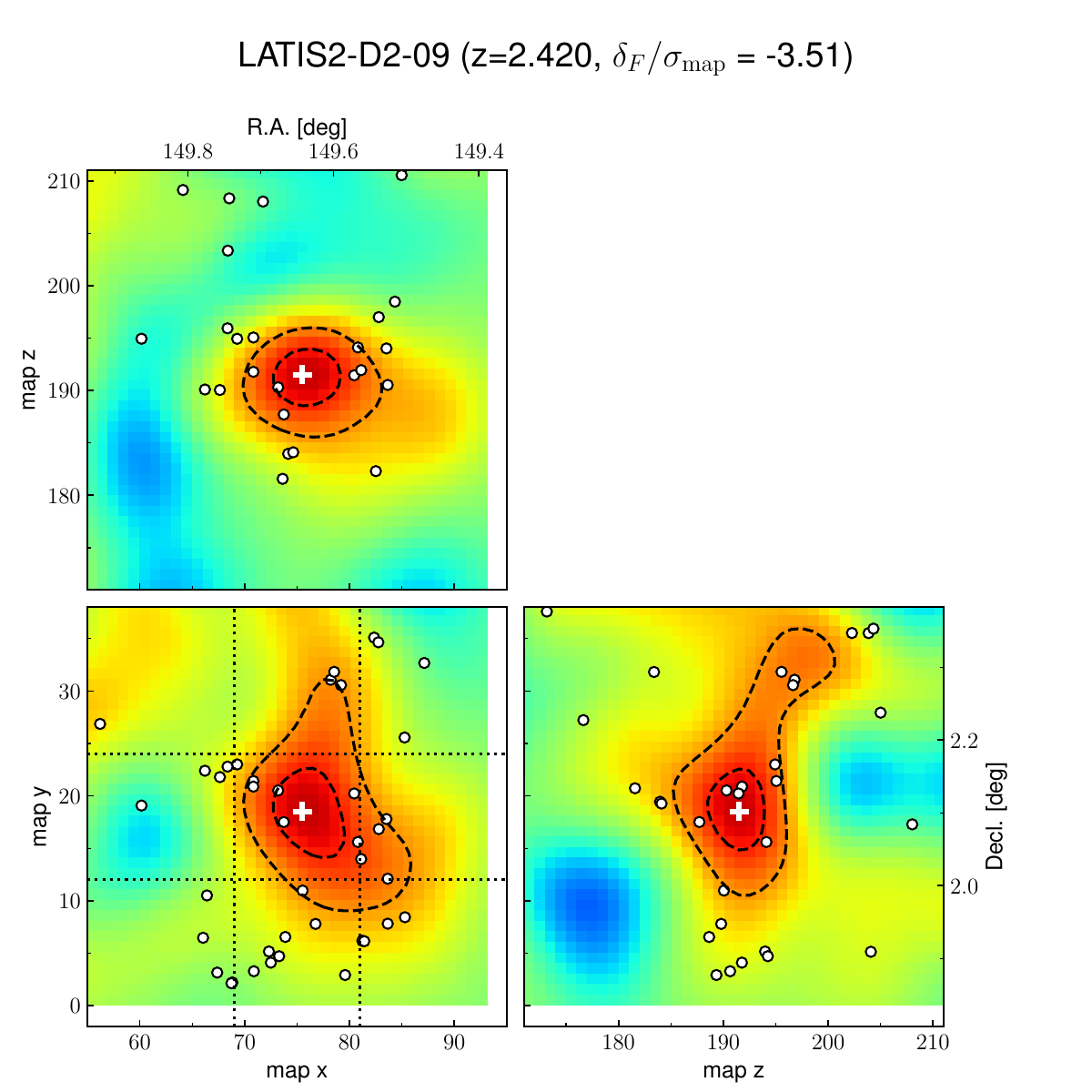} \\
    \includegraphics[width=0.48\linewidth]{minima_dF_colorbar.pdf} 
    \caption{Continuation of Fig.~\ref{fig:cubes3} for the next four strongest Ly$\alpha$ absorption peaks.}
    \label{fig:cubes4}
\end{figure*}

\end{appendix}
\end{document}